\documentclass{aa}  

\usepackage{graphicx}
\usepackage{lscape}
\usepackage{subfigure}
\usepackage{natbib}
\usepackage[varg]{txfonts}
\usepackage{longtable}
\usepackage{booktabs}
\usepackage[normalem]{ulem} 
\usepackage{epstopdf}
\usepackage{xcolor}
\usepackage{soul}
\usepackage{multirow}
\usepackage{amsmath}

\usepackage{xspace}
\usepackage{amssymb}
\usepackage{adjustbox}
\usepackage{arydshln}

\usepackage{scalerel}
\usepackage{tikz}
\usetikzlibrary{svg.path}

\definecolor{orcidlogocol}{HTML}{A6CE39}
\tikzset{
  orcidlogo/.pic={
    \fill[orcidlogocol] svg{M256,128c0,70.7-57.3,128-128,128C57.3,256,0,198.7,0,128C0,57.3,57.3,0,128,0C198.7,0,256,57.3,256,128z};
    \fill[white] svg{M86.3,186.2H70.9V79.1h15.4v48.4V186.2z}
                 svg{M108.9,79.1h41.6c39.6,0,57,28.3,57,53.6c0,27.5-21.5,53.6-56.8,53.6h-41.8V79.1z M124.3,172.4h24.5c34.9,0,42.9-26.5,42.9-39.7c0-21.5-13.7-39.7-43.7-39.7h-23.7V172.4z}
                 svg{M88.7,56.8c0,5.5-4.5,10.1-10.1,10.1c-5.6,0-10.1-4.6-10.1-10.1c0-5.6,4.5-10.1,10.1-10.1C84.2,46.7,88.7,51.3,88.7,56.8z};
  }
}

\newcommand\orcidicon[1]{\href{https://orcid.org/#1}{\mbox{\scalerel*{
\begin{tikzpicture}[yscale=-1,transform shape]
\pic{orcidlogo};
\end{tikzpicture}
}{|}}}}

\usepackage{hyperref} 
\hypersetup{
    colorlinks=true,
    citecolor=blue,
    linkcolor=blue,
    urlcolor=blue,
    }

\newcommand{\ULISSE}{\texttt{ULISSE}\xspace}

\newcommand{\perc}{\,\%\xspace}
\newcommand{\comment}[1]{}
\makeatletter
\renewcommand*\aa@pageof{, page \thepage{} of \pageref*{LastPage}}
\makeatother

\defcitealias{Doorenbos2022}{Paper~I}

\begin{document} 

\title{\ULISSE: Determination of star-formation rate and stellar mass based on the one-shot galaxy imaging technique}

   \author{
   Olena Torbaniuk\inst{1,2,\orcidicon{0000-0003-4465-2564}}, 
   Lars Doorenbos\inst{3,5,\orcidicon{0000-0002-0231-9950}}, 
   Maurizio Paolillo\inst{4,5,6,\orcidicon{0000-0003-4210-7693}},
   Stefano Cavuoti\inst{5,6,\orcidicon{0000-0002-3787-4196}}, \\
   Massimo Brescia\inst{4,5,6,\orcidicon{0000-0001-9506-5680}},
   Giuseppe Longo\inst{4,\orcidicon{0000-0002-9182-8414}} 
   }

\titlerunning{-- \ULISSE: SFR and stellar mass estimations from galaxy images} 
\authorrunning{O.~Torbaniuk et al. 2025}

   \institute{
    Department of Physics and Astronomy  `Augusto Righi', University of Bologna, Via Gobetti 93/2, I-40129 Bologna, Italy\\
    \email{olena.torbaniuk@gmail.com}
    \and
    INAF -- Osservatorio di Astrofisica e Scienza dello Spazio di Bologna, Via Gobetti 101, I-40129 Bologna, Italy
    \and
    AIMI, ARTORG Center, University of Bern, Murtenstrasse 50, CH-3008 Bern, Switzerland
    \and
    Department of Physics `Ettore Pancini', University Federico II, Strada Vicinale Cupa Cintia, 21, 80126 Napoli, Italy
    \and
    INAF -- Astronomical Observatory of Capodimonte, Salita Moiariello 16, I-80131 Napoli, Italy
    \and 
    INFN -- Sezione di Napoli, via Cinthia 9, 80126 Napoli, Italy
             }

   \date{Received XXXX xx, 2024; accepted XXXX xx, 2024}

 
  \abstract
{Modern sky surveys produce vast amounts of observational data, making the application of classical methods for estimating galaxy properties challenging and time-consuming. This challenge can be significantly alleviated by employing automatic machine and deep learning techniques. } 
{We propose an implementation of the \ULISSE algorithm aimed at determining physical parameters of galaxies, in particular star-formation rates (SFR) and stellar masses ($\mathcal{M}_{\ast}$), using only composite-color images.}
{\ULISSE is able to rapidly and efficiently identify candidates from a single image based on photometric and morphological similarities to a given reference object with known properties. This approach leverages features extracted from the {\tt ImageNet} dataset to perform similarity searches among all objects in the sample, eliminating the need for extensive neural network training.}
{Our experiments, performed on the Sloan Digital Sky Survey, demonstrate that we are able to predict the joint star formation rate and stellar mass of the target galaxies within 1\,dex in 60\perc to 80\perc of cases, depending on the investigated subsample (quiescent/star-forming galaxies, early-/late-type, etc.), and within 0.5\,dex if we consider these parameters separately. This is approximately twice the fraction obtained from a random guess extracted from the parent population. Additionally, we find \ULISSE is more effective for galaxies with active star formation compared to elliptical galaxies with quenched star formation. Additionally, \ULISSE performs more efficiently for galaxies with bright nuclei such as AGN.}
{Our results suggest that \ULISSE is a promising tool for a preliminary estimation of star-formation rates and stellar masses for galaxies based only on single images in current and future wide-field surveys (e.g., \textit{Euclid, LSST}), which target millions of sources nightly.}

   \keywords{Catalogs -- Methods: statistical --
                Galaxies: star formation -- Galaxies: spiral -- Galaxies: elliptical and lenticular, cD -- Techniques: image processing
               }

\maketitle

\section{Introduction}

Over the past two decades, our understanding of the Universe has significantly enhanced by exploring vast and deep areas of the sky through multi-wavelength digital imaging surveys such as the Sloan Digital Sky Survey ({\it SDSS}, \citealt{York2000}), the Kilo Degree Survey ({\it KiDS}, \citealt{deJong2015}), the Panoramic Survey Telescope and Rapid Response System ({\it Pan-STARRS}, \citealt{Magnier2020}), the Dark Energy Survey ({\it DES}, \citealt{Abbott2016}), and the Hyper Suprime-Cam Subaru Strategic Program ({\it HSC SSP}, \citealt{Aihara2019}). Looking ahead, new multiband wide-field surveys and projects carried out on such telescopes as the Vera C. Rubin Observatory Large Synoptic Survey Telescope ({\it Rubin-LSST}, \citealt{Ivezic2019}), {\it Euclid} (\citealt{scaramella2022euclid, Mellier24Euclid}), the Nancy Grace Roman Space Telescope (formerly WFIRST, \citealt{Green2012}), and the James Webb Space Telescope ({\it JWST}, \citealt{AlvarezMarquez2019}) are set to further increase the volume of observational data by orders of magnitude. These forthcoming surveys will produce photometric data for millions of sources each night. Given the impracticality of spectroscopic follow-ups for even a small fraction of the sources in these surveys, there is a need for algorithms that can leverage photometric information to either detect, classify, or measure the physical properties (redshift, masses, star formation rate, etc.) of sources in such surveys or at least identify candidates for further investigations. In response to this challenge, there has been a concerted effort in recent years to develop and refine fast, self-adaptive learning methods for data prediction, classification, and visualization. This has led to the adoption of astroinformatics solutions, particularly machine, and deep learning paradigms \citep{Baron2019, Longo2019, Fluke2020, Lecun1998, Disanto2018, Schaefer2018}, which are trying to replace or strengthen the classical methods, which are less efficient for large-scale samples.

Machine learning algorithms can be divided into two main categories: supervised and unsupervised. {\it Supervised} methods, which rely on labelled data for training, are more commonly used due to their ease of interpretation and ability to be tailored to specific problems (e.g., \citealt{Kim2011, Brescia2013, Disanto2018, Kinson2021, Wenzl2021}. {\it Unsupervised} methods, on the other hand, analyze data without prior labels, using labels only for post-analysis, and thus, are less common but have been successfully applied in astrophysics (e.g., \citealt{Baron2017, CastroGinard2018, Razim2021}. In addition, {\it supervised} machine learning requires a training dataset from the real observations or simulations, which often lack the comprehensive coverage of the parameter space and, thus, do not provide a full picture of the physics lying behind (especially for rare and poorly studied objects, see \citealt{Masters2015}). The latter can be solved using {\it unsupervised} methods in the way of the clustering and pre-clustering approaches \citep{Bishop2006}.

As an intermediate approach combining the best from supervised and unsupervised machine learning methods {\it one-shot learning} has been developed \citep{wang2020generalizing}, which uses only a single-labelled sample per the studied class of objects. It allows us to eliminate not only the necessity of the extensive labelled datasets but also address the challenge caused by the presence of rare and under-sampled objects.

Using this approach, we present \ULISSE (aUtomatic Lightweight Intelligent System for Sky Exploration), a one-shot method designed to select objects closely related to a given input by directly applying it to multi-band images. For this, it transforms the image of a given source (which can be called a prototype or a target object) into a set of representative features, which are then used to search for other objects in the feature space sharing similar physical properties to the target one.  A key strength of this approach lies in its flexibility and minimal input requirements: \ULISSE operates purely in image space and does not require photometry, redshifts, or spectroscopic information for the input objects. Instead, it transfers physical properties from a well-characterized reference sample via image-based visual similarity. The power of such an approach is its ability to provide a reliable list of objects possessing similar properties (i.e. neighbours), even in the case of rare and peculiar target objects, which allows the elimination of the need for a large and well-labelled training set essential for supervised methods.  This relative matching strategy enables robust performance even when detailed measurements are unavailable for most sources, making \ULISSE particularly well-suited for large-scale surveys and imaging datasets.

In our previous work \citealt{Doorenbos2022} (hereinafter \citetalias{Doorenbos2022}), we presented the effectiveness of this method for active galactic nucleus (AGN) candidate detection. As one of the results, we observed a correlation between \ULISSE output and galaxy morphology, pointing to the fact that our method seems sensitive enough to distinguish the physical parameters of the studied galaxies. Building on this, we aim to extend the application of our method to probe different galaxy properties, which has the potential to be particularly valuable for large-scale surveys like LSST and Euclid by offering an efficient alternative solution for processing and analyzing the vast amounts of data these surveys will generate in the nearest future. 

In this work, we propose the application of our method to the estimation of such galaxy properties as star formation rate (SFR) and stellar mass ($\mathcal{M}_{\ast}$), which play an important role in the broad range of studies on the galaxy formation and evolution \citep{Conselice:14, Madau:14, Forster:20}, their gas content \citep{Carilli:13, Morganti:18, Maiolino:19} and co-evolution with supermassive black holes in their centers \citep{Fabian2012, Kormendy2013, Heckman2014, Hickox2018, Torbaniuk:24}. Traditionally, the extraction of SFR and $\mathcal{M}_{\ast}$ has relied on spectroscopic data or broadband information for spectral energy distribution (SED) fitting \citep{Calzetti13sfr, Kennicutt12sfr}. However, these methods can be time-consuming and often limited by the availability of the spectra and/or broadband observations.  
Several studies have already proposed an alternative approach to this problem using different machine and deep learning techniques, including supervised \citep{Bonjean:19, DelliVeneri2019, Dominguez:23_SFRmass}, semi-supervised \citep{Humphrey:23_unlabeled} approaches and neural networks pre-trained on the broadband UV, optical and IR photometry and/or images \citep{Surana:20, Chu:24_stelMass, Zeraatgari:24, Zhong:24_images, Euclid:25_MLphysProp}.

The recent work of \citet{Dominguez:23_SFRmass,Zhong:24_images} presents an approach for determining galaxy properties based on images similar to ours presented in this work. However, in contrast to our methods, these works are based on self-supervised pre-trained deep learning techniques.

The paper is structured as follows: In Section\,\ref{sec:method}, we summarize our method, while in Section\,\ref{sec:data} we describe our galaxy sample and present the sample of target objects used for the testing of our method performance in retrieving the galaxy properties. Section\,\ref{sec:exp} is devoted to the presentation of the experiments and discussion of the results. Finally, in Section\,\ref{sec:conclusions} we present our conclusions on the obtained results and its possible application for other samples. Additional supporting material is provided in the Appendices, including the distribution of retrieved neighbours in the SFR–$\mathcal{M}_{\ast}$ plane for different target classes (Appendices\,\ref{sec:appendix} and\,\ref{sec:appendix2}), and the statistical properties of neighbour distributions (Appendix\,\ref{sec:appendix3}).

\section{Method}\label{sec:method} 

The \ULISSE algorithm and its performance have already been presented in \citetalias{Doorenbos2022} in the context of AGN selection. Here, we will provide only a brief summary to introduce the general concept behind the method. Detailed information on the pre-training and feature extraction steps, along with computation times and tests involving single-band images and recursive application, can be found in~\citetalias{Doorenbos2022}.

\ULISSE uses features extracted from a convolutional neural network (CNN, \citealt{Schmidhuber2015}) that was trained on a large-scale dataset. Training a CNN on a large and diverse dataset allows it to learn a broad spectrum of features that can be useful beyond the original task. This concept, widely known as transfer learning, has been successfully applied in numerous areas, including astronomy~\citealt{Awang2020,martinazzo2021self,Cavuoti_DeCicco:24}, malware classification~\citealt{Prima2020}, earth science~\citealt{Zou2018}, and medicine~\citealt{ding2019deep, esteva2017dermatologist,Kim2021tl, menegola2017knowledge}. 
The typical large-scale dataset used for training in this context is {\tt ImageNet}~\citep{deng2009imagenet}. 

These features are subsequently used to identify relevant astronomical objects via a nearest-neighbour search. In order to obtain the features, the fully connected layers of the pre-trained network are discarded, and the feature-extracting component of the network is used directly. This approach removes the need for any neural network training and makes direct application to any new dataset possible. Then, in order to obtain image-level properties, we extract the feature maps from the final convolutional layer of the pre-trained neural network. To reduce the dimensionality of these maps, we average over the spatial dimensions. 
As the deepest layers in the network exhibit the highest level of abstraction~\citep{Goodfellow-et-al-2016}, we assume that objects in the dataset whose images have similar deep features to a target object possess similar morphological properties. 

We use an {\tt EfficientNet-b0} as the CNN \citep{tan2019efficientnet}, resulting in a 1280-dimensional feature descriptor for each image. As these features are derived from natural images, they are not immediately interpretable in an astronomical context, although specific patterns linked to particular features can still be identified, as shown in \citetalias{Doorenbos2022}.

\ULISSE identifies objects with similar properties by performing a similarity search in the pre-trained feature space. Using the given target object (or prototype, as in \citetalias{Doorenbos2022}), it finds the closest objects in this feature space, providing a list of candidate lookalikes. Formally, for the image of a target object $\textbf{x}_q$, the nearest neighbours $\textbf{x}_i$ from a dataset $\{\textbf{x}_i\}_{i=1}^N$ are retrieved by the lowest Euclidean distances $d(\textbf{x}_q, \textbf{x}_i) = ||\textbf{f}_{q} - \textbf{f}_i||^2$ in the pre-trained feature space, where $\textbf{f}_i$ denote the features of image $i$. Although \ULISSE does not require prior information about the dataset used to search candidates with similar properties, having a validation set allows us to evaluate the method's performance by analyzing the behaviour and properties of the returned $n$ closest objects (i.e., neighbours).

\section{Dataset}\label{sec:data}

The estimation of the galaxy properties such as stellar mass $\mathcal{M}_{\ast}$ and SFR is traditionally based on the analysis of certain features in galaxy spectra (e.g. both continuum and emission lines in a wide range of wavelength, see \citealt{Kennicutt12sfr,Calzetti13sfr}) or by fitting the broadband spectral energy distribution (SED). Both methods have their limitations, primarily because they require either spectroscopic information or a comprehensive multiwavelength photometric dataset (spanning from X-rays to the far infrared) for proper SED reconstruction, which is often not available for large samples. Various machine and deep learning methods offer promising alternatives for addressing the aforementioned problems. However, in the initial stage, these algorithms still need to be tested on samples with reliable estimates of SFR and $\mathcal{M}_{\ast}$, derived from well-established traditional methods.

\subsection{Dataset details}\label{sec:primary-sample}

To test the \ULISSE efficiency in the determination of the galaxy characteristics, we used the same {\it galSpec} galaxy catalog\footnote{\url{https://www.sdss.org/dr12/spectro/galaxy_mpajhu/}} as in \citetalias{Doorenbos2022}, which has been produced by the MPA-JHU group as a subsample from the main galaxy catalog of the eighth data release of the Sloan Digital Sky Survey (SDSS DR8, \citealt{Brinchmann2004}). The estimates of total stellar mass ($\mathcal{M}_{\ast}$) in the sample were obtained through Bayesian fitting of the SDSS {\it ugriz} photometry to a grid of models (see details in \citealt{Kauffmann2003a, Tremonti2004}). In the case of star-formation rate determination, two different approaches were used, depending on the object classification according to the BPT-diagram criterion, allowing us to make an assumption on the nature of emission lines (e.g. nebular from star-forming regions or nuclear from AGN) using the ratios of certain emission lines in the galaxy spectrum~\citep{Baldwin1981}. The values of SFR for star-forming galaxies were determined using the H$\alpha$ emission line luminosity, while for all other spectral classes, where the H$\alpha$ emission line is weak (e.g. unclassified objects) or contaminated by emission from non-stellar component (e.g. AGN or composite), the empirical relation between SFR and the Balmer decrement, $D4000$, was used (see details in \citealt{Kauffmann2003b}). 

The detailed description of the sample reduction can be found in \citetalias{Doorenbos2022} (the references therein), where the same sample was used for testing \ULISSE performance in the framework of AGN identification.  
Based on the experiments conducted in our previous \ULISSE paper, the most optimal thumbnail size was chosen as $22\times22$\,arcsec (or $56\times56$ pixels, the SDSS pixel scale is 0.396 arcsec per pixel). However, since the \ULISSE efficiency may vary depending on the angular size of galaxies and its ability to resolve the entire galaxy with its morphological features (e.g. disc, bulge, spiral arms, etc), we also decided to remove galaxies with too large (or too small) angular diameter to the size of the thumbnails. Thus, we limited our sample to objects with Petrosian radius ({\it petroR50} in {\it r}-band) larger than 1.5\,arcsec and smaller than 8\,arcsec. For the same reason, we limited our sample to sources with redshift $z > 0.01$ and up to $z \leq 0.15$, above which the morphological classification of galaxies becomes challenging due to the resolution and depth of the SDSS data. As a result, our sample contains 449\,762 objects. 

In \citetalias{Doorenbos2022} we found that \ULISSE performance seems to differ for galaxies with various appearances, so to examine this point in the current work we selected objects with the available morphological classification in the second release of Galaxy Zoo (GZ2, \citealt{Willett2013}), which resulted in 201\,626 galaxies. It should be noted that the morphology classification in the GZ2 is quite extensive and reflects not only the presence of such features as spiral arms, bulges, and bars but also considers the strength of these features. Since the goal of our work is not to find galaxies that are identical in morphological appearance, but rather to test whether \ULISSE efficiency in SFR and mass prediction varies for galaxies with significantly different appearances, we decided to simplify the original GZ2 classification and instead use only five general morphological classes. The first class, {\it ellipticals} (`E'), contains all galaxies with smooth visual appearance, but without the GZ2 division based on the roundness of the galaxy. { As a result, this class  may include not only genuine early-type `red' galaxies, but also star-forming `bluer' disk galaxies with poorly resolved features that appear smooth due to limited resolution or projection effects.} The next three classes were defined for galaxies with disk or spiral features: {\it spiral} galaxies with a bulge (`S') and/or bar component (`SB') and edge-on spiral galaxies (`Se'). Again, we did not consider the GZ2 division according to the prominence and shape of the galaxy bulge/bar or the number and relative winding of spiral arms. { In addition to the four main morphological categories, we also included galaxies exhibiting visually irregular or atypical features such as rings, lenses and arcs, dust lanes, or irregular/interacting morphologies. These features deviate from the idealized, symmetric morphologies of the classes discussed above but do not necessarily imply the presence of active mergers or major interactions. Nonetheless, the presence of such structural peculiarities can still correlate with deviations in physical properties relative to the more `classical' systems. The galaxies in this class retain their primary morphological classification (i.e. E, S, SB, Se), and we denote them with a (d) suffix (e.g., E(d), S(d), etc.) to indicate the presence of such visual disturbances. For simplicity, we group these systems under their respective disturbed subclasses and refer to them collectively as `E(d),S(d),SB(d),Se(d)' throughout the paper. As demonstrated in Section\,\ref{sec:results-histo}, this grouping has no impact on the overall result and \ULISSE performance is consistent across both regular and structurally disturbed morphologies, with no systematic bias in the predicted stellar masses or SFR}\footnote{ The relatively high number of galaxies with structural irregularities (i.e., E(d), S(d), etc.) in our sample (see also the target selection in Section\,\ref{sec:prototypes}) also arises from our initial selection strategy, in which we aimed at sampling 3–5 galaxies per stellar mass and SFR bin across each morphological subclass, including both regular and visually disturbed systems. This approach was designed to ensure broad coverage across parameter space, rather than to reproduce the actual distribution of galaxy types in the Universe. As a result, the fraction of irregular systems in our sample is artificially elevated compared to their actual occurrence rates in the general galaxy population. We also emphasize that the galaxies marked as E(d), S(d), SB(d) and Se(d), include a wide range of visual features (e.g., rings, dust lanes, tidal structures, etc) and do not solely represent systems with ongoing mergers or strong interactions.}.

As we discussed in \citetalias{Doorenbos2022} \ULISSE performance in AGN identification is linked not only to its ability to retrieve the properties of their host galaxies (e.g. morphology and colour), but it seems also to be able to recognize the presence of a central nuclear source (i.e. AGN). Thus, to test for a possible difference in the \ULISSE efficiency in the retrieving SFR and $\mathcal{M}_{\ast}$ based on whether or not a galaxy hosts AGN, we decided also to add simplified AGN/non-AGN classification for our sample. So, each source in our sample was marked as `AGN' if it has been identified as AGN, low S/N AGN, or composite source (i.e. contribution from both an AGN and star-forming processes) according to the BPT selection criteria (see details in \citealt{Brinchmann2004}), while the other objects like SFG, low S/N SFG and unclassified were accordingly marked as `non-AGN'. 

To assess the effect of dust-related reddening on the performance of our image-based method particularly given its reliance on optical \textit{g-,r-,i-}bands, we also classified galaxies into dust content classes based on the Balmer decrement (H$\alpha$/H$\beta$), which is a widely used tracer of nebular dust extinction~\citep{Brinchmann2004,Koyama:15,Qin:19}. We adopted an intrinsic, dust-free ratio of (H$\alpha$/H$\beta) = 2.86$, consistent with Case~B recombination under typical H\,{\sc ii} region conditions~\citep{Fitzpatrick:99,Lin:24}. Galaxies with (H$\alpha$/H$\beta) \leq 3.2$ (corresponding to extinction of approximately $A_{V} \lesssim  0.5$\,mag) were classified as \textit{low-dust} content. Those with $3.2 < $(H$\alpha$/H$\beta) \leq 5.0$ ($A_{V} = 0.5$-1.5\,mag) were considered as \textit{moderate-dust}, and those with (H$\alpha$/H$\beta) > 5.0$ ($A_{V} \gtrsim 1.5$\,mag) as \textit{high-dust} content. Galaxies with low S/N ratio for one or both lines and providing unphysical values of (H$\alpha$/H$\beta$) which may arise from poor spectral quality or line fitting issues, were excluded from the dust-based subsample analysis (marked as \textit{unknown} dust content).

The numbers of objects (and their fractions in the sample) of different morphology or AGN classes are presented in Table\,\ref{tab:samp-fract}. Since the goal of our work is mainly to show the validity of our method for the galaxy properties extraction and not to obtain results for the whole sample of galaxies, we create a smaller subsample just for more efficient computation. So, similarly to the approach in \citetalias{Doorenbos2022}, we randomly shuffle the coordinates and take the first 100\,000 objects. This subsample is labelled as \textit{Random} in Table\,\ref{tab:samp-fract}, and it can be seen its proportions are practically equal to those of the whole sample. 

\begin{table*}
\caption{The summary of the different subsamples studied in this work.}
\begin{center}
\begin{tabular}{@{\extracolsep{4pt}}cclrrrr@{}}
\hline\hline\\[-1.5ex]
\multicolumn{3}{c}{\multirow{2}{*}{Sample}} & \multicolumn{2}{c}{Entire} & \multicolumn{2}{c}{Random} \\
\cline{4-5}\cline{6-7}\\[-1.5ex]
\multicolumn{3}{l}{} & \multicolumn{1}{c}{$N$} & \multicolumn{1}{c}{Fraction} & \multicolumn{1}{c}{$N$} & \multicolumn{1}{c}{Fraction} \\
\hline\hline\\[-1.5ex]
 & \multirow{5}{*}{{\centering GZ2 Morphology}} & Ellipticals/smooth, E & 73\,075 & 36.2\perc & 36\,173 & 36.2\perc \\
\multirow{12}{*}{\rotatebox[origin=c]{90}{\centering Criteria}}& & Spirals, S & 67\,855 & 33.7\perc & 33\,686 & 33.7\perc \\
& & Spirals with bar, SB & 15\,389 & 7.6\perc & 7\,594 & 7.6\perc \\
& & Edge-on spirals, Se & 19\,980 & 9.9\perc & 9\,865 & 9.8\perc \\
& & { E(d),S(d),SB(d),Se(d)} & 25\,327 & 12.6\perc & 12\,682 & 12.7\perc \\
\cdashline{2-7}\\[-1.5ex]
& \multirow{2}{*}{{\centering BPT-diagram}} & AGN & 56\,536 & 28.0\perc & 28\,039 & 28.0\perc \\
& & non-AGN & 145\,090 & 72.0\perc & 71\,961 & 72.0\perc \\
\cdashline{2-7}\\[-1.5ex]
& \multirow{2}{*}{{\centering SFR--$\mathcal{M}_{\ast}$ diagram}} & Star-forming galaxies, SFG & 109\,664 & 54.4\perc & 54\,301 & 54.3\perc \\
& & Quiescent galaxies, QG & 91\,962 & 45.6\perc & 45\,699 & 45.7\perc \\
\cdashline{2-7}\\[-1.5ex]
& \multirow{3}{*}{{\centering Redshift range}} & $0.01 < z < 0.05$ & 51\,275 & 25.4\perc & 25\,265 & 25.2\perc \\
& & $0.05 < z < 0.1$ & 102\,226 & 50.7\perc & 50\,658 & 50.7\perc \\
& & $0.1 < z < 0.15$ & 48\,125 & 23.9\perc & 24\,077 & 24.1\perc \\
\cdashline{2-7}\\[-1.5ex]
& \multirow{4}{*}{{\centering { Dust content}}} & { low (H$\alpha$/H$\beta) \leq 2.86$} & { 41\,779} & { 20.7\perc} & { 20\,700} & { 20.7\perc} \\
& & { moderate $2.86 < $ (H$\alpha$/H$\beta) \leq 5.0$} & { 110\,423} & { 54.8\perc} & { 54\,908} & { 54.9\perc} \\
& & { high (H$\alpha$/H$\beta) > 5.0$} & { 34\,067} & { 16.9\perc} & { 16\,786} & { 16.8\perc} \\
& & { unknown} & { 15\,337} & { 7.6\perc} & { 7\,606} & { 7.6\perc} \\
\hline\\[-1.5ex]
\multicolumn{3}{l}{Total number of objects} & 201\,626 & \multicolumn{1}{c}{--} & 100\,000 & \multicolumn{1}{c}{--} \\
\hline
\hline
\end{tabular}
\tablefoot{
The fractions represent the percentage of objects in each subsample defined based on GZ2 morphology classification, the presence of AGN according to the BPT-diagram selection criteria, the galaxy position in the SFR-$\mathcal{M}_{\ast}$ diagram, redshift range, { and dust content defined based on the Balmer decrement}, respectively.}
\end{center}
\label{tab:samp-fract}
\end{table*}

\begin{figure}
    \centering
    \includegraphics[width=.5\textwidth]{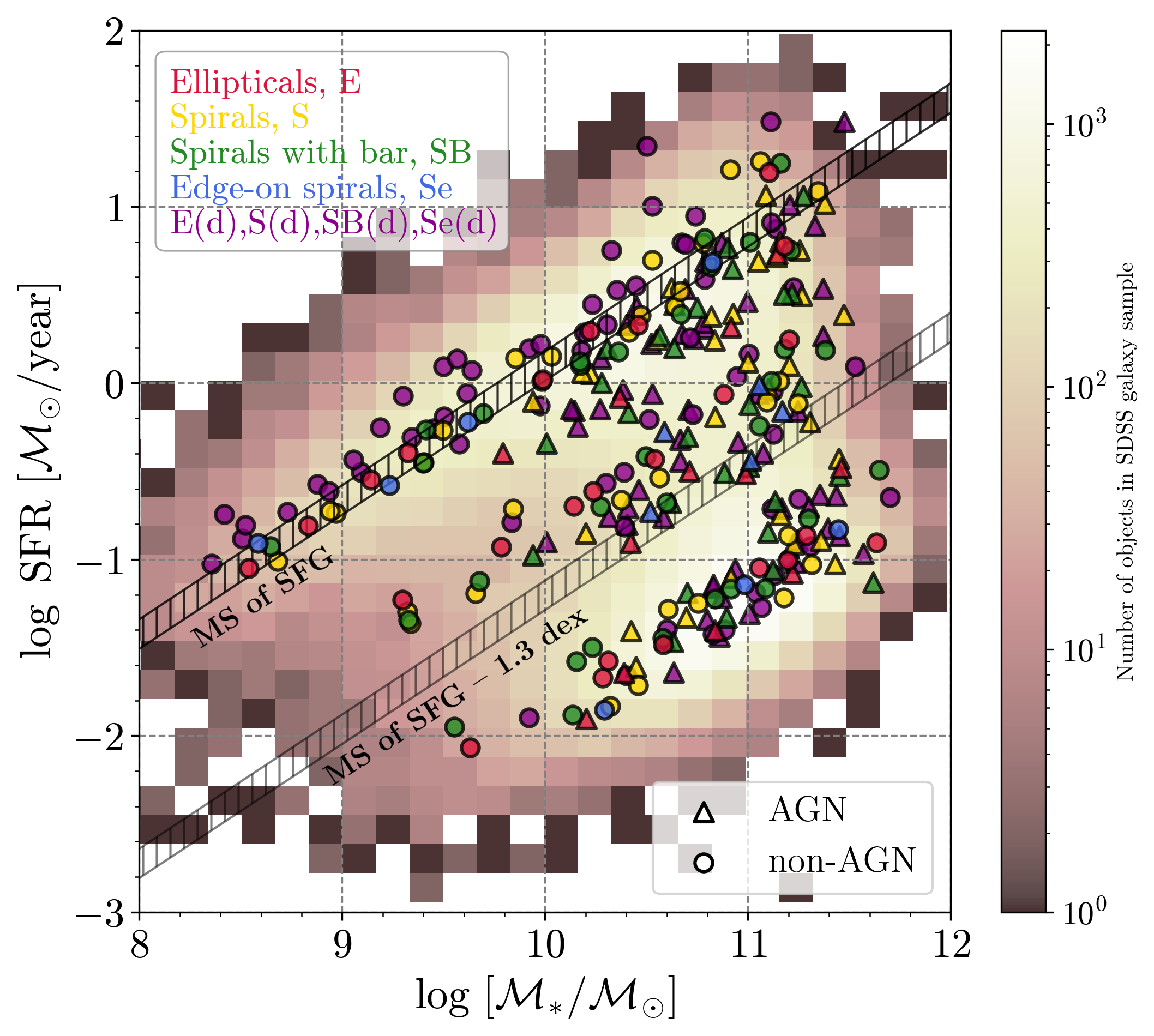}
    \caption{The distribution of SFR versus stellar mass for 100\,000 galaxies in our sample. The individual target objects selected for our study are shown in different colours and shapes depending on GZ2 morphology class (from red to purple) and AGN/non-AGN selection according to the BPT-diagram criteria (triangle and circle), respectively. The black shaded band presents the main sequence of star-forming galaxies (MS of SFG) defined by Eq.\,\eqref{eq:ms-line}. The grey-shaded band shows a cut 1.3\,dex below the MS of SFG applied for the division of the studied galaxies into star-forming and quiescent galaxy populations. Both areas correspond to the studied redshift interval $z = 0.01 - 0.15$.}
    \label{fig:sfr-m-allProt}
\end{figure} 

\subsection{Target objects}\label{sec:prototypes}

To test \ULISSE ability to retrieve galaxies with similar star-formation rates and stellar masses, we need to select the set of target objects within a wide range of SFR and $\mathcal{M}_{\ast}$. Target objects in this study represent galaxies with unknown physical parameters (SFR and $\mathcal{M}_{\ast}$), which we want to measure based on a set of `neighbours' selected by \ULISSE. For this purpose, we plotted our {\it Random} sample in the SFR-$\mathcal{M}_{\ast}$ diagram. The distribution of galaxies in the SFR--$\mathcal{M}_{\ast}$ diagram is known to show the clustering of galaxies into two main populations: the sequence of {\it `star-forming'} galaxies with steady processes of new stars formation (the so-called `main-sequence' of star-forming galaxies; MS of SFG), and {\it `quiescent'} galaxies with passively evolving stellar populations. We marked each galaxy in our sample as star-forming or quiescent based on their position in the SFR--$\mathcal{M}_{\ast}$ diagram relative to the evolving MS of SFG defined by \citet{Aird2017} as: 
\begin{multline}
\log\mathrm{SFR}_{\mathrm{MS}}(z)\,[\mathcal{M}_{\odot}\mathrm{year}^{-1}] =\\
= -7.6 + 0.76\log\,[\mathcal{M}_{\ast}\,/\mathcal{M}_{\odot}] + 2.95\log(1+z).
\label{eq:ms-line}
\end{multline}
The threshold between the two classes was set at 1.3\,dex below the MS of SFG: galaxies that fall below this cut were classified as quiescent, while those above the line as star-forming. Note that the relation in Equation\,\eqref{eq:ms-line} is redshift-dependent, so for the classification, we used the redshift of each object.

\begin{figure}
    \centering
    \includegraphics[width=.47\textwidth]{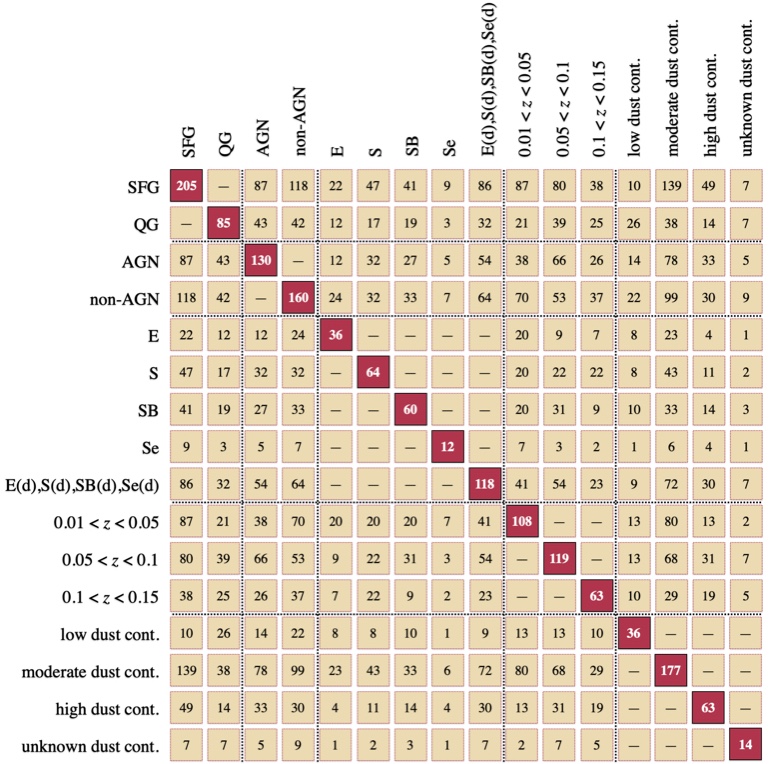}
    \caption{The number of target objects selected among various classes mentioned in Table\,\ref{tab:samp-fract} including classification based on GZ2 galaxy morphology, presence/absence of AGN, location in the SFR-$\mathcal{M}_{\ast}$ diagram, redshift range and { dust content}.}
    \label{fig:prot-N-samples}
\end{figure}

Then, using the same distribution of galaxies in the SFR-$\mathcal{M}_{\ast}$ diagram, we manually selected objects uniformly distributed within the diagram. Selecting targets, we tried to cover all possible morphological classes mentioned in Table\,\ref{tab:samp-fract}, which resulted in 290 objects. The distribution of the selected target objects in the SFR-$\mathcal{M}_{\ast}$ diagram is presented in Fig\,\ref{fig:sfr-m-allProt}. Note, to avoid introducing additional uncertainties to the final results, we did not select target objects on the weakly populated areas of the SFR-$\mathcal{M}_{\ast}$ diagram (i.e. mainly on the edges) where \ULISSE will suffer to probe objects with similar properties due to intrinsically low number of them in our sample. The number of target objects selected among different classes is presented in Fig.\,\ref{fig:prot-N-samples}. It should be noted that the SFR-$\mathcal{M}_{\ast}$ diagram presented in this work has a logarithmic scale (i.e. it is the $\log\,{\rm SFR}$--$\log\,\mathcal{M}_{\ast}$ diagram in reality), but for a matter of simplicity we will refer to both logarithmic quantities ($\log\,{\rm SFR}$ and $\log\,\mathcal{M}_{\ast}$) as SFR and $\mathcal{M}_{\ast}$ throughout the paper.

\section{Experiments and results}\label{sec:exp}

In the current Section, we initially discuss the obtained result using as examples six targets covering all the classes discussed in Section\,\ref{sec:prototypes}. By covering a broad range of galaxy properties, we try to establish whether the performance of our method is sensitive to the particular population of galaxies or depends on their properties, as well as identify possible limitations for some specific class. Then, using the entire set of target objects selected in the previous section, we perform a statistical analysis to define the general efficiency of our method and compare it with a `random guess' approach.

\subsection{Methodology}\label{sec:distances}

As mentioned in the previous section, for each of 290 target objects we have ally estimated values of SFR and $\mathcal{M}_{\ast}$ (see Fig.\,\ref{fig:sfr-m-allProt}) and labels regarding their GZ2 morphology class, AGN/non-AGN nature according to the BPT selection criteria and the position in the SFR-$\mathcal{M}_{\ast}$ diagram (star-forming or quiescent galaxy). Following a similar approach as in \citetalias{Doorenbos2022} we perform our experiment using the {\it g-, r-, i-} color-composite SDSS thumbnails. For visualization of the obtained results, we chose six target objects with different properties and plotted them together with their retrieved nearest neighbours in the SFR--$\mathcal{M}_{\ast}$ diagrams in Fig.\,\ref{fig:obj2}-\ref{fig:obj276}. Each figure also provides the SDSS thumbnail analyzed by \ULISSE as well as the summary of the targets characteristics, including its SDSS name and object ID (objid), celestial coordinates (right ascension RA and declination Dec), spectroscopic redshift $z$, and the object's affiliation to the specific class according to its GZ2 morphology, AGN/non-AGN, and its position in the SFR--$\mathcal{M}_{\ast}$ diagram (see definition in the Section\,\ref{sec:data}). In our work, we decided to set the number of objects $N_{\rm neig}$ retrieved by \ULISSE to the 100 nearest neighbours; however, in practice, the choice of $N_{\rm neig}$ depends on the purpose of the user or the goals of the study, and can be changed to any number.

To quantify the \ULISSE efficiency in retrieving galaxies with similar SFR and $\mathcal{M}_{\ast}$, we define a set of `distances' between the studied target and retrieved neighbours in SFR--$\mathcal{M}_{\ast}$ parameter space. Such distances are divided into two general groups: the first one represents a simple physical distance (in dex) in SFR--$\mathcal{M}_{\ast}$ parameter space, while the second one is the `weighted' version of the first one considering the accuracy of SFR and $\mathcal{M}_{\ast}$ estimates of the target object. For instance, we propose the so-called {\it total} distance, $d_{\rm total}$, which represents the sum of statistical distances (in SFR and $\mathcal{M}_{\ast}$ parameter space) between the target (with $\mathcal{M}_{\ast}^{\rm targ}$ and ${\rm SFR}^{\rm targ}$ in the SFR--$\mathcal{M}_{\ast}$ diagram) and each separate neighbours ($\mathcal{M}_{\ast}^{\rm neig}$ and ${\rm SFR}^{\rm neig}$) normalized by the total number of neighbours retrieved by \ULISSE (i.e. $N_{\rm neig} = 100$), and defined as:\\[-1ex]
$$
d_{\rm total} = \frac{1}{N_{\rm neig}}\sum_{i}^{100} \sqrt{ \left(\mathcal{M}_{\ast}^{\rm targ} - \mathcal{M}_{\ast,i}^{\rm neig}\right)^{2} + \left({\rm SFR}^{\rm targ} - {\rm SFR}_{i}^{\rm neig}\right)^{2} }
$$

Secondly, we introduce the {\it mean} distance, $d_{\rm mean}$, the distance between the position of the target object and the average position among all neighbours retrieved by \ULISSE for a specific target. The latter is presented as the mean values of stellar mass $\langle\mathcal{M}_{\ast}^{\rm neig}\rangle$ and star-formation rate $\langle{\rm SFR}^{\rm neig}\rangle$ calculated for 100 neighbours, and then $d_{\rm mean}$ can be expressed as:\\[-1ex]
$$
d_{\rm mean} = \sqrt{ \left(\mathcal{M}_{\ast}^{\rm targ} - \langle\mathcal{M}_{\ast}^{\rm neig}\rangle \right)^{2} + \left({\rm SFR}^{\rm targ} - \langle{\rm SFR}^{\rm neig}\rangle\right)^{2} }
$$

The third `distance', $d_{\rm norm}$, is defined similarly to $d_{\rm mean}$, but before averaging the distances among all retrieved neighbours the position of each neighbour is normalized by the number of galaxies ($N_{gal}$) in the sample falling within SFR--$\mathcal{M}_{\ast}$ bin where this neighbour is located. In this way, each neighbour obtains a normalization parameter reflecting the `rarity' of this object in the sample and therefore allows to reduce the bias on the average position of the neighbours (and its distance from the target) due to denser sampling of different regions in the SFR--$\mathcal{M}_{\ast}$ diagram, intrinsic to any flux- or volume-limited survey. The expression for $d_{\rm norm}$ can be then written as:\\[-1ex]
$$
d_{\rm norm} = \sqrt{ \left(\mathcal{M}_{\ast}^{\rm targ} - \mathcal{M}_{\ast}^{\rm norm} \right)^{2} + \left({\rm SFR}^{\rm targ} - {\rm SFR}^{\rm norm}\right)^{2} },
$$
where $\mathcal{M}_{\ast}^{\rm norm} = \sum_{i}\mathcal{M}_{\ast,i}^{\rm neig}\cdot w^{\rm neig}_{i} / \sum_{i} w^{\rm neig}_{i}$ and ${\rm SFR}^{\rm norm} = \sum_{i}{\rm SFR}^{\rm neig}_{i}\cdot w^{\rm neig}_{i} / \sum_{i} w^{\rm neig}_{i}$, and $w^{\rm neig}_{i} = 1/N_{gal, i}$, $N_{gal, i}$ is the number of galaxies with each SFR--$\mathcal{M}_{\ast}$ bin for each retrieved neighbour. Note, that the size of each SFR--$\mathcal{M}_{\ast}$ bin in Fig.\,\ref{fig:prot-N-samples} is near $0.17$\,dex.

The second group of `distances' mimics the first one but takes into account the accuracy of spectroscopic SFR and $\mathcal{M}_{\ast}$ estimates of each particular target. For this, each distance defined above is normalised by the SFR and $\mathcal{M}_{\ast}$ uncertainties to obtain the `weighted' distances that reflect the average significance range (in $\sigma$ units) of the target into which the retrieved neighbours fall. 
In this way, the weighted version of the total distance $d_{\rm total}^{\rm weighted}$ can be rewritten as:\\[-1ex]
$$
d_{\rm total}^{\rm weighted} = \frac{1}{N_{\rm neig}}\sum_{i} \sqrt{ \left(\frac{\mathcal{M}_{\ast}^{\rm targ} - \mathcal{M}_{\ast,i}^{\rm neig}}{\sigma_{\mathcal{M}_{\ast}}^{\rm targ}}\right)^{2} + \left( \frac{{\rm SFR}^{\rm targ} - {\rm SFR}_{i}^{\rm neig}}{\sigma_{{\rm SFR}}^{\rm targ}}\right)^{2} },
$$
and the normalised mean distance $d_{\rm mean}^{\rm weighted}$ is as:\\[-1ex]
$$
d_{\rm mean}^{\rm weighted} = \sqrt{ \left(\frac{\mathcal{M}_{\ast}^{\rm targ} - \langle\mathcal{M}_{\ast}^{\rm neig}\rangle}{\sigma_{\mathcal{M}_{\ast}}^{\rm targ}} \right)^{2} + \left(\frac{{\rm SFR}^{\rm targ} - \langle{\rm SFR}^{\rm neig}\rangle}{\sigma_{{\rm SFR}}^{\rm targ}}\right)^{2} }.
$$
And finally, the normalised version of the weighted distance $d_{\rm norm}^{\rm weighted}$ is as:\\[-1ex]
$$
d_{\rm norm}^{\rm weighted} = \sqrt{ \left(\frac{\mathcal{M}_{\ast}^{\rm targ} - \mathcal{M}_{\ast}^{\rm norm} }{\sigma_{\mathcal{M}_{\ast}}^{\rm targ}}\right)^{2} + \left(\frac{ {\rm SFR}^{\rm targ} - {\rm SFR}^{\rm norm}}{\sigma_{{\rm SFR}}^{\rm targ}}\right)^{2} }.
$$

In addition to the distance-based metrics we introduce a set of metrics which can help us to characterize the spatial distribution of the retrieved neighbours with respect to the target's position in the SFR–$\mathcal{M}_{\ast}$ diagram. These metrics may serve as a valuable extension to distance-based measures, offering deeper insight into the efficiency and reliability of the results provided by \ULISSE, particularly in cases where distances alone may fail to capture complex spatial patterns in the retrieved neighbours. First of all, we add the \textit{standard deviations} in stellar mass and SFR (i.e. $\Delta{\mathcal{M}_{\ast}}$ and $\Delta{\mathrm{SFR}}$) to quantify the spread of the neighbours along each axis, along with the \textit{two-dimensional scatter}, computed as $\Delta_{2D} = \sqrt{\Delta{\mathcal{M}_{\ast}}^{2} + \Delta{\mathrm{SFR}}^{2}}$ that represent the individual and joint spread of neighbours serving as a compact diagnostic of how tightly/diffusely they populate the SFR–$\mathcal{M}_{\ast}$ plane.

Additionally, we define the \textit{shape asymmetry} as the ratio $A_{s} = {\Delta{\mathrm{SFR}/\Delta\mathcal{M}_{\ast}}}$, quantifying the relative dispersion of neighbours along $\mathcal{M}_{\ast}$ and SFR axes. In contrast to its common use as a metric for characterising the elongation of a distribution, we apply $A_{S}$ as a diagnostic to assess retrieval imbalance: values significantly deviating from unity indicate disproportionate spread in one parameter relative to the other. In this context, shape asymmetry can be used as an indicator of reduced retrieval efficiency in either $\mathcal{M}_{\ast}$ or SFR, revealing which axis contributes more substantially to the broadening of the neighbour distribution around the target object.

In contrast, the \textit{aspect ratio} ($\mathcal{R}$), defined as the ratio of the principal eigenvalues of the covariance matrix of the neighbour distribution, captures the overall geometric elongation and directional alignment in the SFR–$\mathcal{M}_{\ast}$ plane. While $A_{S}$ quantifies the imbalance of scatter between the two axes, $\mathcal{R}$ reflects the coherence of that scatter, i.e. whether neighbours are distributed isotropically or exhibit a preferred orientation. High values of $\mathcal{R}$ indicate alignment along a dominant direction within the parameter space (e.g., the MS of SFG), suggesting that the retrieved neighbours, despite their possible spread, tend to follow an underlying intrinsic scaling relation of the galaxy population.

Finally, we define the \textit{compactness} ($\mathcal{C}$) as the area of the 1$\sigma$ confidence ellipse enclosing the neighbour distribution, normalized by $\pi$, the area of a reference circle with a radius of 1\,dex in logarithmic space. This dimensionless quantity provides a standardized measure of how concentrated or dispersed the neighbours are relative to a baseline scale (e.g. circle with 1\,dex radius) commonly considered representative of broad distributions in the SFR–$\mathcal{M}_{\ast}$ diagram. Values of $\mathcal{C}$ smaller than unity indicate a distribution more compact than the assumed baseline scale, reflecting tighter clustering around the target, whereas larger values denote greater spread and less localized neighbour retrieval. This reference radius can be adjusted to other values (e.g. 0.5\,dex) depending on the desired sensitivity to clustering scale, the compactness metric naturally rescales with the chosen normalization, allowing flexibility for more stringent or relaxed definitions of concentration.

\subsection{Retrieval efficiency of physical parameter for individual target objects.}\label{sec:results-ind}

\begin{figure*}
    \centering
    \includegraphics[width=.9\textwidth]{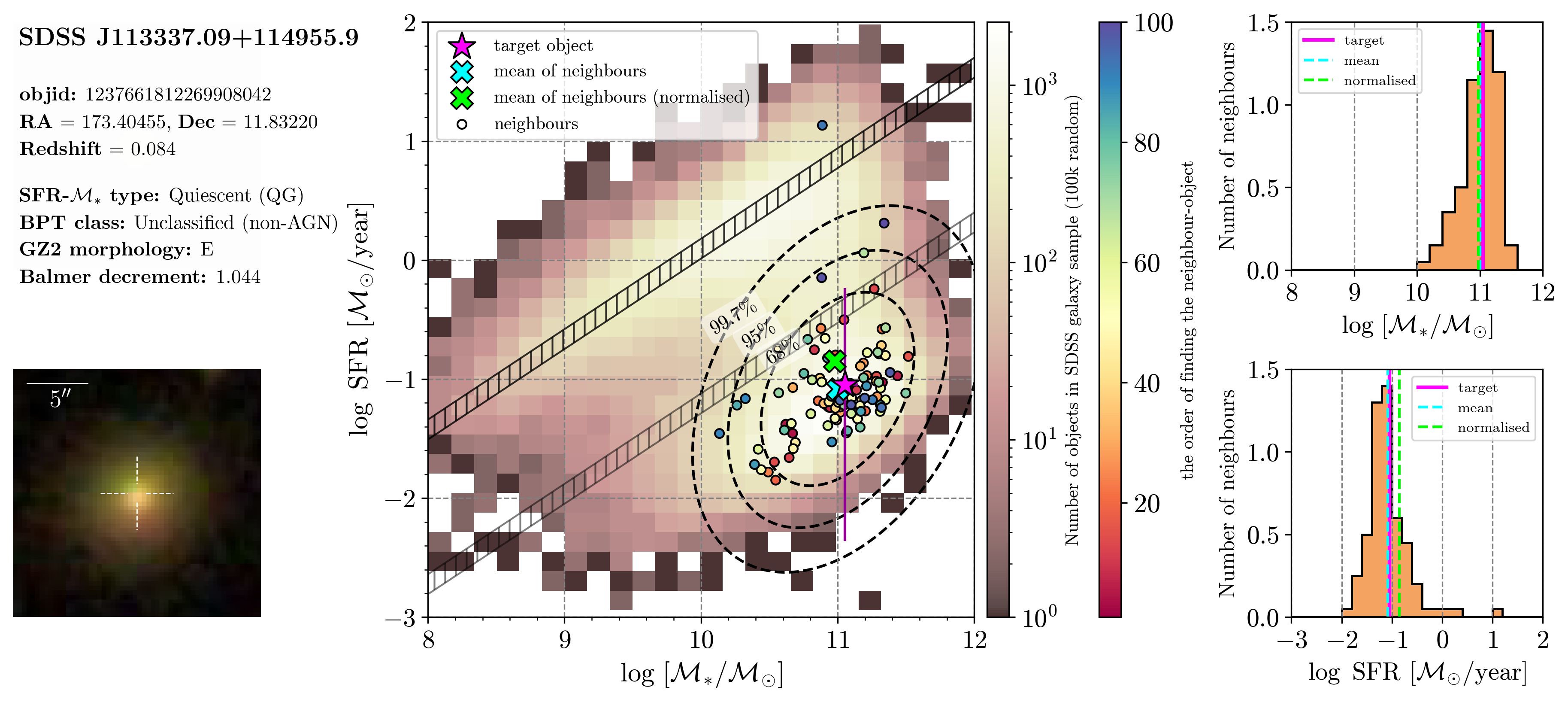}
    \caption{{\it Left:} Properties of target object~\#1 with its classification according to the criteria presented in Section\,\ref{sec:primary-sample} ({\it upper panel}) and the SDSS thumbnail used for analysis by \ULISSE ({\it lower panel}).  {\it  Center:} The distribution of SFR versus stellar mass for 100\,000 galaxies in our sample. The position of the target in the SFR-$\mathcal{M}_{\ast}$ diagram (with 1$\sigma$ errors) is presented as a rev star together with 100 nearest neighbours (circles). The colour of the circles represents the order in which each neighbour was retrieved by \ULISSE.  The cyan and green crosses show the position of the mean and normalised mean position for 100 neighbours, respectively. { Ellipses enclosing 68\%, 95\%, and 99.7\% of the retrieved neighbours are shown by black dashed lines.} The values of distances estimated for this target are presented in Table\,\ref{tab:indiv-targets}. The grey- and black-shaded areas are the same as in Fig.\,\ref{fig:sfr-m-allProt}. {\it Right:} { Distributions of $\mathcal{M}_{\ast}$ ({\it upper panel}) and SFR ({\it lower panel}) for 100 neighbours retrieved by \ULISSE, normalized to the number of neighbours. Vertical lines show the corresponding values of $\mathcal{M}_{\ast}$ or SFR for the target object (rev), the arithmetic mean (cyan), and the normalised mean (green) of the neighbour distributions. }}
    \label{fig:obj2}
\end{figure*}

\begin{figure*}
    \centering
    \includegraphics[width=.9\textwidth]{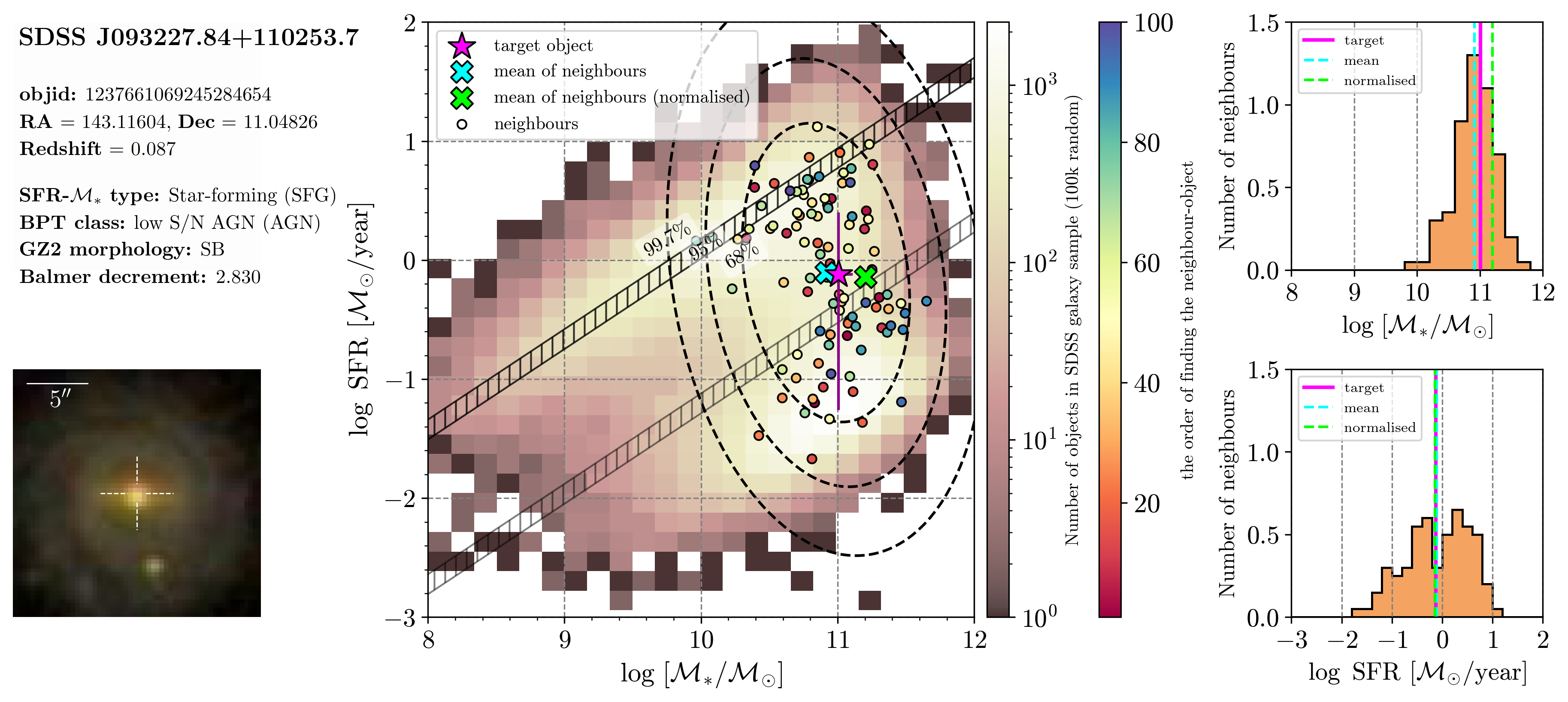}
\caption{The same as in Fig.\,\ref{fig:obj2}, but for the target object~\#2. The values of distances estimated for this target are presented in Table\,\ref{tab:indiv-targets}. }
    \label{fig:obj83}
\end{figure*}

\begin{figure*}
    \centering
    \includegraphics[width=.9\textwidth]{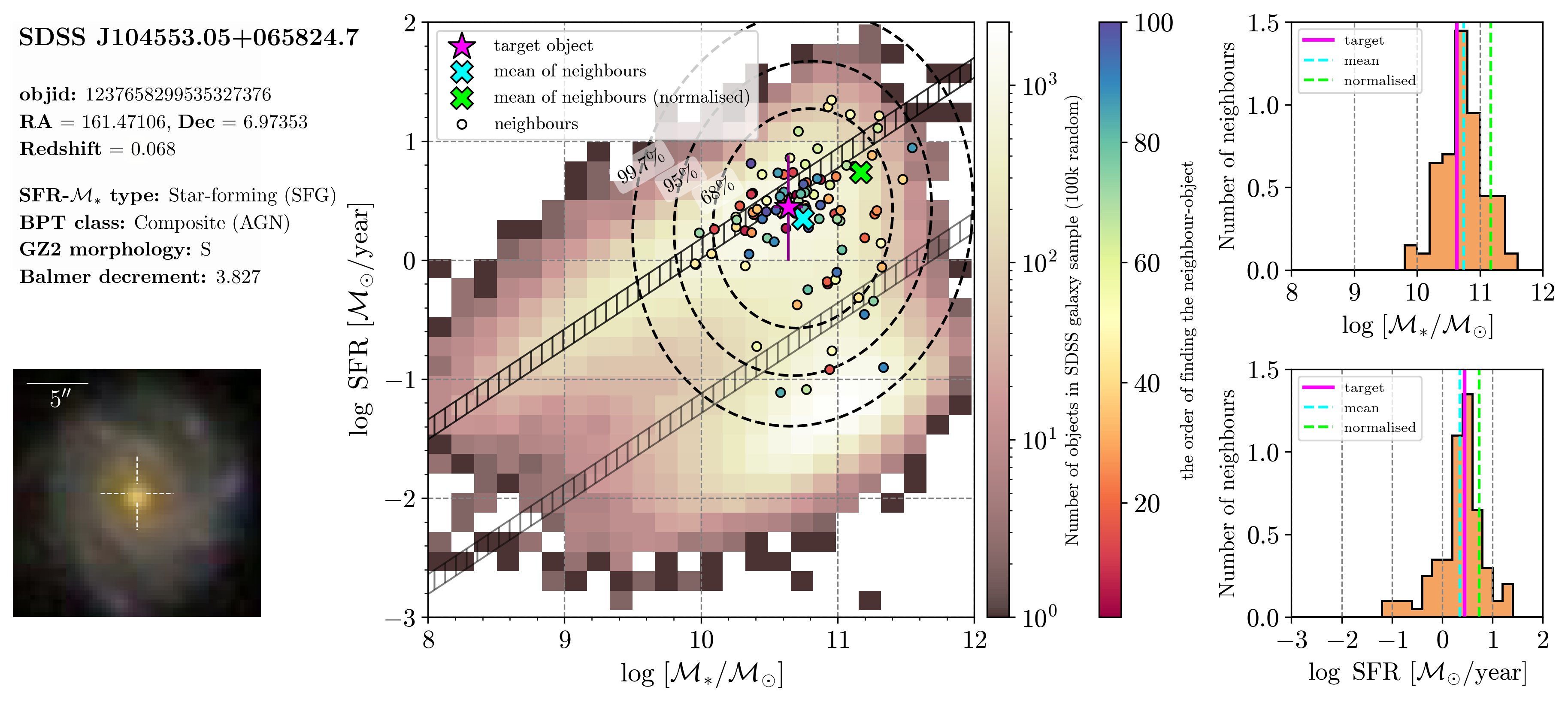}
    \caption{The same as in Fig.\,\ref{fig:obj2}, but for the target object~\#3. The values of distances estimated for this target are presented in Table\,\ref{tab:indiv-targets}.}
    \label{fig:obj184}
\end{figure*}

\begin{figure*}
    \centering
    \includegraphics[width=.9\textwidth]{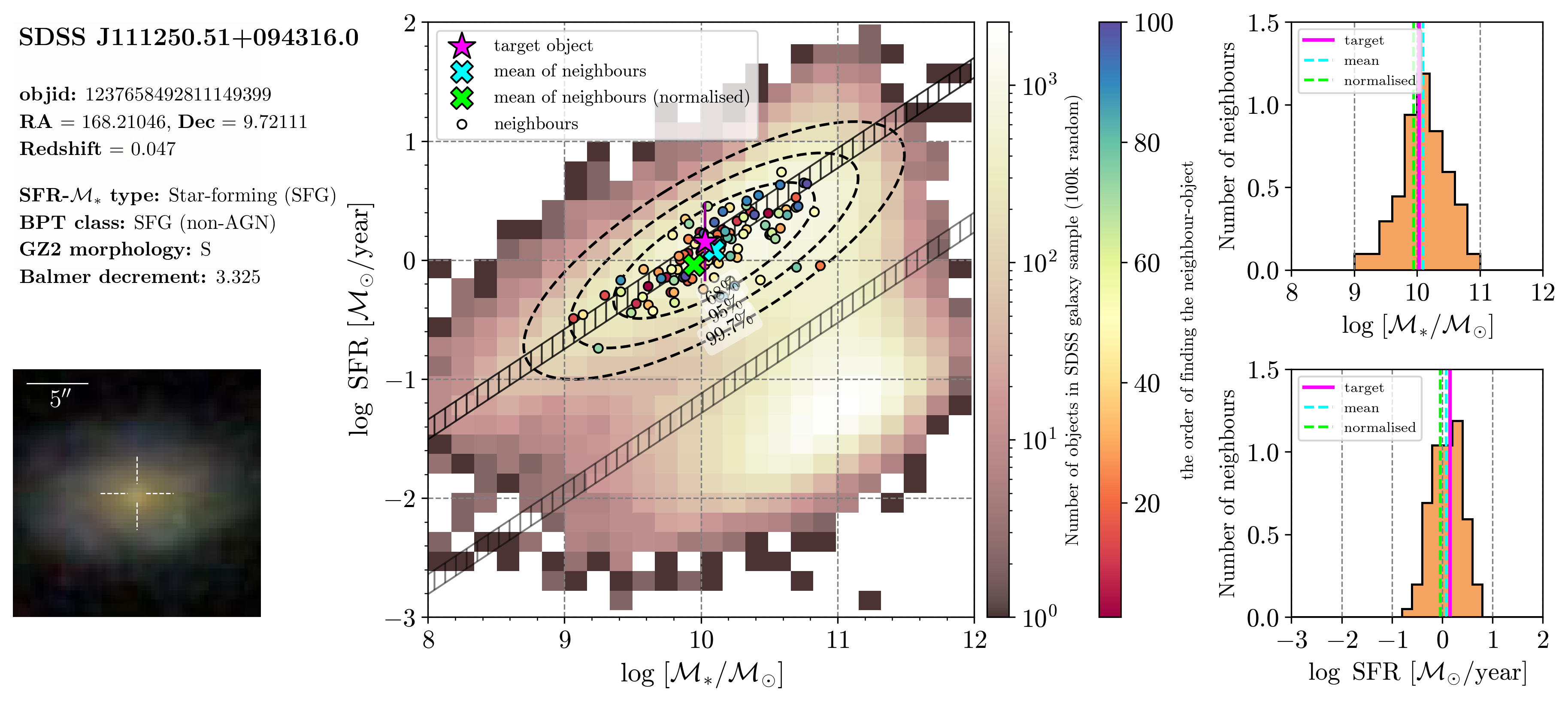}
    \caption{The same as in Fig.\,\ref{fig:obj2}, but for the target object~\#4. The values of distances estimated for this target are presented in Table\,\ref{tab:indiv-targets}.}
    \label{fig:obj217}
\end{figure*}

\begin{figure*}
    \centering
    \includegraphics[width=.9\textwidth]{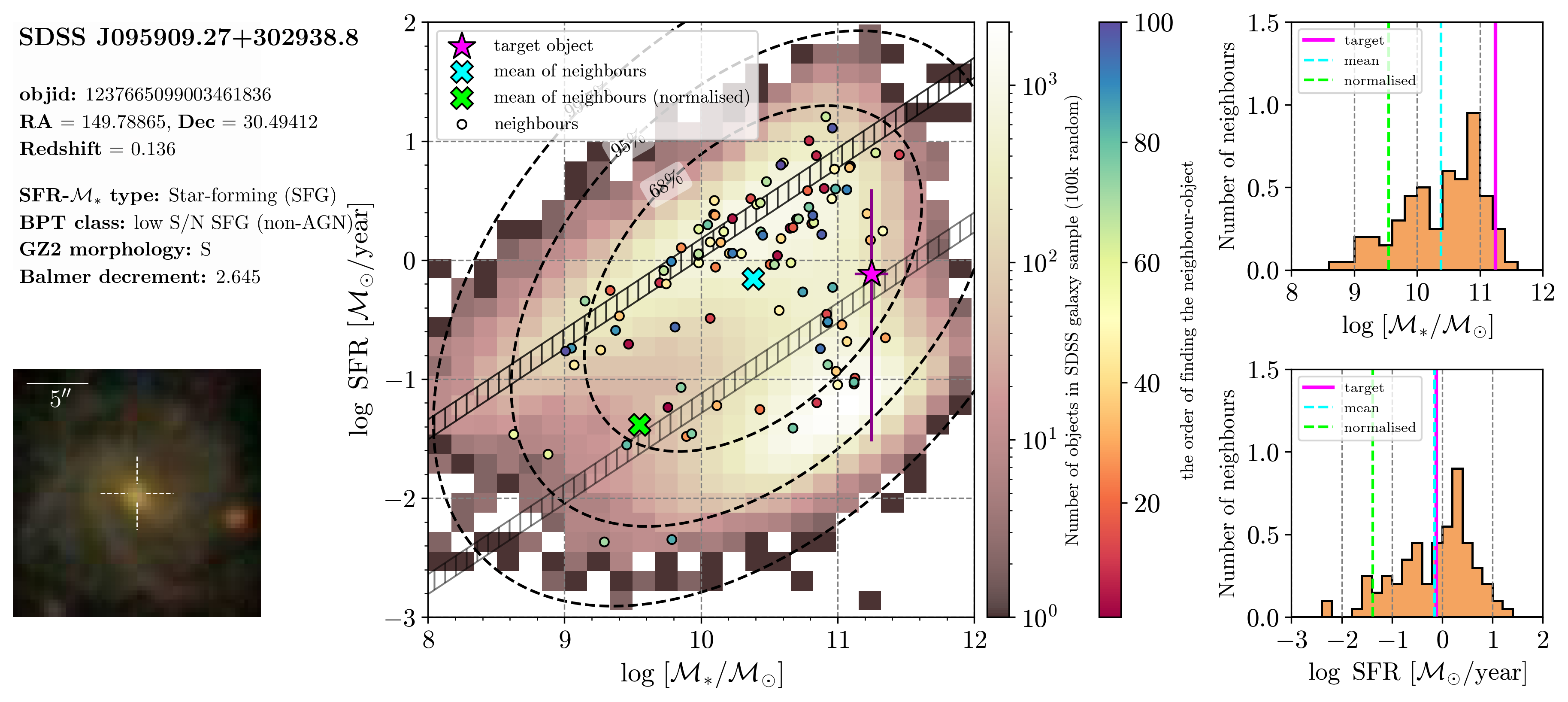}
    \caption{The same as in Fig.\,\ref{fig:obj2}, but for the target object~\#5. The values of distances estimated for this target are presented in Table\,\ref{tab:indiv-targets}.}
    \label{fig:obj215}
\end{figure*}

\begin{figure*}
    \centering
    \includegraphics[width=.9\textwidth]{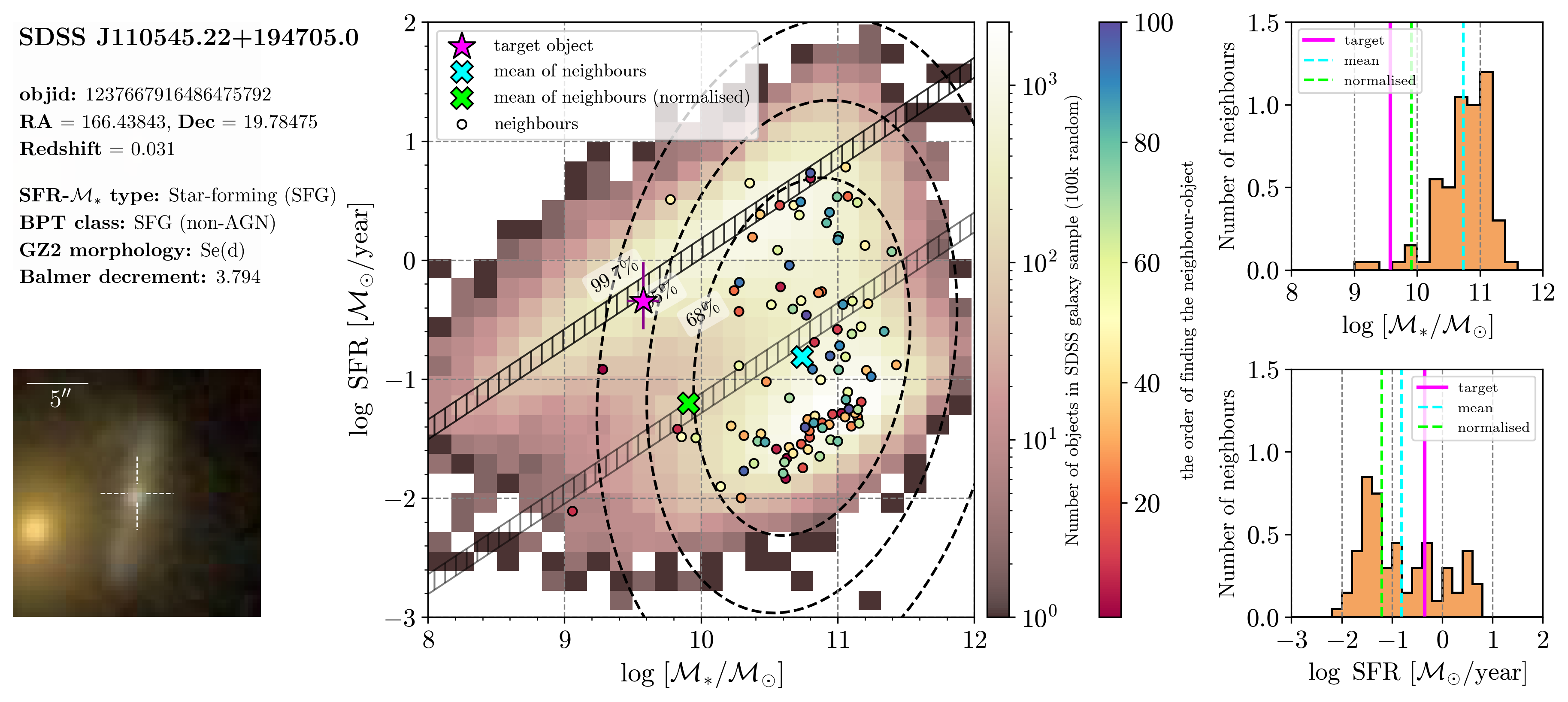}
    \caption{The same as in Fig.\,\ref{fig:obj2}, but for the target object~\#6. The values of distances estimated for this target are presented in Table\,\ref{tab:indiv-targets}.}
    \label{fig:obj276}
\end{figure*}

We calculated the six distances defined above for each of 290 target objects introduced in Section\,\ref{sec:prototypes}. In the following, we start presenting the results for six `template' targets, each one representing a different galaxy class and various positions in the SFR-$\mathcal{M}_{\ast}$ diagram. To simplify the discussion of the results, we refer to each target with the order in which they are presented in Fig.\,\ref{fig:obj2}-\ref{fig:obj276} (i.e. target \#1, \#2, etc).

In Fig.\,\ref{fig:obj2} we show the \ULISSE results for the elliptical galaxy {\bf SDSS\,J113337.09+114955.9} (i.e. target object~\#1), which represents a galaxy with smooth morphology (`E') according to the GZ2 catalogue and shows no evidence of AGN activity (i.e. this object is unclassified based on the BPT selection criterion meaning that its spectrum shows weak or no emission lines). The neighbours retrieved by \ULISSE are well clustered around the target object position in the SFR-$\mathcal{M}_{\ast}$ diagram. The distances presented in Table\,\ref{tab:indiv-targets} are relatively small indicating a close correspondence between the position of the target and averaged values of SFR and $\mathcal{M}_{\ast}$ of the neighbours, which thus represent good tracers of the target properties. { As seen in Table\,\ref{tab:indiv-targets} the scatter in $\mathcal{M}_{\ast}$ and SFR is modest ($\Delta{\mathcal{M}_{\ast}} = 0.29$ and $\Delta{\mathrm{SFR} = 0.42}$), suggesting a relatively homogeneous sample. The compactness  is significantly below unity ($\mathcal{C} = 0.43$), indicating that the neighbours are tightly clustered in the parameter space. The shape asymmetry ($A_{S} = 1.5$) indicates a moderately larger spread in SFR than in stellar mass, though without a strong dominance of one parameter over the other. The aspect ratio $\mathcal{R} = 2.9$ confirms the absence of strong elongation or preferred orientation. Overall, the combination of these parameters indicates a structurally coherent set of retrieved neighbours, reinforcing the robustness of \ULISSE estimates for target~\#1.}

The target object~\#2 presented in Fig.\,\ref{fig:obj83} is the star-forming galaxy {\bf SDSS\,J093227.84+110253.7} with spiral morphology revealing the presence of a bar component (i.e. `SB' morphology) and a low S/N AGN. The retrieved neighbours appear to be located in the narrow range of stellar masses around $\mathcal{M}_{\ast}^{\rm targ}$, while they are significantly scattered in SFR. This scatter can be caused by the `mixed' appearance of the target~\#2 showing `redder' nuclear/bar component of the galaxy (i.e. lower level of star-formation) simultaneously with `blue' (and more star-forming) spiral disk structure of the galaxy, implying that \ULISSE finds similarities with both passive and star-forming sources. However, analysing Table\,\ref{tab:indiv-targets} we can see that both `physical' and weighted distances are small for this target. Moreover, the estimated weighted distances are lying within $2\sigma$ uncertainty of spectroscopic SFR and $\mathcal{M}_{\ast}$ indicating that on average the retrieved SFR and $\mathcal{M}_{\ast}$ still trace well the properties of the target. Furthermore, the large uncertainty in the spectroscopic SFR demonstrates that for such targets the estimate of the stellar formation rate is intrinsically difficult even with spectroscopic SDSS data. The values of $\Delta{\mathcal{M}_{\ast}} = 0.32$ and $\Delta{\mathrm{SFR}} = 0.65$ reflect a relatively tight clustering in stellar mass, while the increased scatter in SFR highlights the intrinsic complexity of the target's SF activity. The compactness of $0.76$ indicates a moderately concentrated neighbour distribution, though less compact than the previous case. The shape asymmetry $A_{S} = 2.1$ and the relatively high aspect ratio $\mathcal{R} = 4.4$ together indicate a neighbour distribution that is elongated predominantly along the SFR axis, reflecting the dual nature of SF in the target's less active nuclear/bar region and more star-forming spiral disk component. 

A similar result was obtained for star-forming galaxy {\bf SDSS\,J104553.05+065824.7} (the target object~\#3) which presents a spiral morphology (`S') and likely hosts an AGN (see Figure\,\ref{fig:obj184}). As in the previous case, we see some scatter of retrieved neighbours toward lower SFR values, most likely due to the presence of the strong red nuclear component. The mean distances for retrieved neighbours are also small (see Table\,\ref{tab:indiv-targets}) pointing to a good agreement between the properties of the target~\#3 and the average retrieved parameters. It should be noted that the normalised mean position for neighbours is slightly overestimated with respect to the position of the target (and out of its 5$\sigma$ significance range), although it remains along the MS of SFG. Such a shift toward higher stellar masses can be possibly brought by the presence of the strong nuclear component (i.e. AGN) in the discussed galaxy or by the fact that \ULISSE found a similarity with some `rare' sources and the normalisation overemphasises their contribution. In the latter case, the remedy would be to use fewer neighbours, as discussed later. The neighbour distribution shows moderate spreads in both $\mathcal{M}_{\ast}$ and SFR ($\Delta{\mathcal{M}_{\ast}} = 0.34$ and $\Delta{\mathrm{SFR}} = 0.48$) indicating a reasonably consistent clustering around the target galaxy. The compactness parameter ($\mathcal{C} = 0.60$) reflects a fairly concentrated spatial distribution in the SFR–$\mathcal{M}_{\ast}$ plane. The shape asymmetry ($A_{S} = 1.4$) points to a near-equilibrium dispersion between SFR and $\mathcal{M}_{\ast}$.

Such assumption can be tested on the target object~\#4, star-forming galaxy {\bf SDSS\,J111250.51+094316.0} belonging to the same morphological class `S', but without the presence of a bright nucleus or an AGN according to the BPT diagram (see Fig.\,\ref{fig:obj217}). As a result, the positions of the retrieved neighbours are less scattered than for the two previous objects (\#2 and \#3) and are concentrated along the MS. As can be seen in Table\,\ref{tab:indiv-targets} the mean distances are small, indicating an excellent agreement between the SFR and $\mathcal{M}_{\ast}$ of the target and those derived from the neighbours. We also point out that, for all previous targets (\#1-4) the weighted distances are within 3$\sigma$ of the target (except $d_{\rm mean}^{\rm norm}$ for \#3 as discussed above), indicating good performance of our method when also considering the uncertainties in the spectroscopic estimates of SFR and $\mathcal{M}_{\ast}$. The neighbour distribution shows moderate spreads in $\mathcal{M}_{\ast}$ and SFR ($\Delta{\mathcal{M}_{\ast}} = 0.38$ and $\Delta{\mathrm{SFR}} = 0.30$) both indicating good retrieval efficiency. The relatively large aspect ratio ($\mathcal{R} = 7.3$) indicates a pronounced elongation of the neighbour distribution, consistent with visual alignment along the MS of SFG (see Fig.\,\ref{fig:obj217}). Despite this elongation, the compactness parameter ($\mathcal{C} = 0.28$) demonstrates that the neighbours remain highly concentrated overall, reflecting a tightly clustered distribution around the target.

Nevertheless, for some selected targets the performance of our method was not as effective as shown above. For instance, for the star-forming galaxy {\bf SDSS\,J095909.27+302938.8} with a spiral (`S') morphology and a relatively bright nucleus (the target \#5 in Figure\,\ref{fig:obj215}) the neighbours are spread over a broad range of stellar masses and SFR with a slight concentrating around the MS. The distances presented in Table\,\ref{tab:indiv-targets} are also large, indicating a significant mismatch between the average position of the neighbours and the target object. Such poor performance may be the result of the more complex morphology of target \#5 and the fact that it fills the thumbnail, but could also indicate that the spectroscopic values of the mass and SFR (which has a large uncertainty) are biased toward the nuclear properties (possibly due to the SDSS fibre size). In fact, most neighbours cluster around the main sequence, suggesting a higher SFR more consistent with the observed spiral morphology). The relatively high dispersions $\Delta{\mathcal{M}_{\ast}} = 0.64$ and $\Delta{\mathrm{SFR} = 0.75}$ show that the neighbours are widely scattered across the parameter space, reflecting a less homogeneous grouping. This is supported by the large $\mathcal{C} = 1.61$, which quantifies the broad spatial distribution of neighbours, likely influenced by the complex morphology of the target and potential biases from limited spectroscopic aperture.

Another example of \ULISSE poor performance is presented in Fig.\,\ref{fig:obj276}. Target object~\#6 is an edge-on spiral galaxy {\bf SDSS\,J110545.22+194705.0}, which { shows the presence of disturbed morphology} in the GZ2 catalogue (`Se(d)' class in Fig.\,\ref{fig:obj276}). The location of the target is close to the MS at relatively low masses, while the neighbours are mostly retrieved as quiescent galaxies with higher stellar masses in the SFR-$\mathcal{M}_{\ast}$ diagram. The reason for such discrepancy is most likely due to the elliptical galaxy located near our target within our thumbnail. Moreover, this elliptical `companion' appears brighter than the target object, which may lead \ULISSE to focus on its properties rather than those of the target. This confirmed also by the substantial scatter in $\Delta{\mathrm{SFR} = 0.78}$, while the stellar mass dispersion remains moderate ($\Delta{\mathcal{M}_{\ast}} = 0.41$). Together with a relatively broad neighbour distribution ($\mathcal{C} = 1.19$) these parameters shows that a retrieval efficiency is influenced by environmental complexity.

The issues highlighted above, which may significantly degrade the performance of our method, can be partly addressed by using larger thumbnails or masking nearby and/or overlapping sources.

\begin{figure*}
    \centering
    \includegraphics[width=.95\textwidth]{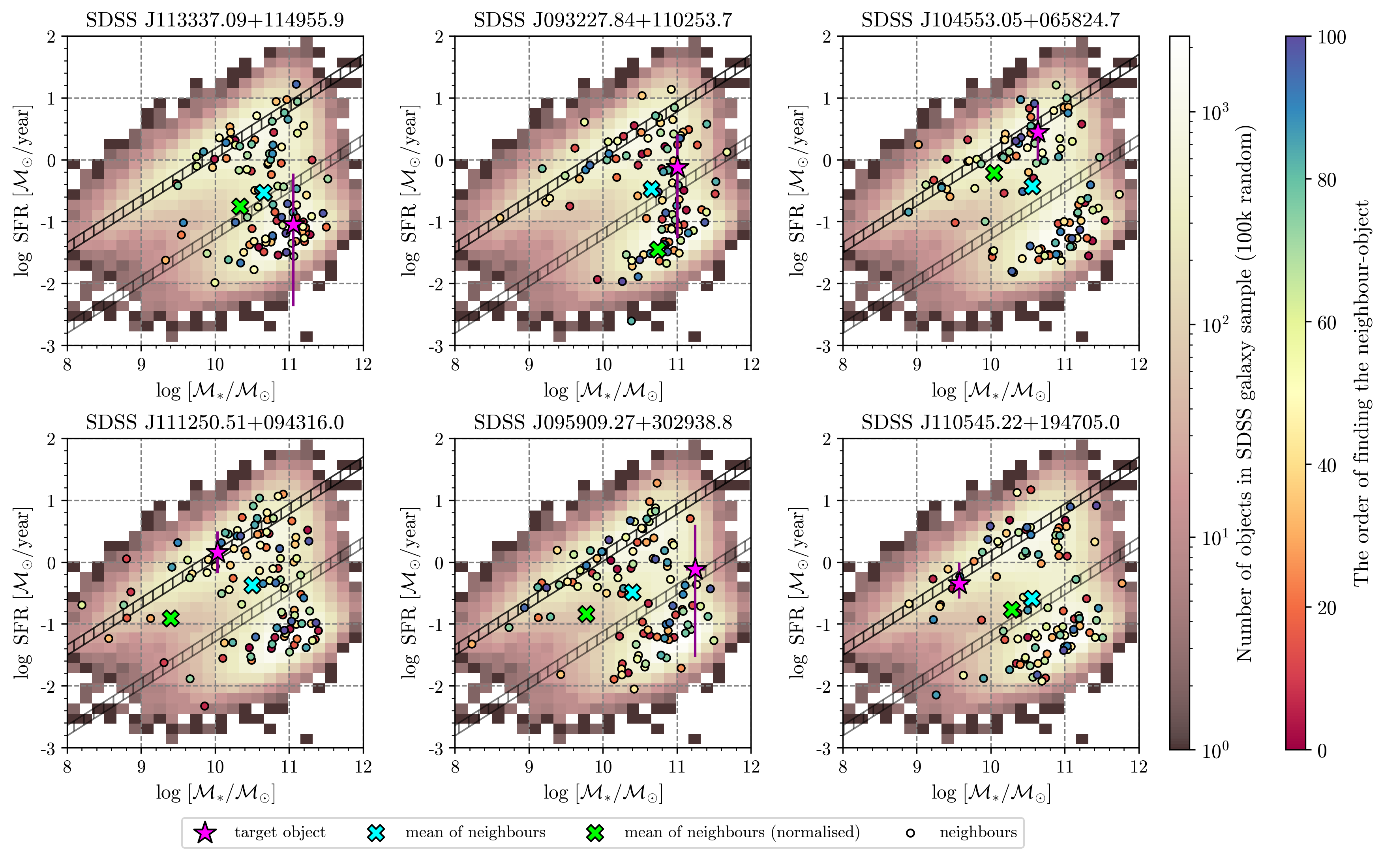}
    \caption{ The similar SFR vs $\mathcal{M}_{\ast}$ distribution for target objects presented in Fig.\,\ref{fig:obj2}-\ref{fig:obj276}, but 100 neighbours have been retrieved randomly from the sample.}
    \label{fig:random-sfr-m}
\end{figure*}

In order to compare the efficiency of our method with a purely random guess, for each target object we randomly selected 100 sources from our sample. The distribution of the randomly-selected sources in the SFR-$\mathcal{M}_{\ast}$ diagram and their total, mean and normalised mean distances for targets~\#1-6 are presented in Fig.\,\ref{fig:random-sfr-m} and Table\,\ref{tab:indiv-targets}. As expected, the random sources are distributed in the most populated areas of the SFR-$\mathcal{M}_{\ast}$ diagram, and therefore, their means lie in the `valley' between the populations of the star-forming and quiescent galaxies with $10^{9}-10^{11}\,\mathcal{M}_{\odot}$. While Table\,\ref{tab:indiv-targets} shows that the distances remain within 1\,dex from the target, this trend is mainly due to targets originally located near the `center' of the SFR-$\mathcal{M}_{\ast}$ diagram. In fact, the distances derived for the random samples are generally larger than those estimated for actual neighbours and are comparable only in the case of the objects \#5 and \#6, for which our method has shown a lower performance (see discussion above). 

\begin{table*}
\caption{The total, mean and normalised mean distances (and their weighted version) for the target objects~\#1-6 received by our method and random guess (presented in the squared parenthesis). The columns also include the standard deviations in $\mathcal{M}_{\ast}$ and SFR ($\Delta{\mathcal{M}_{\ast}}$ and $\Delta{\mathrm{SFR}}$), along with shape asymmetry ($A_{S}$), aspect ratio ($\mathcal{R}$), and compactness ($\mathcal{C}$) of the neighbour distributions for these target objects.}
\begin{center}
\fontsize{8}{10}\selectfont
\setlength{\tabcolsep}{5pt}
\begin{tabular}{ccccccccccccc}
\hline\hline\\[-1.7ex]
\# & SDSS name & $d_{\rm total}$ & $d_{\rm mean}$ & $d_{\rm norm}$ & $d_{\rm total}^{\rm weighted}$ & $d_{\rm mean}^{\rm weighted}$ & $d_{\rm norm}^{\rm weighted}$ & { $\Delta{\mathcal{M}_{\ast}}$} &  {$\Delta{\mathrm{SFR}}$} & {$A_{S}$} & {$\mathcal{R}$} & {$\mathcal{C}$} \\[1.3ex]
\hline\hline\\[-1.3ex]
1 & J113337.09+114955.9 & 0.41 [0.95] & 0.07 [0.65] & 0.21 [0.78] &  2.61 [5.41]  &  0.60 [4.46] &  0.87 [8.02] & {0.29} & {0.42} & {1.5} &  {2.9} & {0.43}\\
2 & J093227.84+110253.7 & 0.66 [0.97] & 0.10 [0.50] & 0.20 [1.35] &  2.91 [4.99]  &  0.99 [3.78] &  2.07 [3.29] & {0.32} & {0.65}  & {2.1} & {4.4} & {0.76} \\
3 & J104553.05+065824.7 & 0.49 [1.17] & 0.14 [0.88] & 0.61 [0.88] &  3.23 [6.07]  &  1.15 [2.18] &  5.64 [6.42] & {0.34} & {0.48}  & {1.4} & {2.0} & {0.60} \\
4 & J111250.51+094316.0 & 0.42 [1.13] & 0.10 [0.70] & 0.21 [1.24] &  3.94 [9.52]  &  0.83 [5.99] &  1.20 [8.45] & {0.38} & {0.30}  & {0.8} & {7.3} & {0.28} \\
\hdashline
5 & J095909.27+302938.8 & 1.18 [1.25] & 0.87 [0.93] & 2.11 [1.64] &  7.91 [7.81]  &  7.76 [7.61] & 15.27 [13.22] & {0.64} & {0.75} &{1.2} &{ 2.7} & {1.61} \\ 
6 & J110545.22+194705.0 & 1.50 [1.39] & 1.25 [1.01] & 0.92 [0.82] & 11.67 [10.61] & 11.05 [9.25] &  4.46 [6.82] & {0.41} & {0.78}  & {1.9} & {3.8} & {1.19} \\[0.8ex]
\hline
\hline
\end{tabular}
\end{center}
\label{tab:indiv-targets}
\end{table*}

\subsection{General efficiency of the method}\label{sec:results-histo}

To evaluate the general \ULISSE efficiency we studied the cumulative distribution of the `distances' defined in Sec.\ref{sec:distances} for all targets in our sample (see Fig.\,\ref{fig:hist-dist0}-\ref{fig:hist-dist3} in Appendix\,\ref{sec:appendix}). Using these distributions the \ULISSE efficiency can be defined as the fraction of target objects (of a certain class) having the corresponding `physical' distance smaller than a certain value, i.e. the position in the SFR--$\mathcal{M}_{\ast}$ diagram estimated based on the retrieved neighbours, is within some threshold value from the position of the target estimated based on spectroscopic data. A similar approach can be applied to the weighted distances in order to obtain the significance threshold of the offset based on the uncertainties on $\mathcal{M}_{\ast}$ and SFR of the target. In this work, we chose the values 1\,dex and 5$\sigma$ as cuts to define the \ULISSE efficiency in retrieving galaxy properties; however, the threshold can be adapted to the specific science case of interest (favouring, e.g., completeness versus accuracy) using the distributions in Fig.\,\ref{fig:hist-dist0}-\ref{fig:hist-dist3}. Table\,\ref{tab:dist-dex-fractions} presents the fractions of targets falling below the defined thresholds.

Analysing the results reported in Tables\,\ref{tab:dist-dex-fractions} (and figures in Appendix\,\ref{sec:appendix}) we found that our algorithm, based on the average properties of the neighbours found by \ULISSE, is able to provide an estimate of the target galaxy properties within 1\,dex of the expected value for 60-80\perc of the studies galaxies. At the same time, a disjoint analysis of star-forming and quiescent galaxies reveals a slightly increased efficiency for quiescent galaxies (78\perc on average for the three distances) compared to SFG (72\perc), which can be due to the fact that quiescent galaxies represent a homogeneous population of massive and old galaxies in the SFR-$\mathcal{M}_{\ast}$ diagram, while SFG exhibit the variety of morphological features (e.g. spirals, bars, etc) and a wider range of physical parameters. At the same time,  the spectroscopically estimated SFR for SFG with a more complex appearance (i.e. the presence of multiple stellar components) can be easily miscalculated due to the limitations in the SDSS fiber size and its positioning on the particular galaxy component. However, it should be also pointed out higher efficiency for quiescent galaxies must be interpreted with caution due to the limitation of our primary sample. The quiescent sample primarily includes high-mass, red-sequence galaxies with very low SFRs, occupying a relatively narrow region in the SFR–$\mathcal{M}_{\ast}$ diagram (typically 1.5–2.5\,dex in both axes). In such a constrained parameter space, even simple similarity-based retrievals are likely to succeed. Therefore, part of the observed performance may reflect the limited diversity of our quiescent sample, rather than indicating universally high efficiency of the method across all early-type systems, and as a result require more detailed studies on a broader and more diverse sample of quiescent galaxies in the future. This is also evident from the average values of the scatter parameters presented in Table\,\ref{tab:dist-scatter-fractions}. For instance, the average spread in $\mathcal{M}_{\ast}$ is slightly higher for SFGs ($\Delta\mathcal{M}_{\ast} = 0.40$) than for QGs ($\Delta\mathcal{M}_{\ast} = 0.33$), while the SFR scatter is comparably large for both populations ($\Delta{\mathrm{SFR}} \approx 0.6$).

At the same time, the presence of an AGN in the galaxy core seems to increase only slightly the efficiency of our method (77.9\perc for AGN versus 71.2\perc for non-AGN target objects). The average scatter in stellar mass ($\Delta\mathcal{M}_\ast$) is comparable between AGN and non-AGN target objects. In contrast, $\Delta$SFR appears to be larger for AGN hosts, possibly due to contamination from AGN emission affecting the galaxy’s integrated colors and thereby biasing the inferred star formation properties.

Comparing the fractions of galaxies with different morphology classes we can see that elliptical galaxies (E) seem to have relatively lower efficiency (near 68.5\perc on average for three distances) compared to all other classes (76.0\perc, 77.2\perc, 75.0\perc, and 72.6\perc on average for objects with S, SB, Se and disturbed morphology, respectively). This result shows the apparent contradiction with the results presented above for the case of quiescent/star-forming galaxies defined by their position below the MS of SFG (in Section\,\ref{sec:prototypes}), where quiescent galaxies appears to have higher efficiency with respect to SFG (see Table\,\ref{tab:dist-dex-fractions}). However, as we pointed out in Section\,\ref{sec:prototypes} the elliptical (`E') class in the GZ catalogue {is based purely on smooth visual appearance and does not strictly correspond to classical elliptical morphology. As a result, this group includes not only true early-type galaxies but also a fraction of blue, star-forming systems whose features are poorly resolved due to limited image resolution} (which can be also seen from the distribution of `E' galaxies in the SFR-$\mathcal{M}_{\ast}$ diagram in Fig.\,\ref{fig:prot-N-samples}). At the same time, both $\Delta\mathcal{M}_{\ast}$ and $\Delta\mathrm{SFR}$ appear to be similar across target objects with different morphological classes.

The further investigation of \ULISSE performance against dust-related reddening shows that 
the retrieving efficiency remains relatively constant for low-, moderate-, and high-dust content subsamples of galaxies (72.2\perc, 73.8\perc, and 70.9\perc on average, respectively). The same trend is seem for the average scatter of retrieved stellar mass ($\Delta\mathcal{M}_{\ast}$ in Table\,\ref{tab:dist-scatter-fractions}) for different dust content samples, while $\Delta\mathrm{SFR}$ scatter is showing a modest increase with dust content of galaxies (from 57\perc for low dust content to 65\perc for high dust). This can be likely due to the fact that \ULISSE selection of similar features also accounts for typical dust-induced reddening features present in galaxies within the sample. Consequently, dusty galaxies tend to be matched with analogues of comparable dust content and morphological appearance, reducing biases caused by dust obscuration and age-metallicity degeneracies. And as a result, it allows \ULISSE to maintain stable performance even when relying solely on optical \textit{g-,r-,i-}band images.

In addition, we see that the efficiency is dependent on the redshift, with the highest efficiency (near 80.4\perc on average) for the central $0.05 < z < 0.1$ redshift interval with respect to 75.0\perc and 59.3\perc for $0.01 < z < 0.05$ and $0.1 < z < 0.15$ intervals, respectively. 
This is a by-product of the dependence of the angular size of galaxies from redshift, considering that we use a fixed thumbnail size. This leads to situations where at low redshifts we may be missing part of the information as we are cutting the edges of the galaxy, while at higher $z$ the resolution of our images may not be high enough to reveal the same number of morphological features. Across different redshift ranges, $\Delta\mathrm{SFR}$ remains relatively constant, while $\Delta\mathcal{M}_{\ast}$ shows a tendency to increase with redshift. This trend may reflect the greater sensitivity of $\mathcal{M}_{\ast}$ estimates to image resolution and structural detail, while SFR estimates, which are more closely tied to integrated colours, appear less affected across the redshift range considered.

The relations between $\Delta\mathcal{M}_{\ast}$ and $\mathcal{M}_{\ast}$, as well as between $\Delta\mathrm{SFR}$ and SFR, are presented in the supplementary material (see Fig.\,\ref{fig:dev_Mass_SFR} in Appendix\,\ref{sec:appendix3}). 

A summary of the neighbour distributions retrieved by \ULISSE across different subsamples is presented in Table\,\ref{tab:dist-scatter-fractions}. Across all classes (defined by galaxy morphology, AGN contribution, dust content, and redshift) the distributions tend to be compact, with compactness values $\mathcal{C} < 1$, indicating tightly clustered neighbour distribution in the SFR–$\mathcal{M}_{\ast}$ space. The shape asymmetry $A_S$, which characterizes whether scatter is more dominant in SFR or stellar mass, varies between subsamples: it is generally higher for quiescent galaxies compared to star-forming ones, reflecting larger uncertainties in SFR estimates for passive systems. A similar trend is seen for AGN-hosting galaxies and spiral morphologies relative to non-AGN and smooth (elliptical-like) systems, respectively, suggesting greater physical diversity or complexity in their star formation histories, structural features, or spectral energy distributions. The shape ratio $\mathcal{R}$ shows moderate values across all subsamples (typically $\sim$4), although its interpretation as elongation along the MS is only valid for targets located near the MS of star-forming galaxies. In off-sequence regions, $\mathcal{R}$ still provides a useful tracer of the neighbour distribution's geometry, but its physical interpretation may differ. A full distribution of these parameters for all 290 target objects is presented in Fig.\,\ref{fig:comp-As-R-morph} in Appendix\,\ref{sec:appendix3} and can serve as an additional diagnostic to assess the performance of \ULISSE on an object-by-object basis.

\begin{table*}
\caption{The \ULISSE efficiency in retrieving SFR and $\mathcal{M}_{\ast}$ for different classes of galaxies. The efficiency is defined as the fraction of target objects with the `physical' distance (total, mean, or normalised mean) smaller than 1\,dex or with the weighted distances smaller than 5$\sigma$ (see details in the text). }
\begin{center}
\fontsize{8}{10}\selectfont
\setlength{\tabcolsep}{5pt}
\begin{tabular}{lcccccccc}
\hline\hline\\[-2ex]
Target object & \multirow{2}{*}{$N$} & \multicolumn{3}{c}{within 1\,dex} & & \multicolumn{3}{c}{within 5$\sigma$}\\
\cline{3-5}\cline{7-9}\\[-2ex]
sample & & $d_{\rm total}$ & $d_{\rm mean}$ & $d_{\rm norm}$ & & $d_{\rm total}^{\rm weighted}$ & $d_{\rm mean}^{\rm weighted}$ & $d_{\rm norm}^{\rm weighted}$ \\[0.5ex]
\hline\hline
All & 290 & 69.0\perc [14.8\,\%] &  80.0\perc [60.3\,\%] & 72.4\perc [34.8\,\%]  & & 54.5\perc [15.2\,\%] & 65.2\perc [44.1\,\%] &  65.2\perc [29.0\,\%] \\
\hline
SFG & 205 & 66.8\perc [16.1\,\%] & 76.6\perc [54.6\,\%] & 72.7\perc [39.5\,\%]  & & 46.8\perc [12.2\,\%] &  60.5\perc [41.0\,\%] & 58.5\perc [29.3\,\%] \\
QG & 85 & 74.1\perc [11.8\,\%] & 88.2\perc [71.8\,\%] & 71.8\perc [23.5\,\%] & &  69.4\perc [12.9\,\%] & 76.5\perc [51.8\,\%] & 80.0\perc [24.7\,\%] \\
\hdashline
AGN & 130 & 73.8\perc [22.3\,\%] & 86.2\perc [76.2\,\%] & 73.8\perc [33.8\,\%]  & & 63.8\perc [9.2\,\%] &   74.6\perc [57.7\,\%] & 73.1\perc [26.2\,\%] \\
non-AGN & 160 & 65.6\perc [10.6\,\%] & 75.0\perc [47.5\,\%] & 73.1\perc [35.6\,\%]  & & 43.8\perc [13.1\,\%] & 57.5\perc [33.1\,\%] & 58.5\perc [28.1\,\%] \\
\hdashline
E & 36 & 63.9\perc [16.7\,\%] & 66.7\perc [58.3\,\%] & 75.0\perc [41.7\,\%]  & & 41.7\perc [5.6\,\%] & 61.1\perc [41.7\,\%] & 69.4\perc [22.2\,\%] \\
S & 64 & 70.3\perc [10.9\,\%] & 84.4\perc [59.4\,\%] & 73.4\perc [18.8\,\%]  & & 53.1\perc [7.8\,\%] & 70.3\perc [46.9\,\%] & 60.9\perc [17.2\,\%] \\
SB & 60 & 68.3\perc [16.7\,\%] & 83.3\perc [61.7\,\%] & 80.0\perc [45.0\,\%]  & & 58.3\perc [11.7\,\%] & 66.7\perc [50.0\,\%] & 68.3\perc [28.3\,\%] \\
Se & 12 & 66.7\perc [25.0\,\%] & 75.0\perc [66.7\,\%] & 83.3\perc [41.7\,\%]  & & 50.0\perc [16.7\,\%] & 50.0\perc [50.0\,\%] & 66.7\perc [25.0\,\%] \\
{ E(d), S(d), SB(d), Se(d)} & 118 & 67.8\perc [16.9\,\%] & 79.7\perc [59.3\,\%] & 70.3\perc [36.4\,\%]  & & 55.1\perc [16.1\,\%] & 66.9\perc [44.1\,\%] & 62.7\perc [32.2\,\%] \\
\hdashline
$0.01 < z < 0.05$ & 108 & 65.7\perc [12.0\,\%] & 76.9\perc [50.0\,\%] & 82.4\perc [43.5\,\%]  & & 35.2\perc [6.5\,\%] & 50.9\perc [38.9\,\%] & 57.4\perc [36.1\,\%] \\
$0.05 < z < 0.1$ & 119 & 77.3\perc [23.5\,\%] & 88.2\perc [71.4\,\%] & 75.6\perc [39.5\,\%]  & & 68.1\perc [20.2\,\%] & 84.0\perc [64.7\,\%] & 71.4\perc [31.9\,\%] \\
$0.1 < z < 0.15$ & 63 & 52.4\perc [4.8\,\%] & 68.3\perc [52.4\,\%] & 57.1\perc [12.7\,\%]  & & 46.0\perc [4.8\,\%] &  61.9\perc [15.9\,\%] & 61.9\perc [3.2\,\%] \\
\hdashline
{ $({\rm H}\alpha/{\rm H}\beta) \leq 2.86$} & { 36} & { 63.9\perc [11.1\,\%]} & { 77.8\perc [58.3\,\%]} & { 75.0\perc [19.4\,\%]}  & & { 58.3\perc [0.0\,\%]} & { 61.1\perc [44.4\,\%]} & { 58.3\perc [22.2\,\%]} \\
{ $2.86 < ({\rm H}\alpha/{\rm H}\beta) \leq 5.0$} & { 177 } & { 68.4\perc [14.1\,\%]} & { 80.8\perc [56.5\,\%]} & { 72.3\perc [38.4\,\%]}  & & { 49.2\perc [11.9\,\%]} & { 67.8\perc [40.7\,\%]} & { 64.4\perc [28.2\,\%]} \\
{ $({\rm H}\alpha/{\rm H}\beta) > 5.0$} & { 63} & { 68.3\perc [15.9\,\%]} & { 73.0\perc [61.9\,\%]} & { 71.4\perc [39.7\,\%]}  & & { 52.4\perc [7.9\,\%]} &  { 61.9\perc [49.2\,\%]} & { 60.3\perc [30.2\,\%]} \\
{ unknown dust content} & { 14} & { 78.6\perc [35.7\,\%]} & { 92.9\perc [85.7\,\%]} & { 71.4\perc [28.6\,\%]}  & & { 92.9\perc [42.9\,\%]} &  { 92.9\perc [71.4\,\%]} & { 85.7\perc [21.4\,\%]} \\
\hline\hline
\end{tabular}
\end{center}
\label{tab:dist-dex-fractions}
\end{table*}

\begin{table}
\caption{Average values of standard deviations in stellar mass ($\Delta{\mathcal{M}_{\ast}}$) and SFR ($\Delta{\mathrm{SFR}}$), and two-dimensional scatter ($\Delta_{2D}$), as well as shape asymmetry ($A_{S}$), aspect ratio ($\mathcal{R}$), compactness ($\mathcal{C}$) for the distributions of the retrieved neighbours of different classes of target galaxies. Definitions of all parameters are provided in Section\,\ref{sec:distances}.}
\begin{center}
\fontsize{8}{10}\selectfont
\setlength{\tabcolsep}{4pt}
\begin{tabular}{lcccccc}
\hline\hline\\[-2ex]
{ Target object sample} & { {$\langle\Delta{\mathcal{M}_{\ast}}\rangle$}} & { {$\langle\Delta{\mathrm{SFR}}\rangle$}} & { {$\langle\Delta_{\mathrm{2D}}\rangle$}} & { {$\langle A_{S} \rangle$}} & { {$\langle \mathcal{R} \rangle$}} & { {$\langle\mathcal{C}\rangle$}} \\[0.5ex]
\hline\hline
{ All} & { 0.38} & {0.61} & { 0.47} & { 1.7} & { 4.6} & { 0.79} \\
\hline
{ SFG} & { 0.40} & { 0.60} & { 0.48} & { 1.6} & { 4.8} & { 0.80}  \\
{ QG} & { 0.33} & { 0.62} & { 0.45} & { 1.9} & { 4.4} & { 0.75} \\
\hdashline
{ AGN} &  { 0.36} & { 0.66} & { 0.48} & { 1.9} & { 4.3} & { 0.85}  \\
{ non-AGN} & { 0.39} & { 0.57} & { 0.47} & { 1.5} & { 4.9} & { 0.74} \\
\hdashline
{ E} &  { 0.37} & { 0.55} & { 0.45} & { 1.5} & { 3.9} & { 0.72} \\
{ S} & { 0.38} & { 0.60} & { 0.47} & { 1.7} & { 4.8} & { 0.77}  \\
{ SB} & { 0.36} & { 0.62} & { 0.46} & { 1.8} & { 4.5} & { 0.76} \\
{ Se} & { 0.40} & { 0.66} & { 0.51} & { 1.7} & { 4.5} & { 0.91} \\
{ E(d), S(d), SB(d), Se(d)} & { 0.39} & { 0.62} & { 0.48} & { 1.7} & { 4.9} & { 0.82}  \\
\hdashline
{ $0.01 < z < 0.05$} & { 0.41} & { 0.57} & { 0.47} & { 1.5} & { 5.2} & { 0.75} \\
{ $0.05 < z < 0.1$} & { 0.37} & { 0.63} & { 0.48} & { 1.8} & { 4.2} & { 0.82} \\
{ $0.1 < z < 0.15$} & { 0.35} & { 0.63} & { 0.46} & { 1.9} & { 4.4} & { 0.79} \\
\hdashline
{ $({\rm H}\alpha/{\rm H}\beta) \leq 2.86$} & { 0.36} & { 0.57} & { 0.44} & { 1.7} & { 4.3} & { 0.72} \\
{ $2.86 < ({\rm H}\alpha/{\rm H}\beta) \leq 5.0$} & { 0.38} & { 0.60} & { 0.47} & { 1.7} & { 5.0} & { 0.77} \\
{ $({\rm H}\alpha/{\rm H}\beta) > 5.0$} & { 0.38} & { 0.65} & { 0.49} & { 1.8} & { 3.8} & { 0.88} \\
{ unknown dust content} & { 0.34} & { 0.65} & { 0.46} & { 2.0} & { 4.9} & { 0.78} \\
\hline\hline
\end{tabular}
\end{center}
\label{tab:dist-scatter-fractions}
\end{table}

\begin{table*}
\caption{The fraction of target objects with the mean, $\langle\mathcal{M}_{\ast}^{\rm neig}\rangle$ or $\langle{\rm SFR}^{\rm neig}\rangle$, or normalised mean, $\mathcal{M}_{\ast}^{\rm norm}$ or ${\rm SFR}^{\rm norm}$, values for the retrieved neighbours within 1\,dex\;$/\sqrt{2}$ from the $\mathcal{M}_{\ast}^{\rm targ}$ or ${\rm SFR}^{\rm targ}$ estimations of the target. }
\begin{center}
\fontsize{9}{12}\selectfont
\setlength{\tabcolsep}{5pt}
\begin{tabular}{lcccc}
\hline\hline\\[-2ex]
Target object sample & $\mathcal{M}_{\ast}^{\rm targ} - \langle\mathcal{M}_{\ast}^{\rm neig}\rangle$ & $\mathcal{M}_{\ast}^{\rm targ} - \mathcal{M}_{\ast}^{\rm norm}$ & ${\rm SFR}^{\rm targ} - \langle{\rm SFR}^{\rm neig}\rangle$ & ${\rm SFR}^{\rm targ} - {\rm SFR}^{\rm norm}$ \\[2pt]
\hline\hline
All & 94.1\perc [81.4\,\%] &  93.8\perc [61.0\,\%] & 81.0\perc [70.7\,\%] & 73.4\perc [57.2\,\%] \\ 
\hdashline
SFG & 92.2\perc [79.5\,\%] & 91.2\perc [62.9\,\%] & 80.0\perc [70.2\,\%] & 76.1\perc [54.1\,\%] \\
QG & 98.8\perc [85.9\,\%] & 100.0\perc [56.5\,\%] & 83.5\perc [71.8\,\%] & 67.1\perc [63.5\,\%] \\
\hline\hline
\end{tabular}
\end{center}
\label{tab:dist-dex-fractions-M-SFR}
\end{table*}

Analysing the distributions of distances for the random sample (see dashed lines in Fig.\,\ref{fig:hist-dist0}-\ref{fig:hist-dist3}) we found that the fraction of targets with distances below 1\,dex are in general half of those obtained by our method, meaning that \ULISSE is at least twice as efficient than a random guess. However, we point out that using the mean distance ($d_{\rm mean}$), the efficiency for the random sample is significantly larger compared to the other distances. This is mostly due to the fact that this distance indicator is less sensitive to outliers, and since most sources in the SFR-$\mathcal{M}_{\ast}$ diagram are clustered around the MS or in the quiescent galaxies region, (about 1.5$\times$1.5\,dex in size), even a random guess will often return close neighbours. Thus, the mean distance is not the best estimator to quantify the efficiency of our approach and we recommend using the normalised mean distance instead, which considers the varying density of objects in our sample across the SFR-$\mathcal{M}_{\ast}$ diagram, and also performs well using lower distance thresholds (e.g. 0.5\,dex instead of 1\,dex, Fig.\,\ref{fig:hist-dist0}-\ref{fig:hist-dist3}). 

The distribution of the weighted distances presents a similar trend, but the fractions of neighbours retrieved within 5$\sigma$ significance range are generally lower compared to the fractions obtained using a threshold of 1\,dex. As can be seen in Table\,\ref{tab:dist-dex-fractions}, independently of the galaxy subsample, our method always returns more than 50\perc of neighbours within 5$\sigma$ range for SFR and $\mathcal{M}_{\ast}$ of the targets. This is mostly due to the relatively small formal errors on the spectroscopic mass estimates (see Fig.\,\ref{fig:obj2}-\ref{fig:obj276}), so that in practical applications one may relax this constrain depending on the quality of the reference spectroscopic sample.

Our tests so far have considered the position of the neighbours of each target in the SFR-$\mathcal{M}_{\ast}$ diagram. However, as highlighted above and in Section\,\ref{sec:results-ind}, the uncertainties in both the expected mass of the target (based on SDSS spectroscopy) and the retrieved one (based on the neighbour dispersion) are usually much smaller than the one on the SFR. Thus, to further characterize our efficiency, in Fig.\,\ref{fig:hist-dist-M-SFR} (see Appendix\,\ref{sec:appendix2}) and Table\,\ref{tab:dist-dex-fractions-M-SFR} we present the retrieval fractions within 1\,dex$/\sqrt{2}$\footnote{Assuming a uniform distribution; However, assuming a Rayleigh distribution, i.e. normal distribution in both quantities, we get a slightly lower factor of 0.66.} considering separately the SFR and $\mathcal{M}_{\ast}$. Taking into account that the results presented in Table\,\ref{tab:dist-dex-fractions} are relatively similar for different galaxy subsamples, we decided to present results only for all, star-forming and quiescent galaxies. 

As expected, we find that our approach is more efficient in retrieving $\mathcal{M}_{\ast}$ than SFR, i.e. the fraction of neighbours returned by our method for the entire sample of target objects (i.e. `all galaxies') is near 94\perc within 1\,dex\;$/\sqrt{2}$ range (with $\sim$71\perc for the random approach), and it even reaches 100\perc applying the method for the subsample of quiescent galaxies (with $\sim$71\perc for the random approach). On the contrary, for the SFR the efficiency of our method is close to 77\perc on average. The efficiency of SFR determination for quiescent galaxies is compatible with the random method which is a consequence of the definition of the class, clustering at very low SFR within less than 1\,dex\;$/\sqrt{2}$ from the average value. In fact, we retrieve fractions corresponding to 1\,dex in the SFR-$\mathcal{M}_{\ast}$ diagram within 0.5\,dex in each individual quantity making the method even more robust.

\subsection{Applications and limitations of the \ULISSE method}\label{sec:discussion}

A central advantage of \ULISSE lies in its ability to estimate key galaxy properties such as stellar mass and star formation rate using only single-band images, without requiring any additional photometric, spectroscopic, or redshift information. This makes the method particularly well suited for application to current and upcoming wide-field imaging surveys (e.g., LSST, Euclid etc.), where some data may be incomplete, expensive to acquire, or entirely unavailable in early stages. Unlike most machine learning pipelines that rely on redshift-dependent training sets and regress directly on physical properties, \ULISSE operates by retrieving visually and structurally similar galaxies from a pre-characterized reference sample. This image-similarity-based approach captures morphological, structural, and colour features simultaneously, linking a galaxy’s appearance directly to physical properties in a data-driven way.

It is important to emphasize that \ULISSE does not perform absolute regression or classification, but rather a relative matching of each input galaxy to a reference sample whose properties are derived through photometric or spectroscopic measurements. This template-based comparison strategy allows \ULISSE to effectively transfer the accuracy and reliability of these measurements to larger imaging-only datasets. However, the precision of the stellar mass and SFR estimates is inherently tied to the quality and representativeness of the reference sample, because the method relies on the properties of the matched target objects. In addition to estimating physical properties, \ULISSE can also be used in a reverse mode as a similarity-based retrieval tool. For instance, given a small sample of rare or atypical galaxies (e.g., post-mergers, ring galaxies, or extreme starbursts) \ULISSE can identify other objects with similar appearance within large surveys. This capability allows for systematic identification of analogues to specific galaxy populations and supports the construction of statistically robust or morphologically uniform samples of rare systems, thereby providing a foundation for targeted follow-up observations and population analyses.

It worth to mention that the presence of underlying physical correlations between galaxy properties can enhance the performance of \ULISSE. A clear example is the star-forming main sequence, a well-established relation between stellar mass ($\mathcal{M}_\ast$) and star formation rate (SFR) for star-forming galaxies. Along this sequence, structural and colour properties tend to co-evolve, meaning that galaxies with similar visual appearance also share similar physical characteristics. This is seen in the left panel of Fig.\,\ref{fig:comp-As-R-morph}, where neighbour distributions that are compact ($\mathcal{C} \approx 1$) and symmetric ($A_S \approx 1$) also exhibit high shape ratios $\mathcal{R}$, indicative of elongated distributions aligned with the MS of SFG. Conversely, in regions of parameter space that are less well-populated in the reference sample (such as the green valley, rare morphologies, or low-mass systems) the lack of a consistent underlying relation and limited statistical support can reduce the reliability of the retrieved parameters.

To explore the limitations of \ULISSE in handling targets located in sparsely populated regions of the SFR–$\mathcal{M}\ast$ diagram, we examined the relationship between the distance ratio and various characteristics of the neighbour distributions. The distance ratio is defined as the ratio between the `unnormalised' mean distance of neighbours and the normalised mean distance ($d_{\rm mean}/d_{\rm norm}$), where the latter incorporates a weighting factor that reflects the rarity of the target object location within the SFR–$\mathcal{M}_\ast$ plane (see definition in Section\,\ref{sec:distances}). In this context, values of $d_{\rm mean}/d_{\rm norm}$ significantly greater than unity indicate targets located in sparsely populated regions of parameter space, where the `unnormalised' mean distance $d_{\rm mean}$ is biased by the limited number of similar objects.  As a result, the normalised mean distance $d_{\rm norm}$ provides a more reliable measure of the neighbour retrieval efficiency in these cases.

Intuitively, it might be expected that targets with high distance ratios (i.e. those in low-density or edge regions of the parameter space) would display neighbour distributions that are more scattered or asymmetric, reflecting lower efficiency of \ULISSE. However, our analysis (see right panel of Fig.\,\ref{fig:comp-As-R-morph}) reveals no significant correlation between the distance ratio and two-dimensional scatter $\Delta_{2D}$ (or shape asymmetry $A_S$) of the neighbour distributions. This result implies that \ULISSE retrieval efficiency is not determined solely by the density of the reference sample. Instead, other factors (e.g. image quality, resolution or other observational limitations) are likely to influence the fidelity of neighbour retrieval. This complexity highlights the intrinsic difficulty of correlating image-based similarity metrics with underlying physical galaxy properties, and indicates that simplistic dispersion measures may be insufficient to reliably flag targets prone to less accurate parameter estimation.

As shown in Table\,\ref{tab:dist-scatter-fractions}, \ULISSE achieves typical uncertainties of approximately 0.3–0.4\,dex in $\mathcal{M}_{\ast}$ and 0.5–0.6\,dex in SFR (see Fig.\,\ref{fig:dev_Mass_SFR}). Conventional techniques such as spectral energy distribution (SED) fitting~\citep{Bruzual:03, Salim:07, Conroy:13} or spectroscopic line diagnostics (e.g., \citealt{Brinchmann2004}) typically achieve uncertainties on the order of 0.1–0.3\,dex for $\mathcal{M}_{\ast}$ and less than 0.3–0.5\,dex for SFR, but these methods require high-quality multi-wavelength photometry or spectroscopy and are subject to systematic uncertainties associated with assumptions about star formation histories, dust attenuation, and stellar population models. Moreover, spectroscopic SFR indicators can be unreliable in quiescent galaxies or AGN-dominated systems due to weak or contaminated emission lines. While \ULISSE precision does not match that of physically motivated methods, its efficiency remains significant given that it relies solely on imaging data. Moreover, it provides robust estimates across diverse galaxy populations, including quiescent, star-forming, and AGN-hosting systems, and is relatively insensitive to dust effects or spectral contamination, making it a computationally efficient, minimal-input alternative for physical parameter estimation.

Another distinctive strength of \ULISSE is its independence from the presence of redshift information, which is an uncommon feature among modern machine learning (ML) approaches for estimating galaxy properties. Most traditional and ML-based techniques rely on luminosity-sensitive observables, requiring either spectroscopic redshifts or distance proxies as inputs~\citep{Bonjean:19, Dominguez:23_SFRmass, Zeraatgari:24}. In contrast, \ULISSE operates purely in image space, leveraging structural and colour information without requiring availability of redshift. This makes it particularly well suited for application in early stages of wide-field imaging surveys and datasets where distance information may be incomplete or unavailable. Nonetheless, while redshift is not an explicit input to our method, its performance is not entirely redshift-independent, as indicated by subtle differences in \ULISSE results across redshift ranges shown in Table\,\ref{tab:dist-dex-fractions}. This is expected because features such as angular size, surface brightness, and apparent morphology are affected by redshift-driven effects like cosmological dimming and resolution loss.

This also raises the question of whether \ULISSE, although not designed for this purpose, might implicitly capture some redshift information through its neighbour selection process. To explore this, we conducted a simple test comparing the spectroscopic redshift of each target galaxy with the average redshift of its retrieved neighbours (see Fig.\,\ref{fig:mean-Z} in Appendix\,\ref{sec:appendix3}). The obtained average absolute redshift differences shows to be relatively small, typically in the range of 0.02–0.03. Although this level of agreement is promising, it requires careful interpretation due to potential underlying uncertainties. The narrow redshift range of our dataset ($0.01 < z < 0.15$) limits morphological evolution and restricts the parameter space, which likely contributes to the observed consistency. Interestingly, Fig.\,\ref{fig:mean-Z} reveals a correlation between \ULISSE performance and redshift consistency: targets with larger offsets between their redshift and the mean redshift of their neighbours tend to exhibit larger deviations in estimated $\mathcal{M}_{\ast}$ and SFR (i.e., higher $\Delta \mathcal{M}_\ast$ and $\Delta \mathrm{SFR}$). This suggests that the accuracy of the method can be improved by applying \ULISSE within narrower redshift intervals or by incorporating a redshift prior (when available) into the neighbour selection process. Such a refinement would be analogous to known practices in SED fitting, where including redshift priors is known to improve the robustness of derived galaxy properties~\citep{Bolzonella:00, Ilbert:06, Conroy:13}.

Additionally, the evolving nature of galaxy populations with redshift, particularly regarding their SFR and $\mathcal{M}_\ast$, further complicates the interpretation. As galaxies evolve, the distribution of them in the SFR-$\mathcal{M}_\ast$ plane changes, meaning a reference sample drawn from a limited redshift range may not fully represent these variations at higher redshifts. Observational limitations, such as survey sensitivity and instrument capabilities, can influence this effect by preferentially excluding low-SFR or low-mass galaxies at higher redshifts. These factors reduce the representativeness of the reference set in some regions of parameter space, potentially impacting the reliability of neighbour matching and any implicit redshift inference.

Furthermore, redshift-dependent degeneracies in image appearance (e.g. surface brightness dimming and limited angular resolution) complicate the interpretation of similarity-based matching in this context. Since \ULISSE selects analogues based on visual similarity, features correlated with redshift may bias neighbour selection, making it challenging to disentangle intrinsic physical resemblance from redshift-driven projection effects.

Consequently, while these findings suggest that \ULISSE may exhibit some degree of redshift consistency through structural matching, it is not designed to serve as a redshift estimator. A more systematic assessment of \ULISSE potential in this role would require a dedicated study covering a broader redshift range and incorporating appropriate calibration techniques. More detailed comparisons with established photometric redshift methods~\citep{Ilbert:06, Way:09_GPS, Hildebrandth:10_PHAT, Cavuoti:17_METAPHOR, Soo:18_MorphoZ, Pasquet:19_CNN, Euclid:20_photoZ, Pathi:25+ANNZ} would be also necessary to rigorously evaluate its applicability for redshift estimation.


\section{Conclusions}\label{sec:conclusions}

In this work, we present the application of \ULISSE (aUtomatic Lightweight Intelligent System for Sky Exploration) to the prediction of stellar mass and star-formation rate of resolved galaxies.  
Our method relies on a single composite-color image of a target galaxy of unknown physical properties, and based on a pre-trained convolutional neural network, extracts from the image a set of representative features without requiring any specific astrophysical knowledge. Applied to a sample of galaxies with known properties (derived, e.g., from spectroscopy), \ULISSE sorts all objects according to the distance in this feature space (i.e. from the most to the least similar) from the target object and thus allows to retrieve a specified set of sources with similar properties. 

In \citetalias{Doorenbos2022} we have already applied our method to the selection of AGN candidates based on a sample of SDSS galaxies and their composite-color images. In this work, we test the performance of our method in predicting the galaxy $\mathcal{M}_{\ast}$ and SFR, assuming that the average properties of the retrieved neighbours are representative of those of the target galaxy. Based on the results obtained running \ULISSE on a sample of 290 sources with different stellar masses, SFR, morphologies (e.g. elliptical, spiral with/without bar structure based on the GZ2 survey), and the lack/presence of AGN signatures, we reached the following conclusions:

\begin{itemize}
\setlength\itemsep{0.5ex}
    \item The efficiency of our method in estimating the physical properties of the target galaxies, defined as a fraction of objects whose predicted position in the SFR--$\mathcal{M}_{\ast}$ diagram is within 1\,dex from the `true' position (estimated from spectroscopy) ranges from $\sim$60\perc and up to 88\perc depending on the typology of a galaxy (quiescent/star-forming, morphology, AGN/non-AGN). In general, the method is at least twice as efficient as using a random guess { and provides typical scatters of $\sim$0.3–0.4\,dex in $\mathcal{M}_{\ast}$ and $\sim$0.5–0.6\,dex in SFR.}
    \item Our tests show that the performance of the method is mainly dependent on the colour and morphology of the target galaxy. For instance, the efficiency for galaxies with low/high stellar formation (i.e. quiescent/SFG) are respectively $\sim$78\perc and $\sim$72\perc (averaged for all studied distances), while for targets with relatively featureless and `smooth' appearance is on average $\sim$68\perc and increases to $\sim$76\perc for galaxies with spiral and bulge/bar structures. 
    \item The presence of a bright nucleus also affects the performance of our method showing an increased efficiency for galaxies identified as `AGN' according to the BPT criteria ($\sim$77\perc in average among all distances) with respect to the `non-AGN' target objects ($\sim$71\perc), which appears in agreement with the results of \citetalias{Doorenbos2022};
    \item The analysis of the separate efficiencies in predicting just $\mathcal{M}_{\ast}$ or SFR reveals a higher efficiency in $\mathcal{M}_{\ast}$ retrieval (near 94\perc) in comparison to SFR (near 77\perc); this is expected given the fact that the mass depends mainly on the total luminosity while SFR is harder to measure, both photometrically and spectroscopically, and is mainly linked to the colour of the galaxy. 
\end{itemize}

We conclude that, while traditional methods for estimating galaxy properties, such as SFR and $\mathcal{M}_{\ast}$, require time-consuming spectroscopic observations or multi-band photometry for SED fitting, the use of artificial intelligence algorithms represents a viable and faster alternative although somewhat less accurate. Considering the results presented in \citetalias{Doorenbos2022} and this work, \ULISSE emerges as a promising approach for either selecting specific classes of objects (e.g. AGN, quiescent, star-forming galaxies) as well as predicting their properties in current and upcoming wide-field surveys, such as Euclid and LSST, that target millions of sources every single night. 


\begin{acknowledgements}

L.D. acknowledges support from research grant 200021\_192285 `Image data validation for AI systems' funded by the Swiss National Science Foundation (SNSF). M.P. and O.T. also acknowledge financial support from the agreement ASI-INAF n.2017-14-H.O. M.P. also acknowledge the financial contribution from PRIN-MIUR 2022 and from the Timedomes grant within the ``INAF 2023 Finanziamento della Ricerca Fondamentale''. L.D., S.C. and M.B. also acknowledge financial contributions from the agreement ASI/INAF 2018-23-HH.0, Euclid ESA mission - Phase D.  

Funding for the Sloan Digital Sky Survey V has been provided by the Alfred P. Sloan Foundation, the Heising-Simons Foundation, the National Science Foundation, and the Participating Institutions. SDSS acknowledges support and resources from the Center for High-Performance Computing at the University of Utah. SDSS telescopes are located at Apache Point Observatory, funded by the Astrophysical Research Consortium and operated by New Mexico State University, and at Las Campanas Observatory, operated by the Carnegie Institution for Science. The SDSS web site is \url{www.sdss.org}.

SDSS is managed by the Astrophysical Research Consortium for the Participating Institutions of the SDSS Collaboration, including Caltech, The Carnegie Institution for Science, Chilean National Time Allocation Committee (CNTAC) ratified researchers, The Flatiron Institute, the Gotham Participation Group, Harvard University, Heidelberg University, The Johns Hopkins University, L’Ecole polytechnique f\'{e}d\'{e}rale de Lausanne (EPFL), Leibniz-Institut f\"{u}r Astrophysik Potsdam (AIP), Max-Planck-Institut f\"{u}r Astronomie (MPIA Heidelberg), Max-Planck-Institut f\"{u}r Extraterrestrische Physik (MPE), Nanjing University, National Astronomical Observatories of China (NAOC), New Mexico State University, The Ohio State University, Pennsylvania State University, Smithsonian Astrophysical Observatory, Space Telescope Science Institute (STScI), the Stellar Astrophysics Participation Group, Universidad Nacional Aut\'{o}noma de M\'{e}xico, University of Arizona, University of Colorado Boulder, University of Illinois at Urbana-Champaign, University of Toronto, University of Utah, University of Virginia, Yale University, and Yunnan University.

This research has made use of Galaxy Zoo classifications (\url{https://www.zooniverse.org/projects/zookeeper/galaxy-zoo/}), so the authors thank the Galaxy Zoo team and the tens of thousands of volunteers who have contributed to the project.

\end{acknowledgements}

\bibliographystyle{aa}
\bibliography{references} 
\onecolumn
\begin{appendix}

\appendix 
\section{The distribution of the retrieved neighbours in the SFR-$\mathcal{M}_{\ast}$ diagram with respect to the target objects of different classes.}\label{sec:appendix}
\begin{figure*}[bh!]
    \centering
    \includegraphics[width=.48\textwidth]{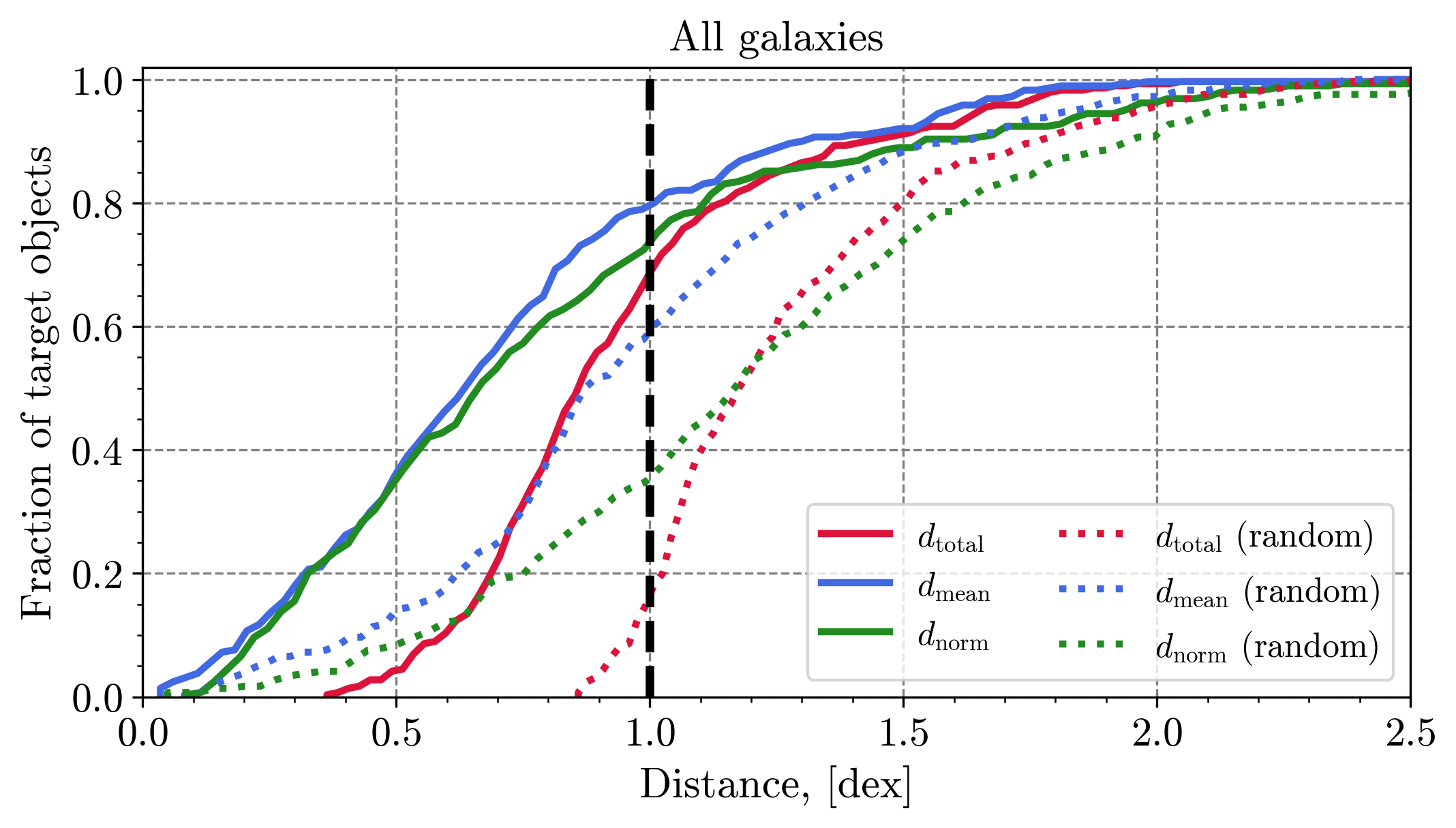}
    \includegraphics[width=.48\textwidth]{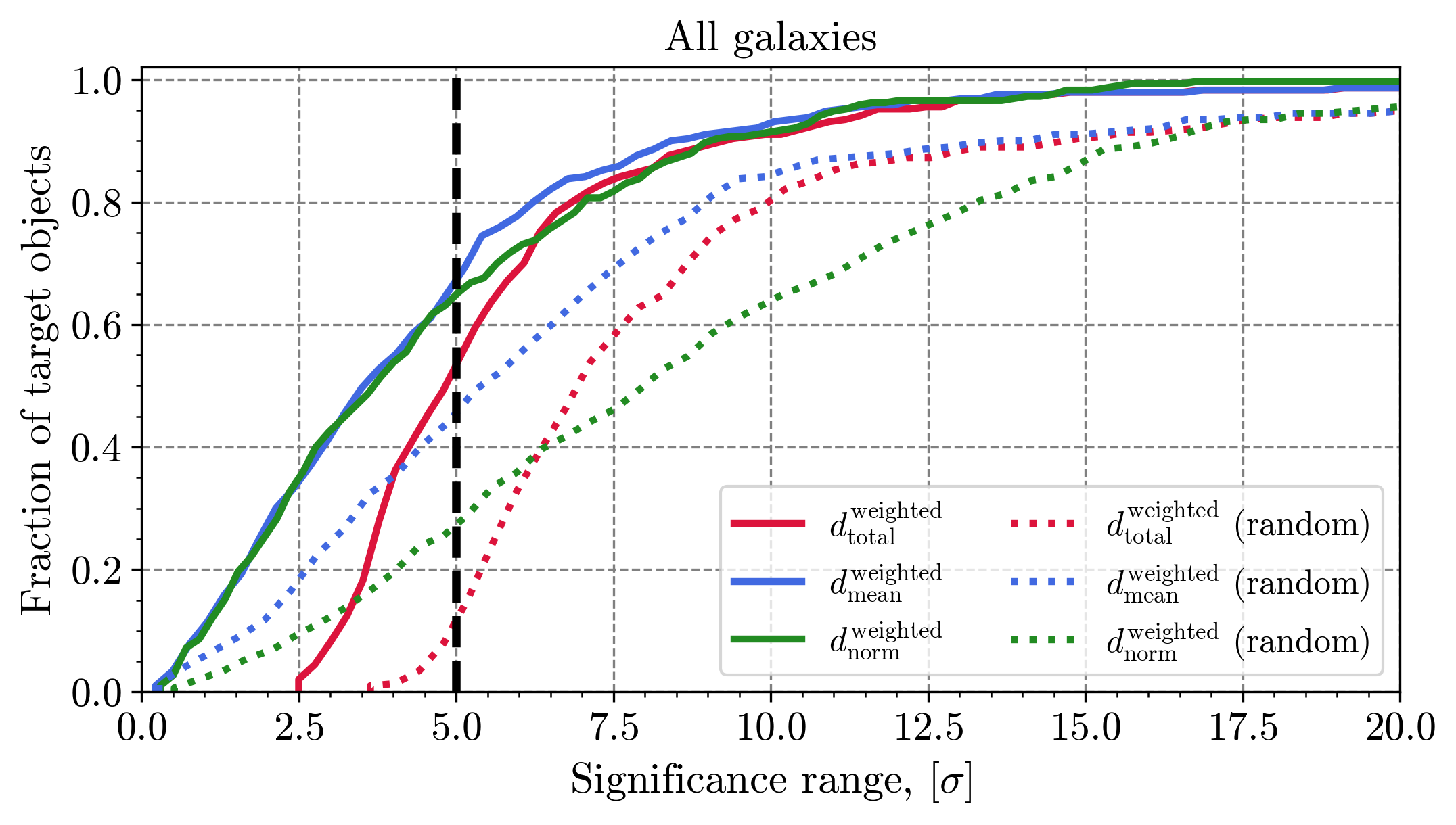}\\

    \includegraphics[width=.48\textwidth]{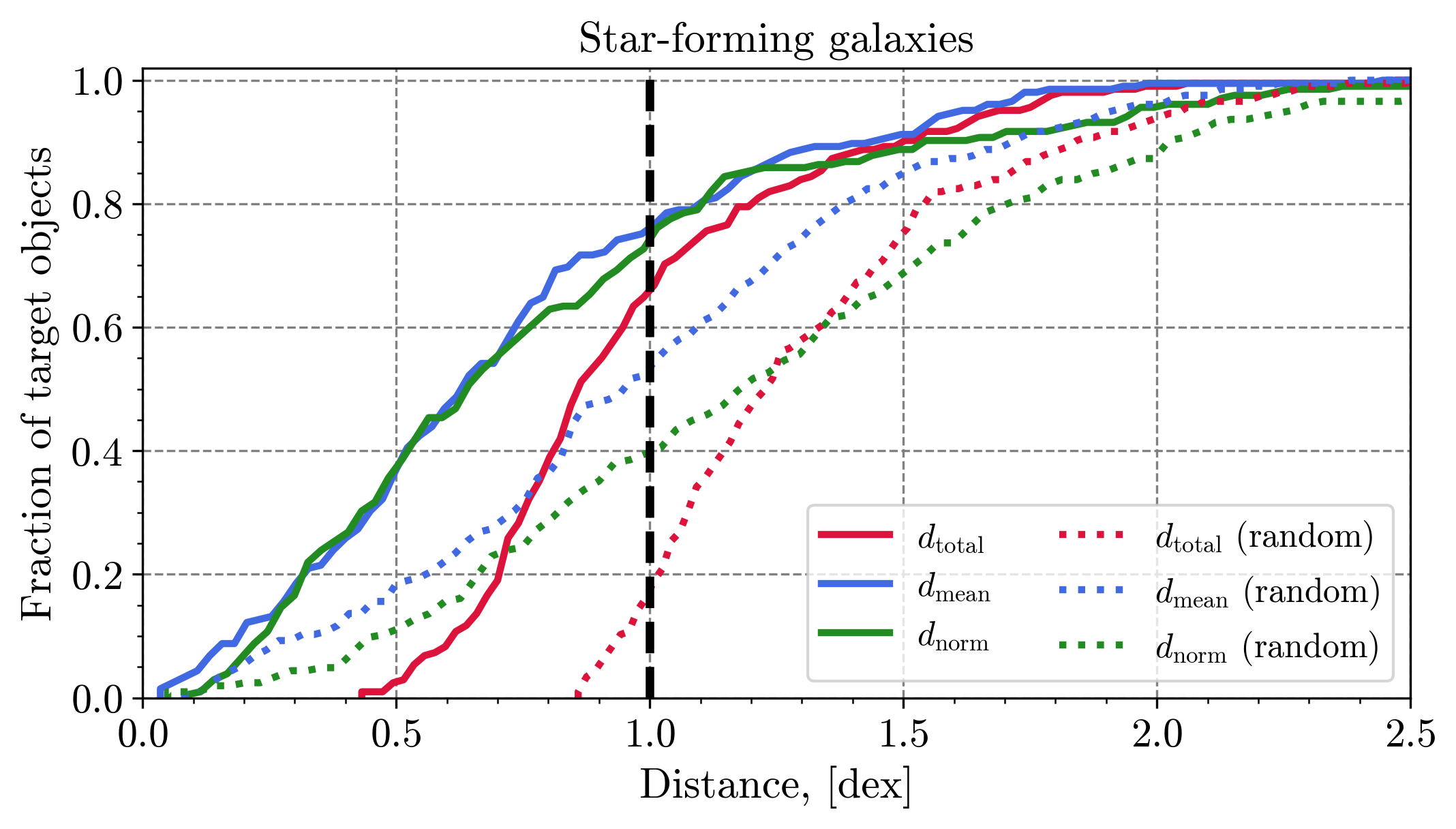}
    \includegraphics[width=.48\textwidth]{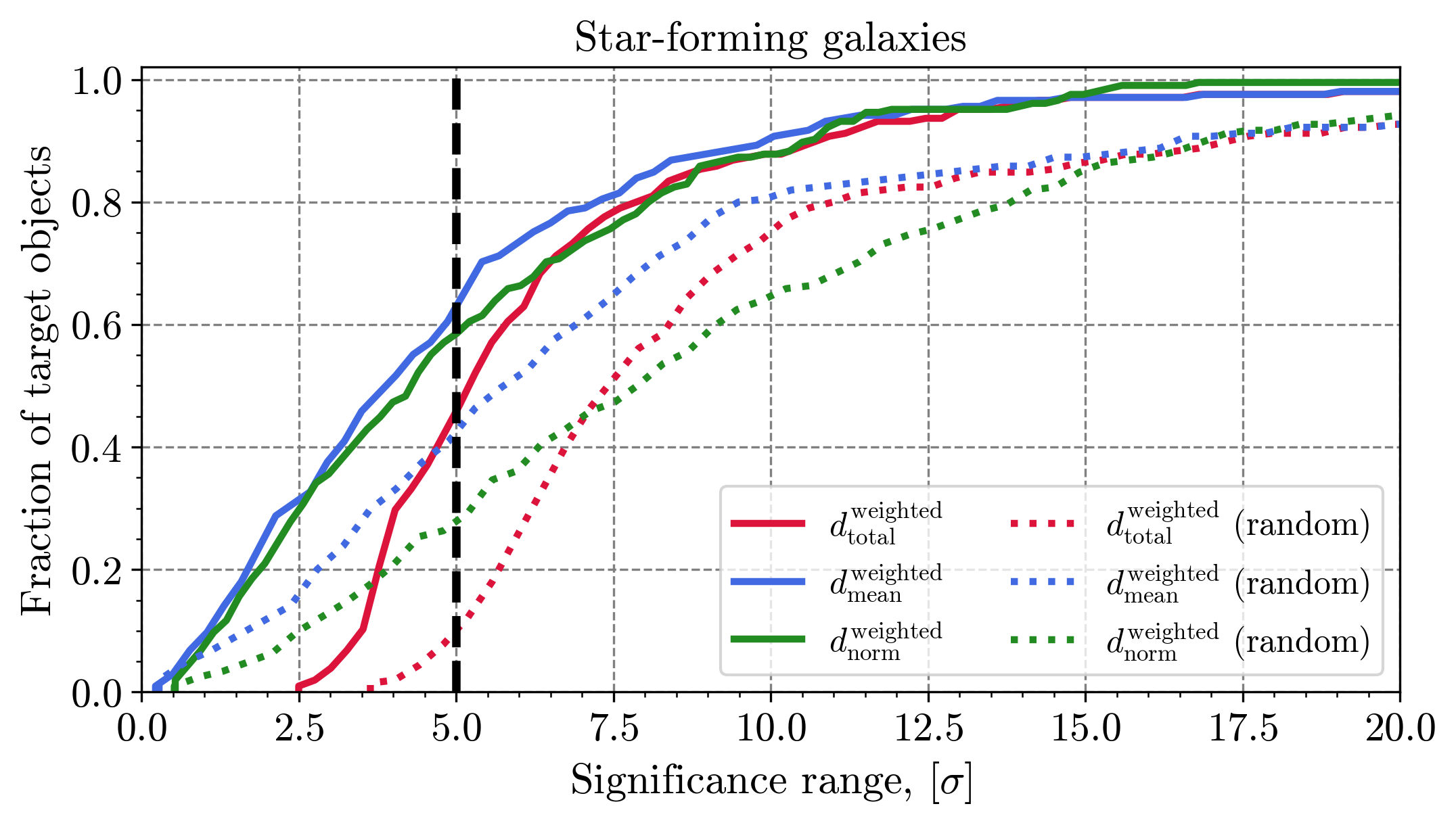}\\

    \includegraphics[width=.48\textwidth]{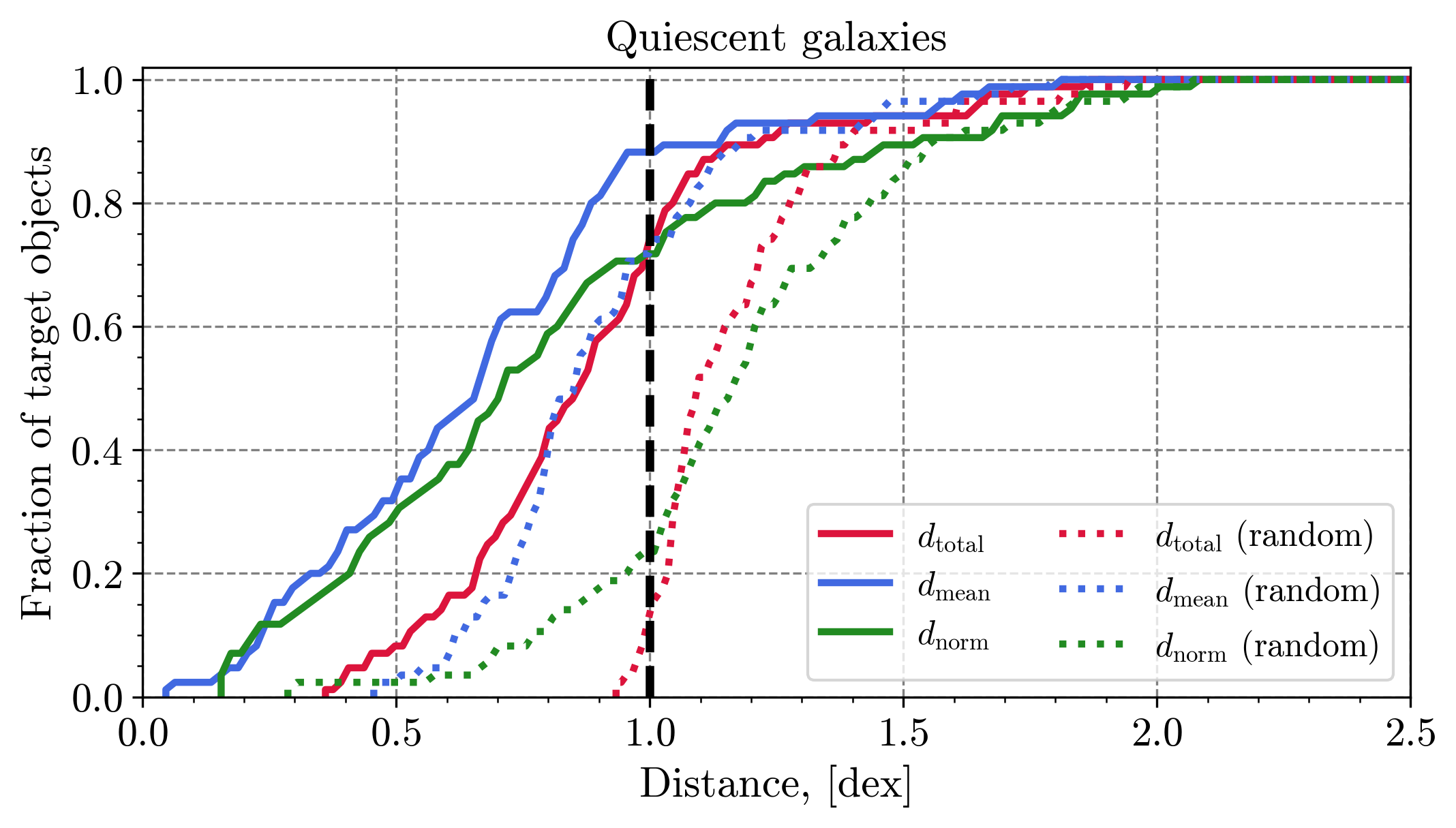}
    \includegraphics[width=.48\textwidth]{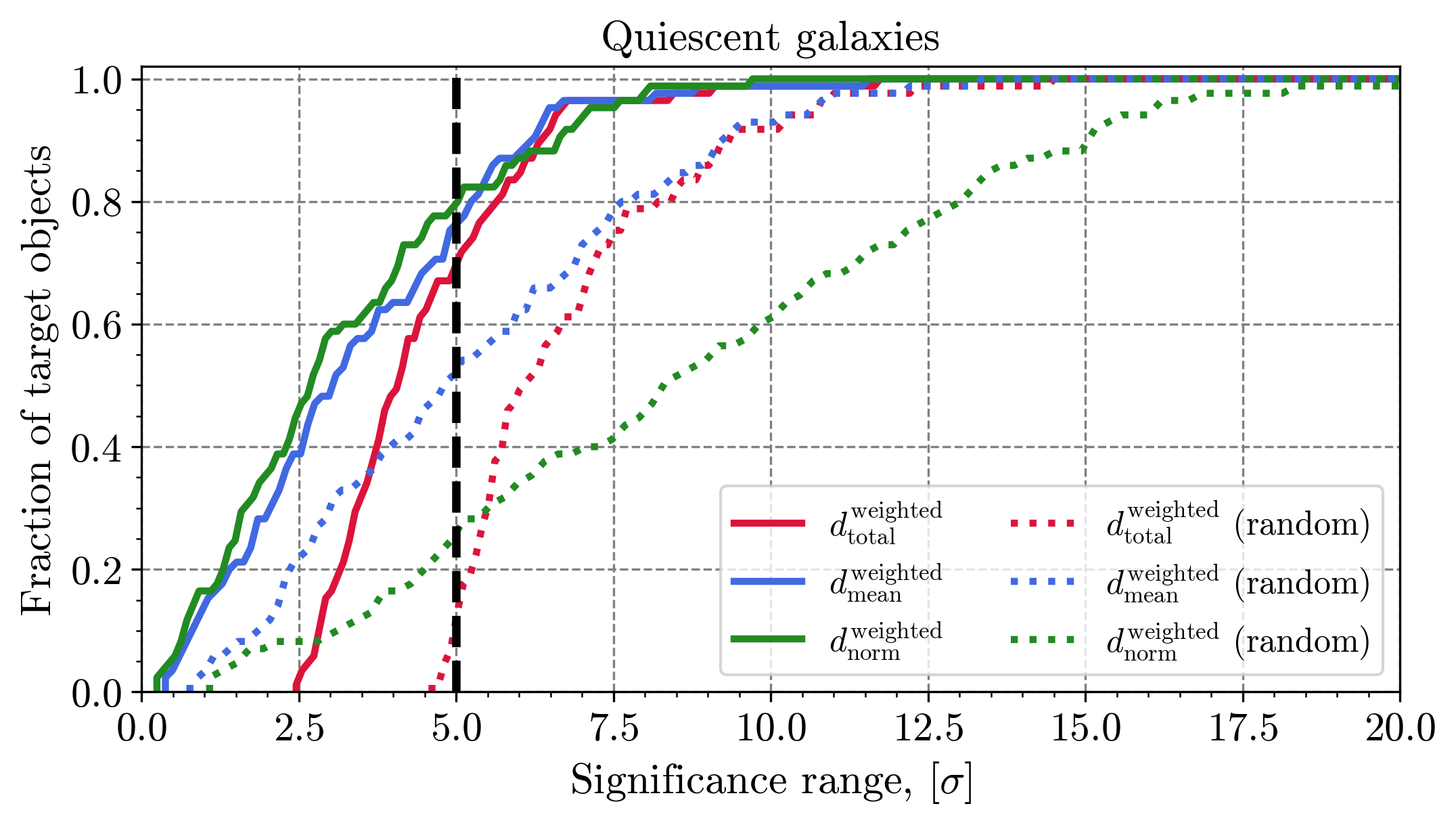}\\

    \caption{{\it Left panel:} The distribution of the total ($d_{\rm total}$ by red color), mean ($d_{\rm mean}$ by blue color) and normalised mean ($d_{\rm norm}$ by green color) distances for {\it all, star-forming} and {\it quiescent} subsamples of target objects. The black dashed line represents the cut at 1\,dex used to define \ULISSE efficiency in retrieving galaxy properties (see description in Section\,\ref{sec:results-histo}). {\it Right panel:} The similar distribution as on the left panel, but for the total, mean and normalised mean distances weighted for the uncertainties of SFR and $\mathcal{M}_{\ast}$ parameters of the target objects. The black dashed line is the cut at 5$\sigma$.}
    \label{fig:hist-dist0}
\end{figure*}

\begin{figure*}
    \centering
    \includegraphics[width=.45\textwidth]{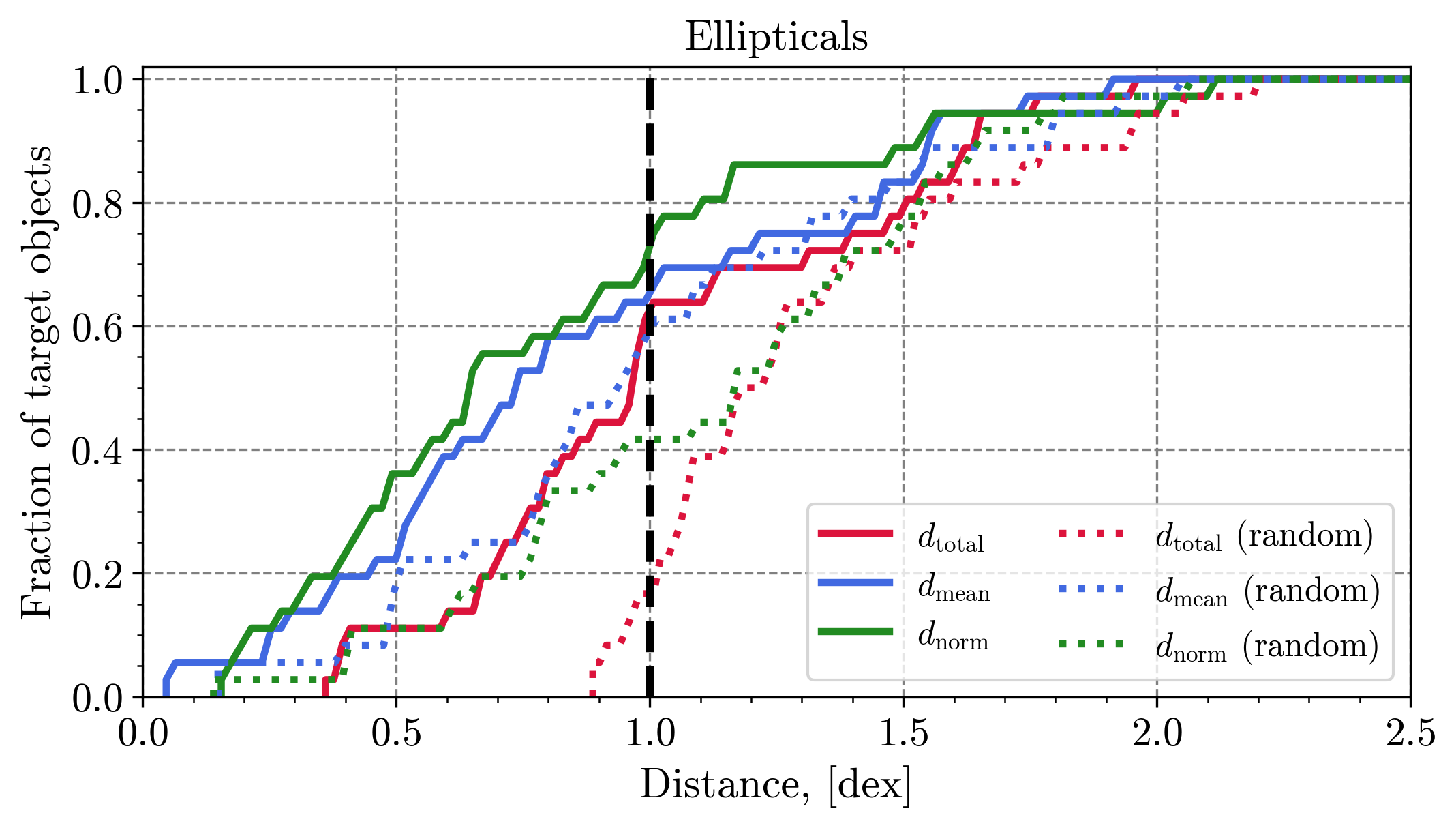}
    \includegraphics[width=.45\textwidth]{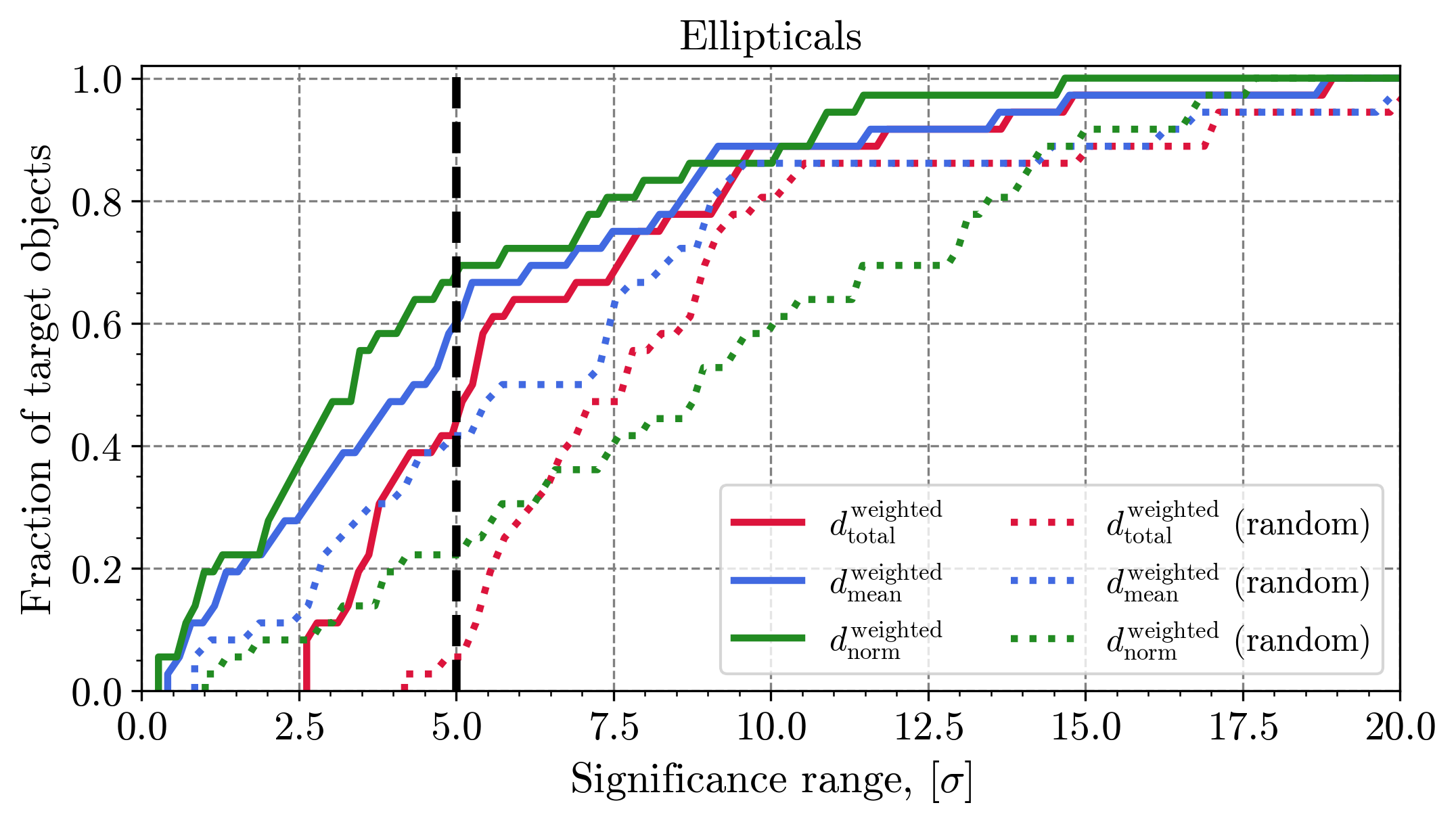}\\

    \includegraphics[width=.45\textwidth]{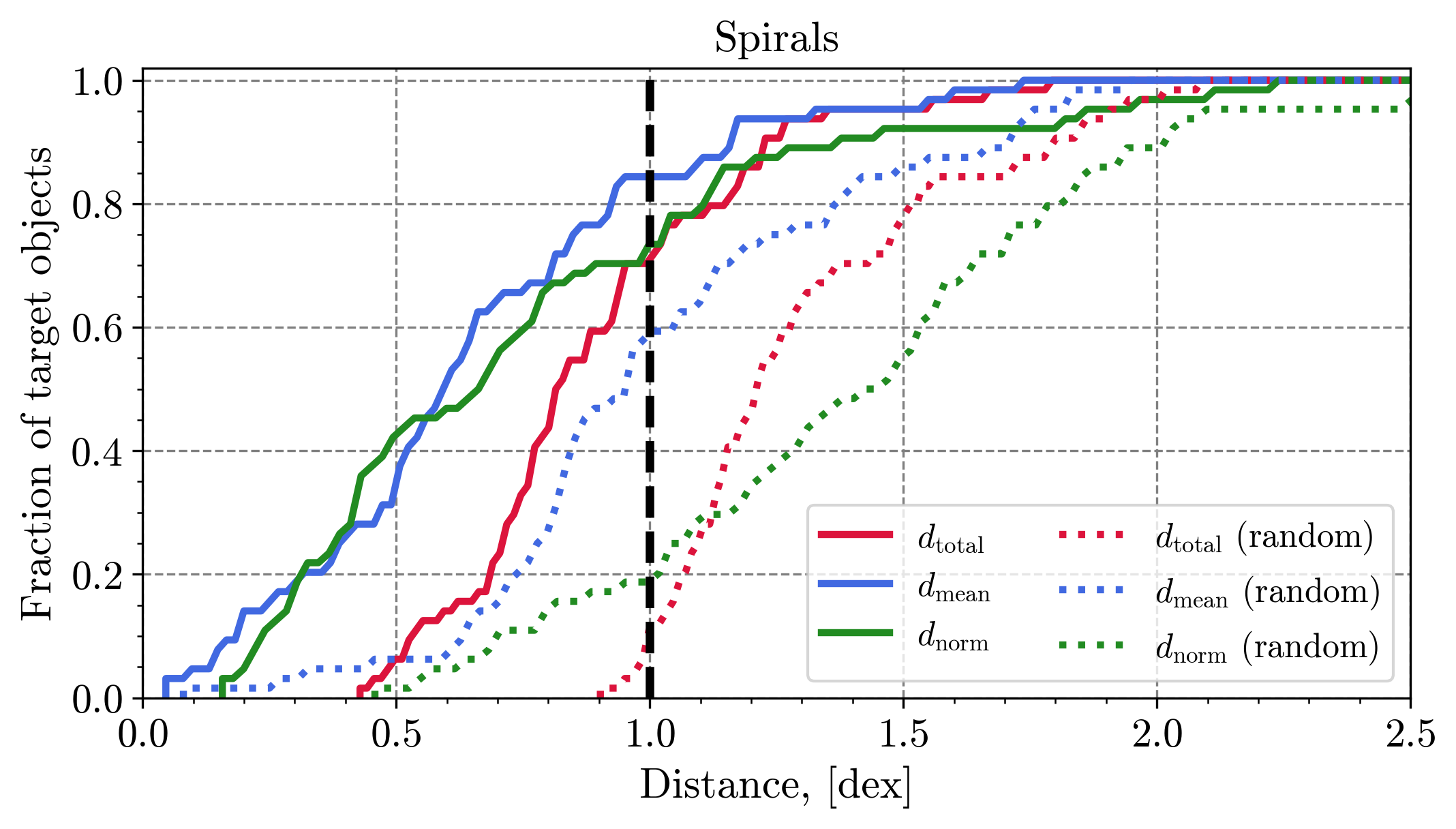}
    \includegraphics[width=.45\textwidth]{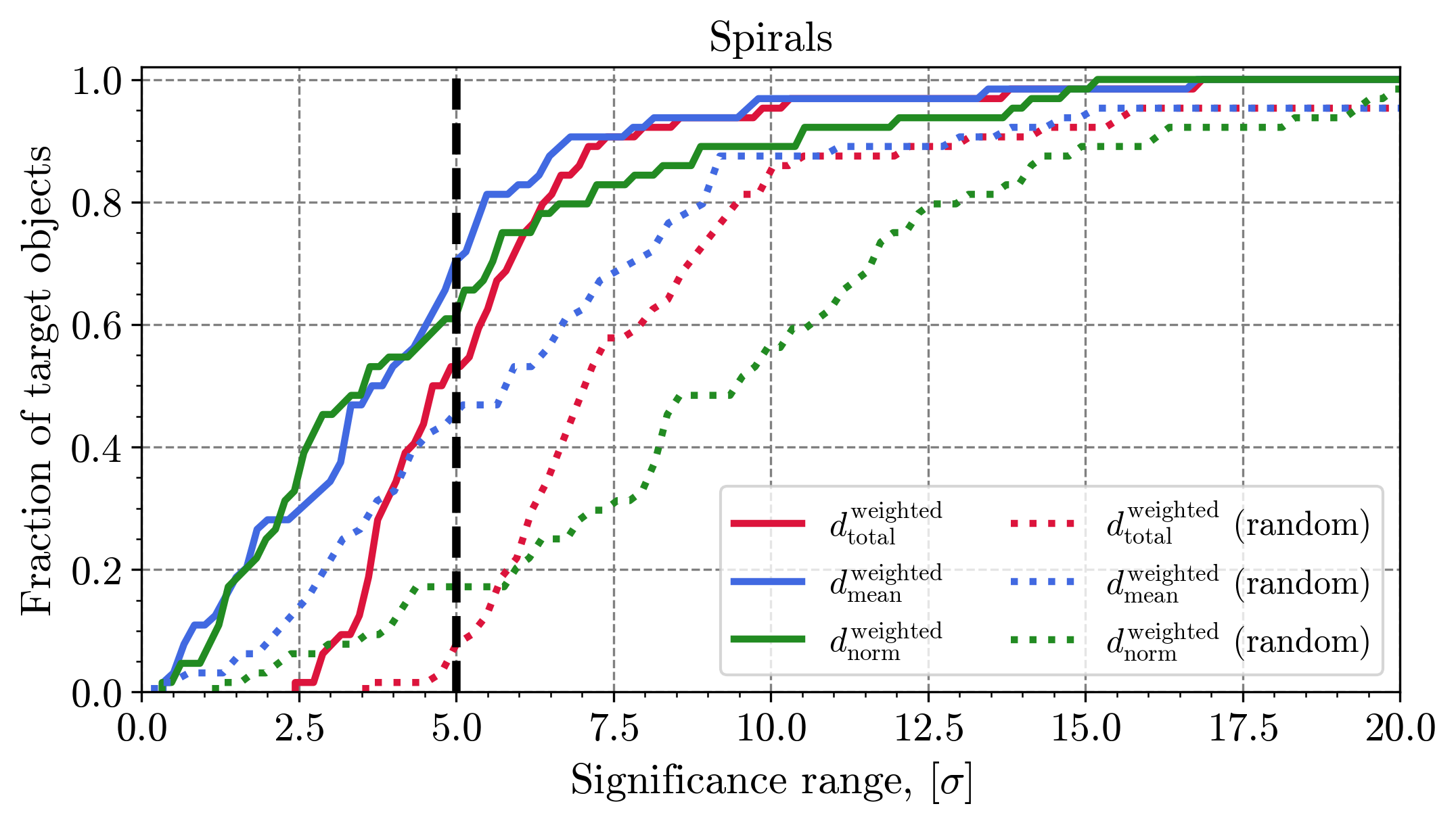}\\

    \includegraphics[width=.45\textwidth]{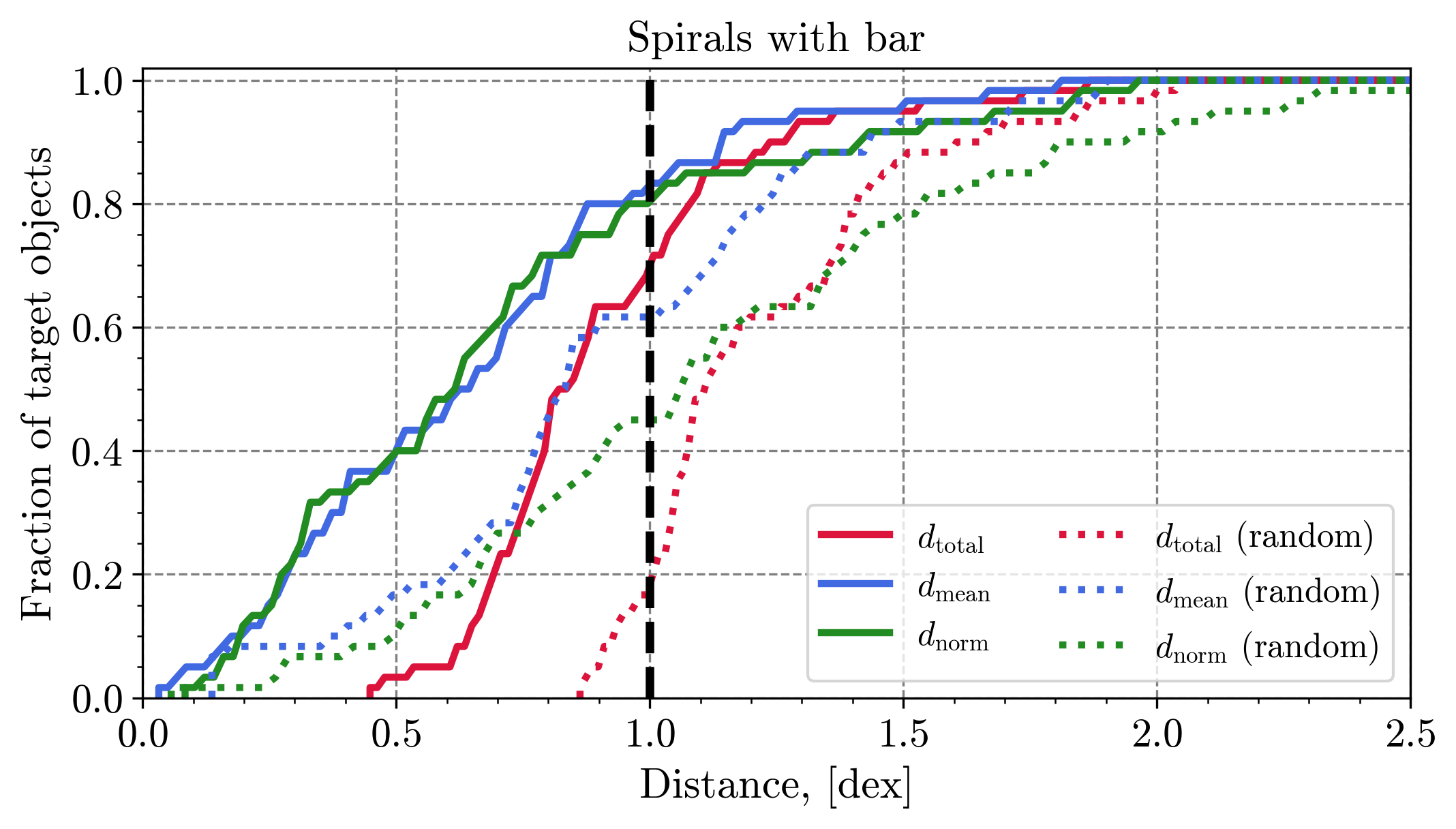}
    \includegraphics[width=.45\textwidth]{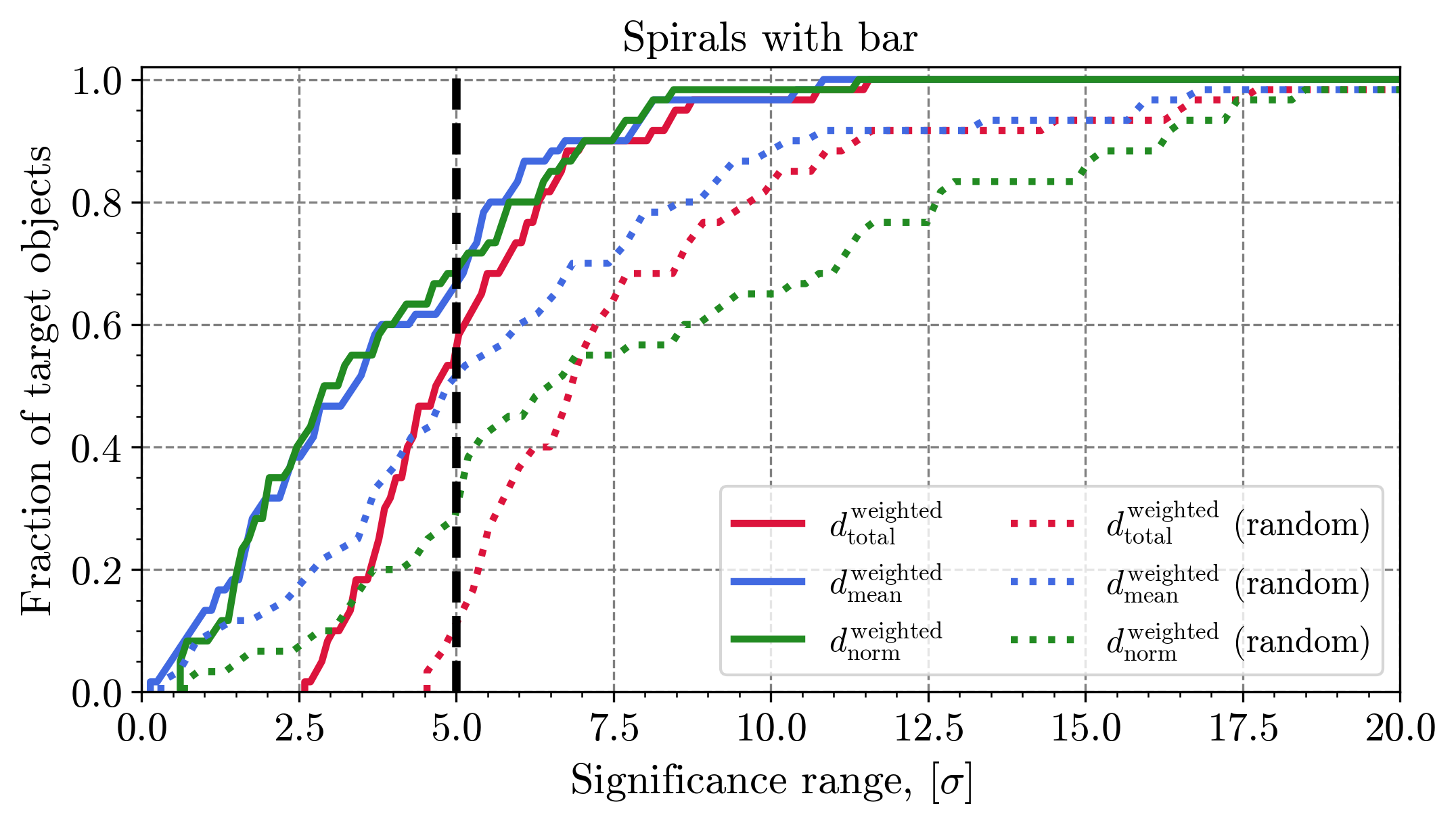}\\

    \includegraphics[width=.45\textwidth]{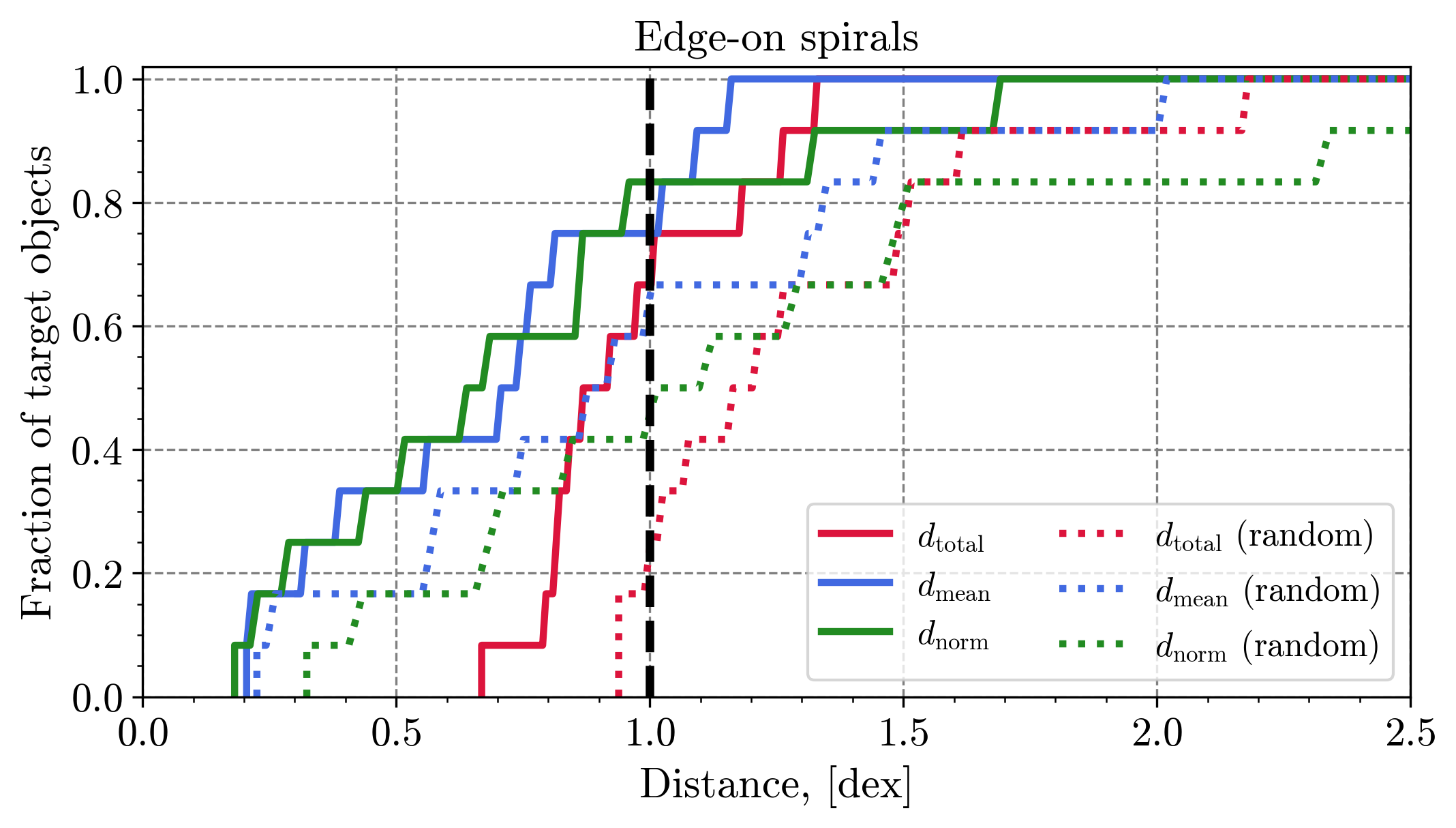}
    \includegraphics[width=.45\textwidth]{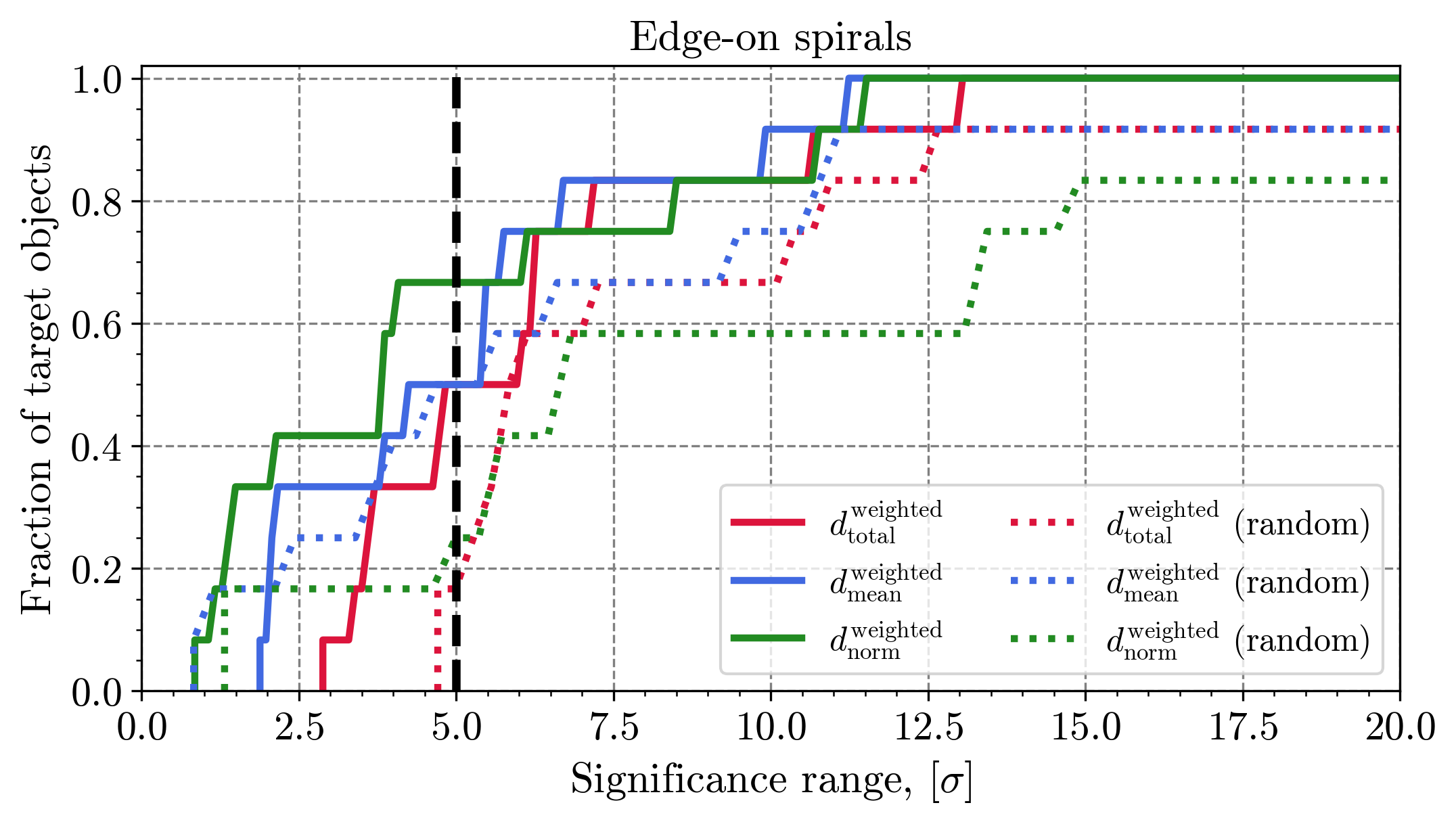}\\

    \includegraphics[width=.45\textwidth]{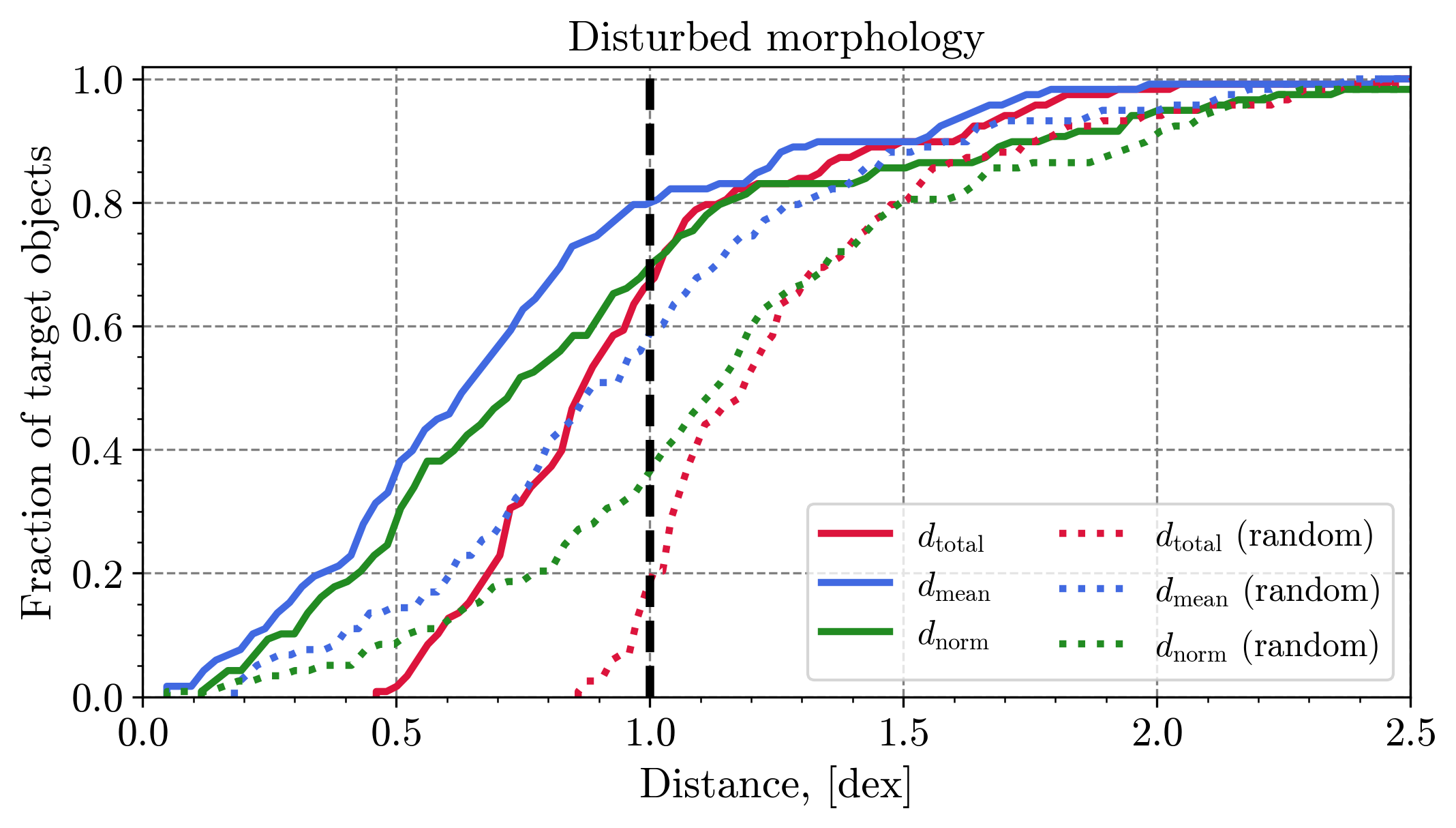}
    \includegraphics[width=.45\textwidth]{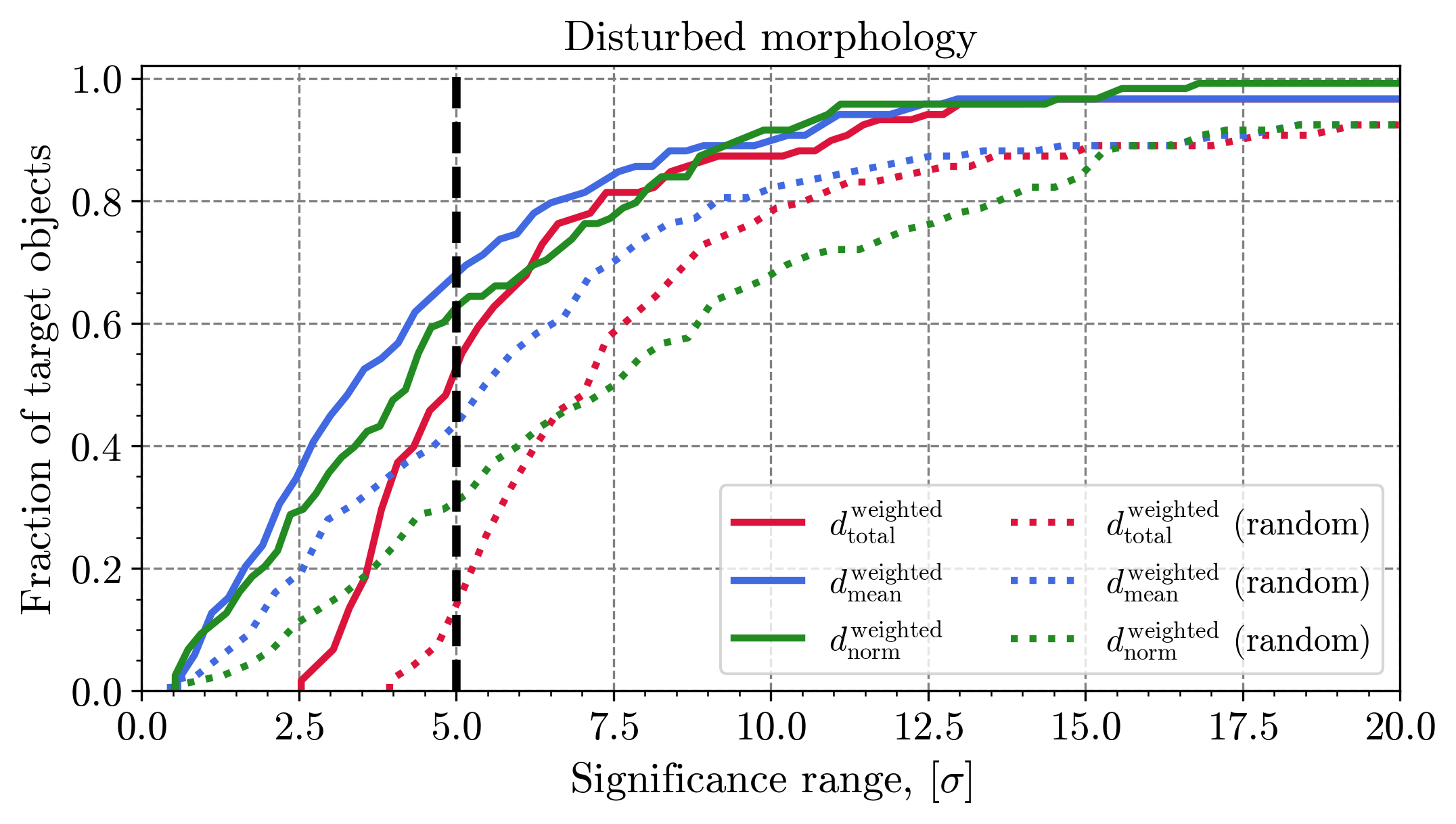}\\
    \caption{The same as in Fig.\,\ref{fig:hist-dist0} for subsamples of target objects with {\it elliptical} (E), {\it spiral with bulge} (S) and {\it bar} (SB) structure, {\it spiral edge-on} (Se) and with { E(d), S(d), SB(d), Se(d)} `{\it disturbed}' morphology. }
    \label{fig:hist-dist1}
\end{figure*}

\begin{figure*}
    \centering
    \includegraphics[width=.44\textwidth]{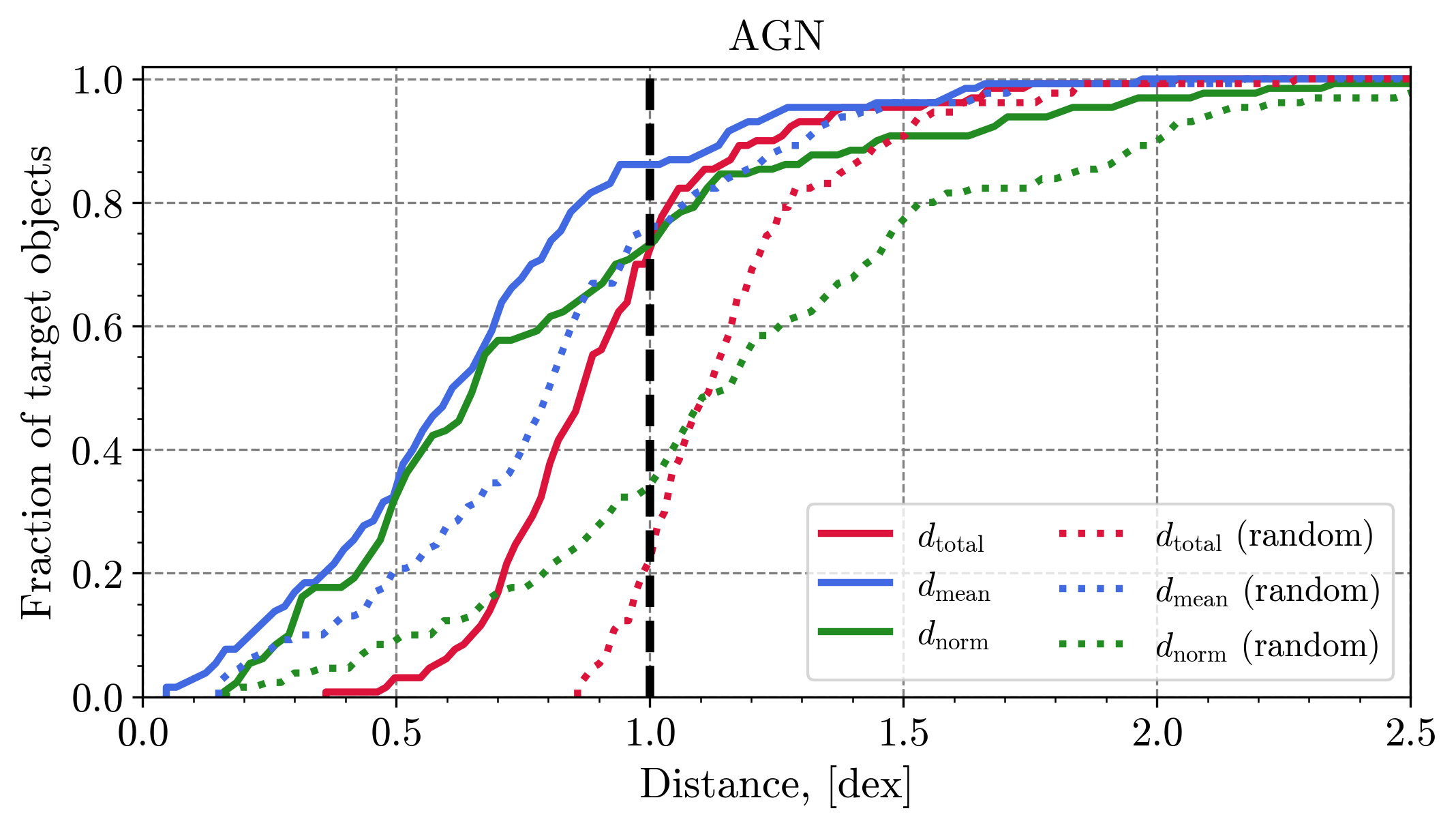}
    \includegraphics[width=.44\textwidth]{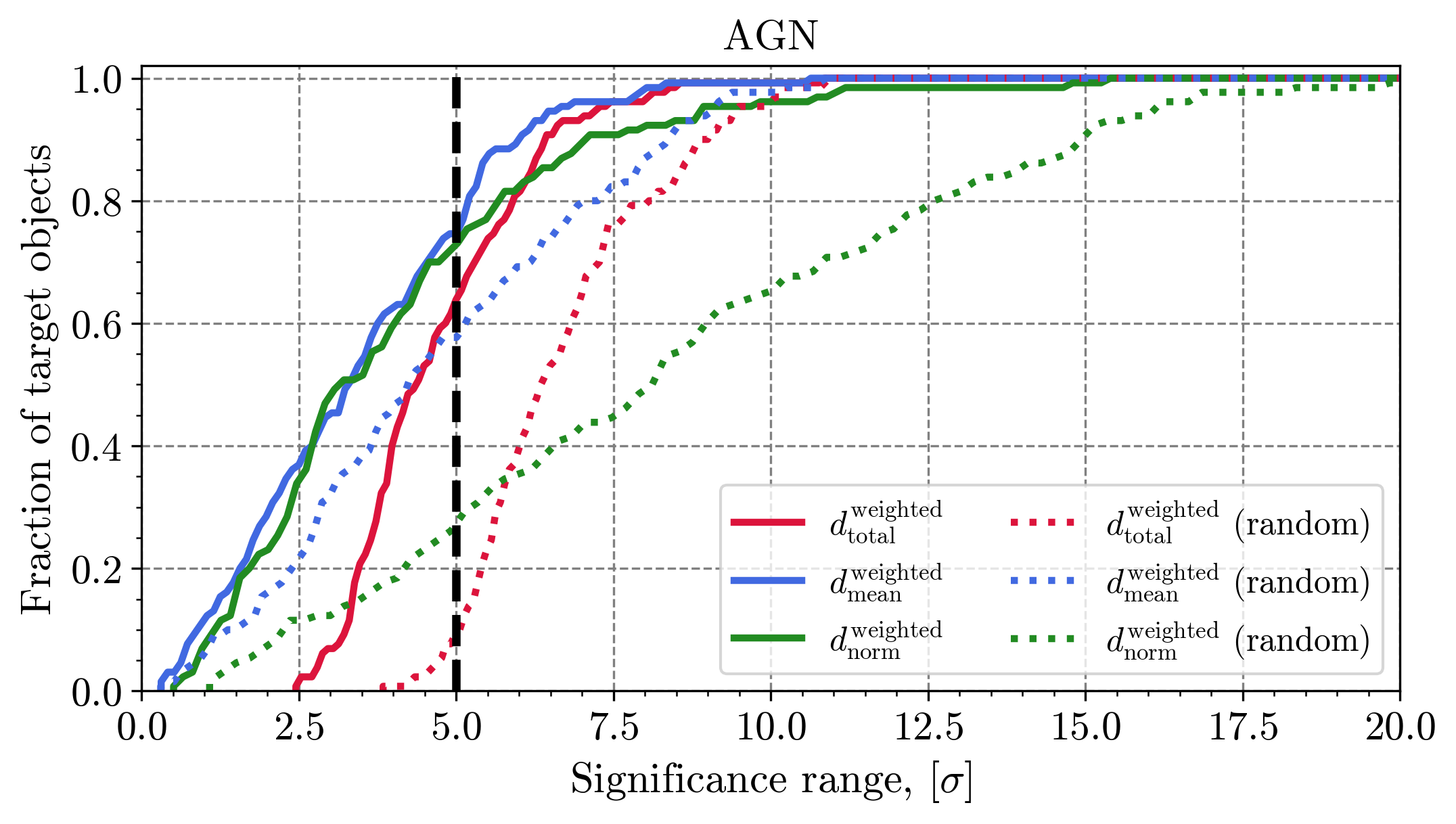}\\

    \includegraphics[width=.44\textwidth]{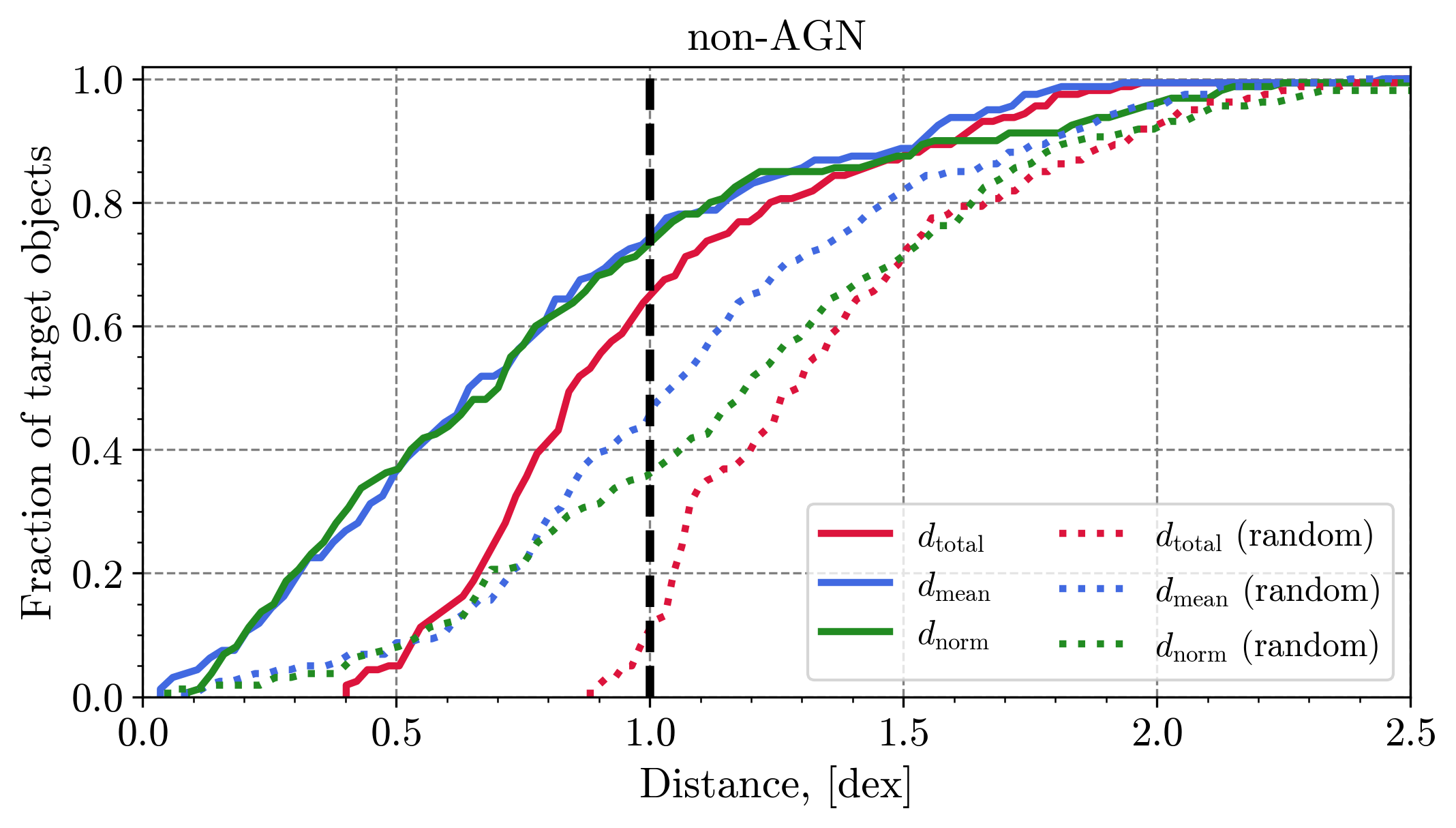}
    \includegraphics[width=.44\textwidth]{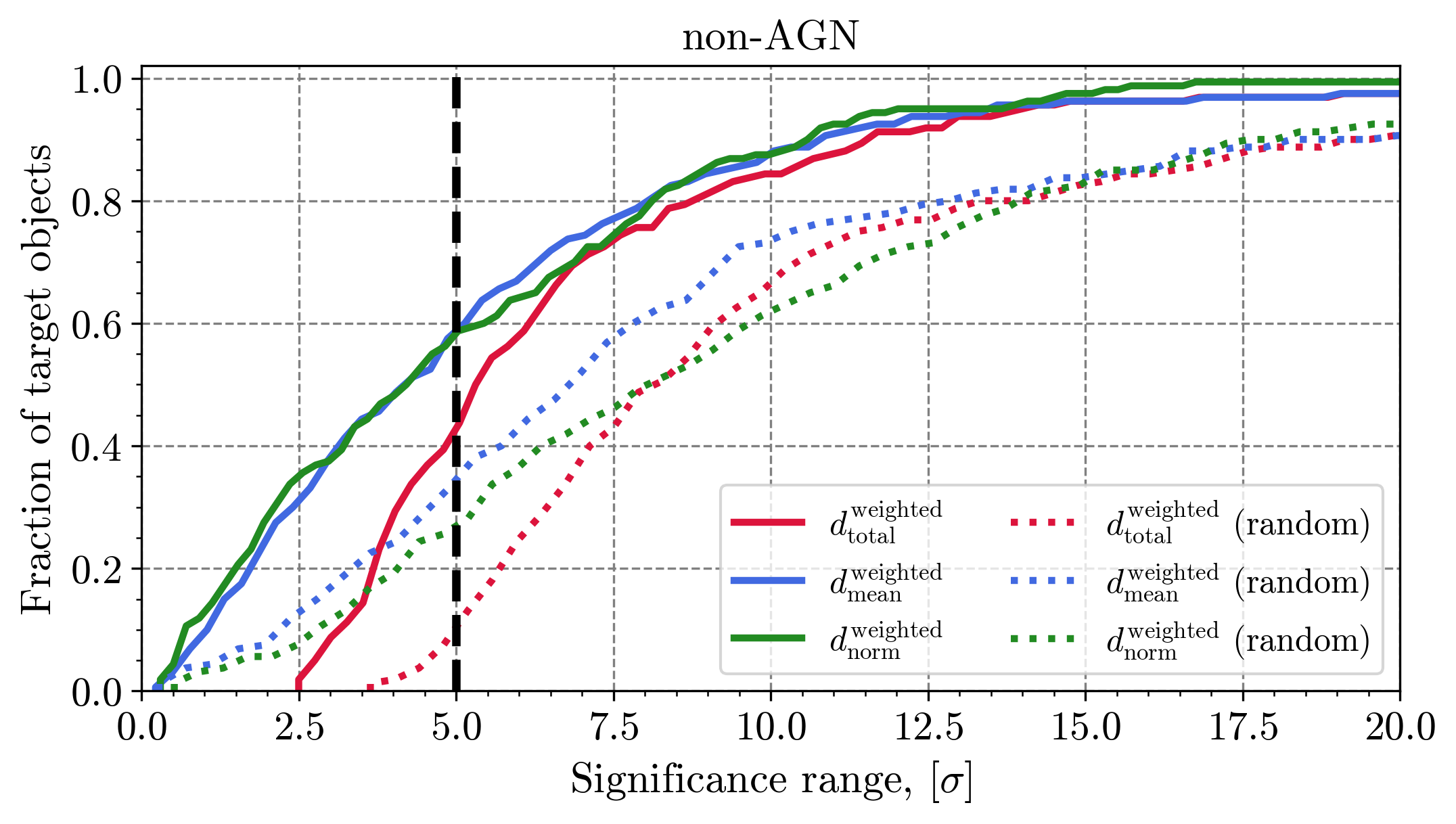}\\

    \caption{The same as in Fig.\,\ref{fig:hist-dist0} for target objects identified AGN or non-AGN according to the BPT selection criteria.}
    \label{fig:hist-dist2}
\end{figure*}

\begin{figure*}
    \centering
    \includegraphics[width=.44\textwidth]{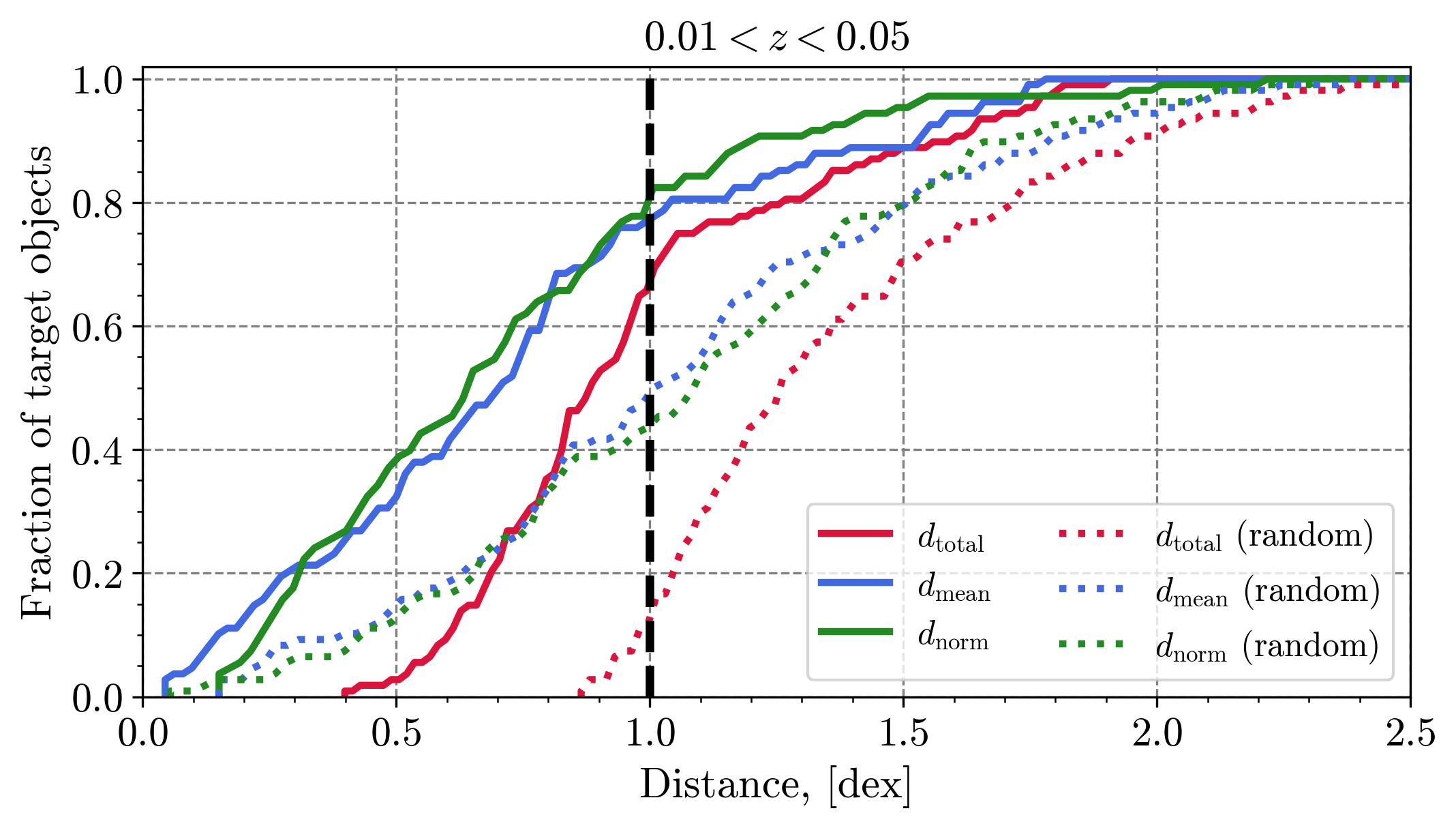}
    \includegraphics[width=.44\textwidth]{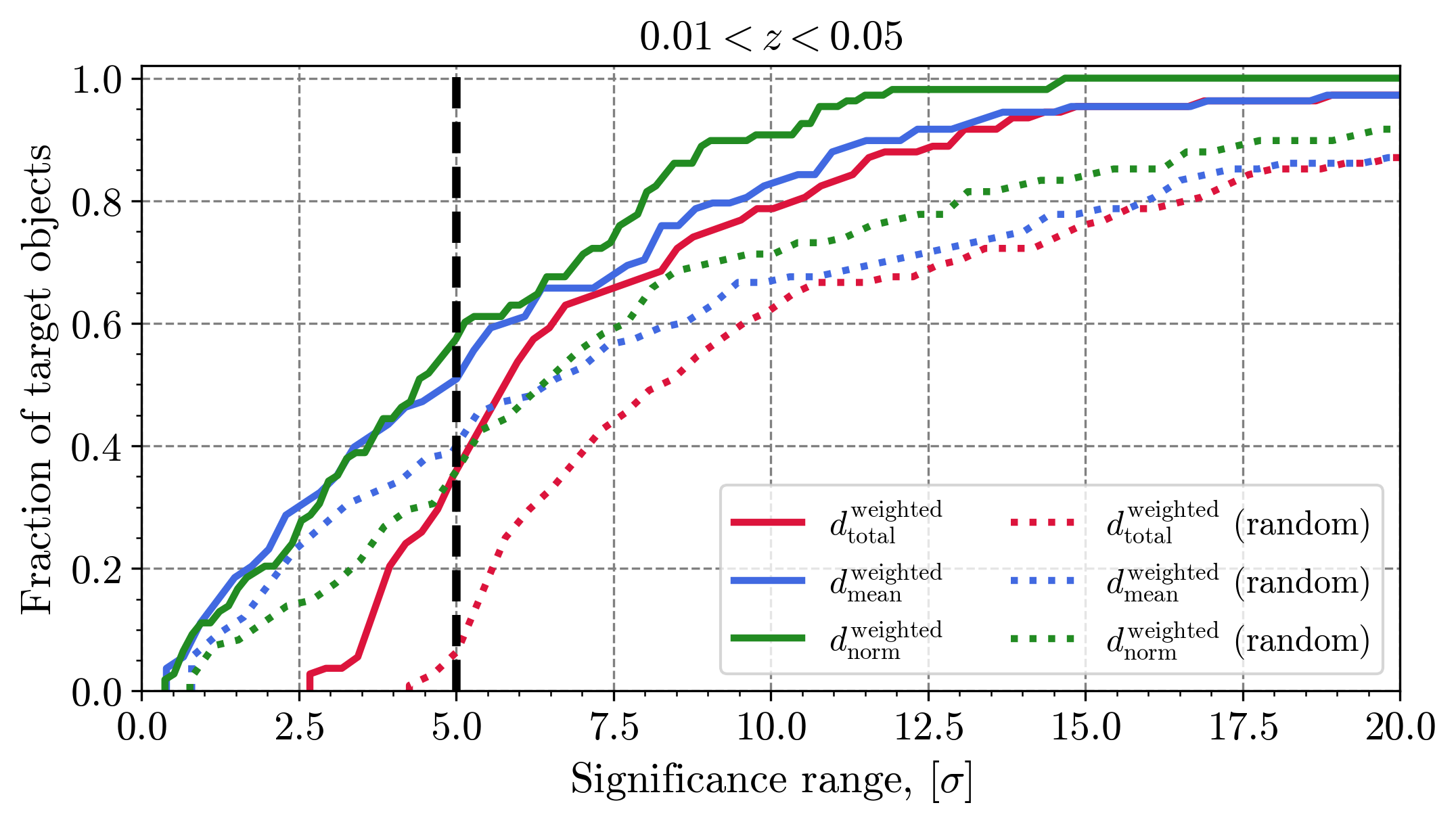}\\

    \includegraphics[width=.44\textwidth]{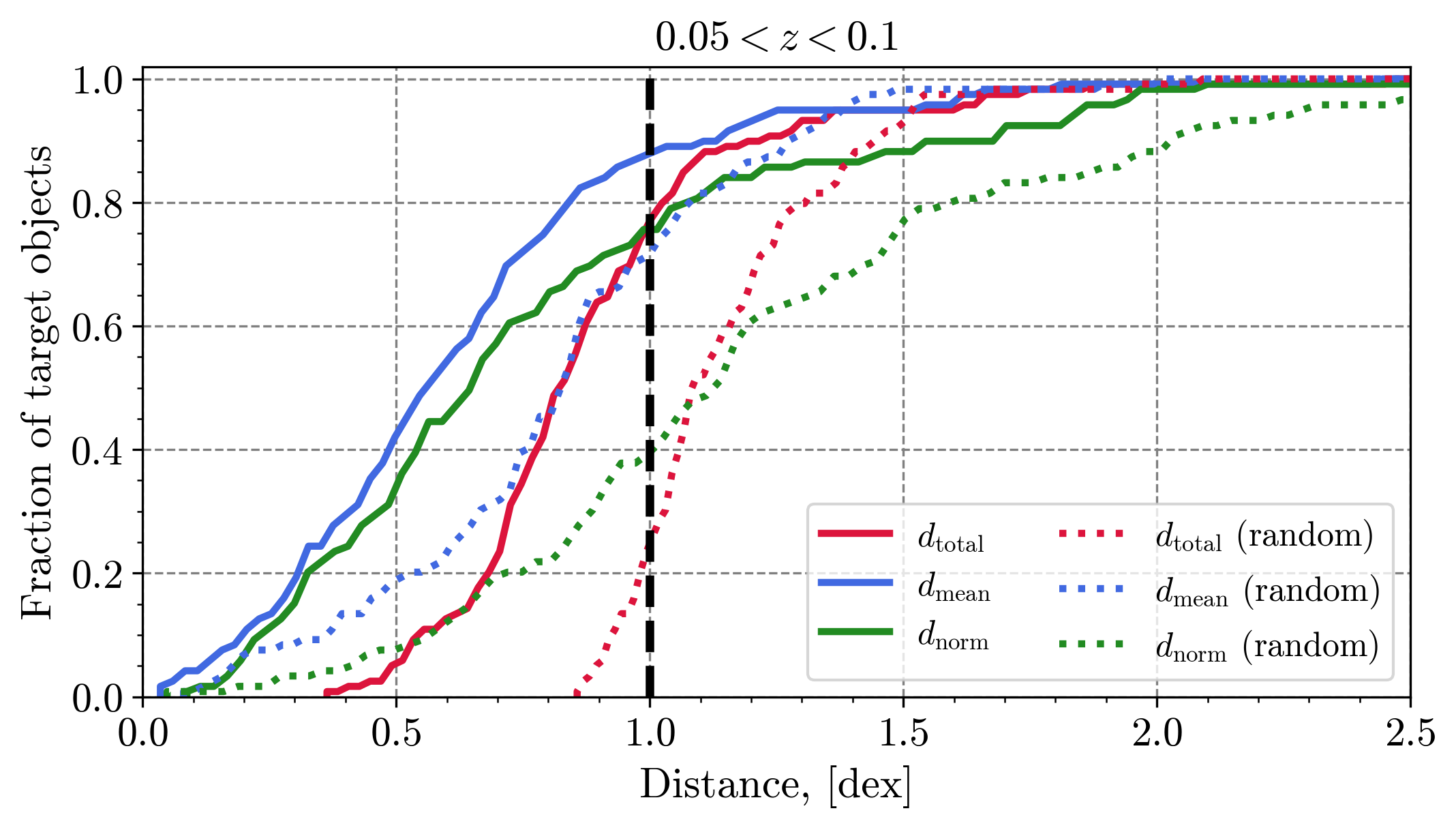}
    \includegraphics[width=.44\textwidth]{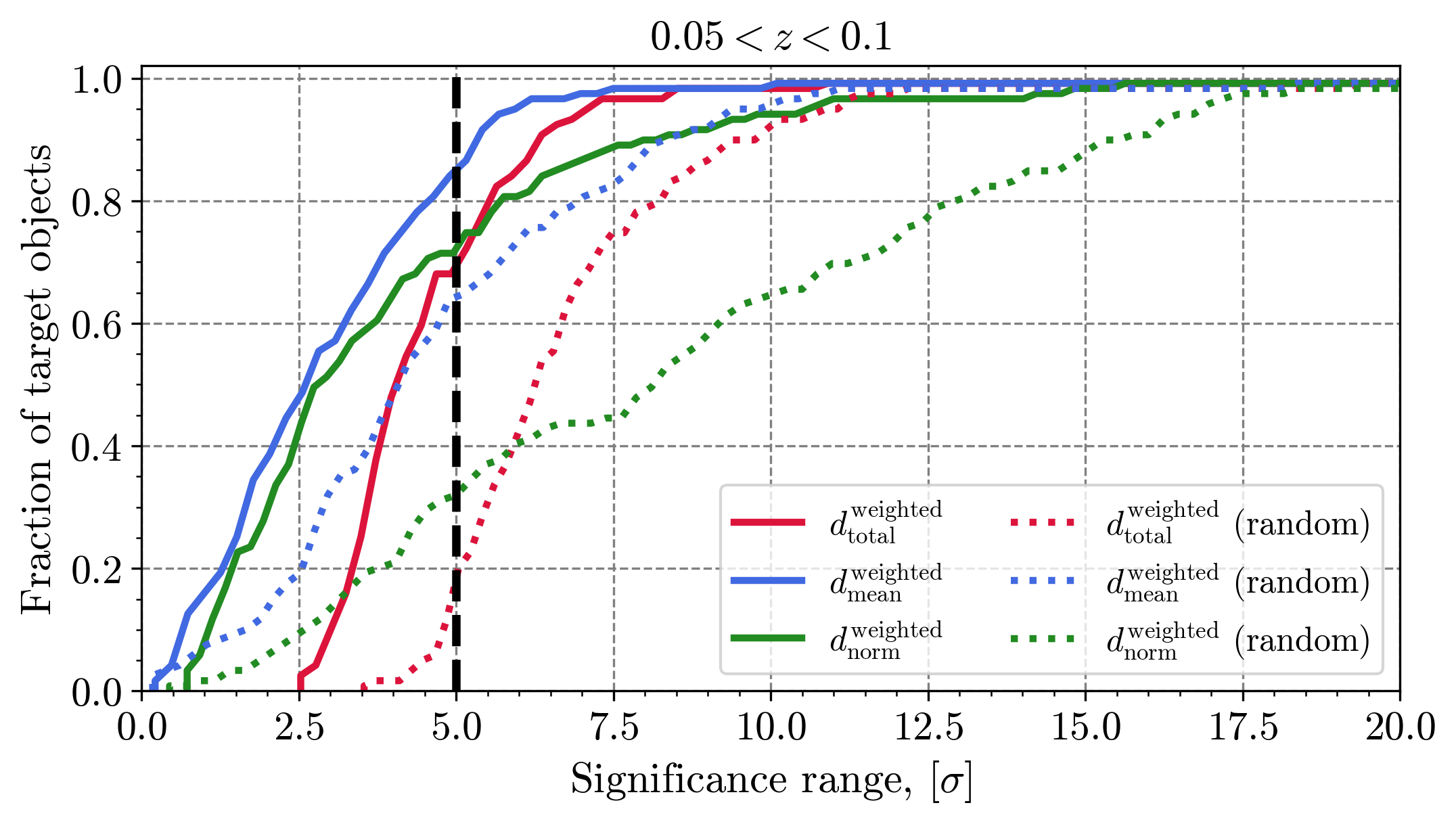}\\

    \includegraphics[width=.44\textwidth]{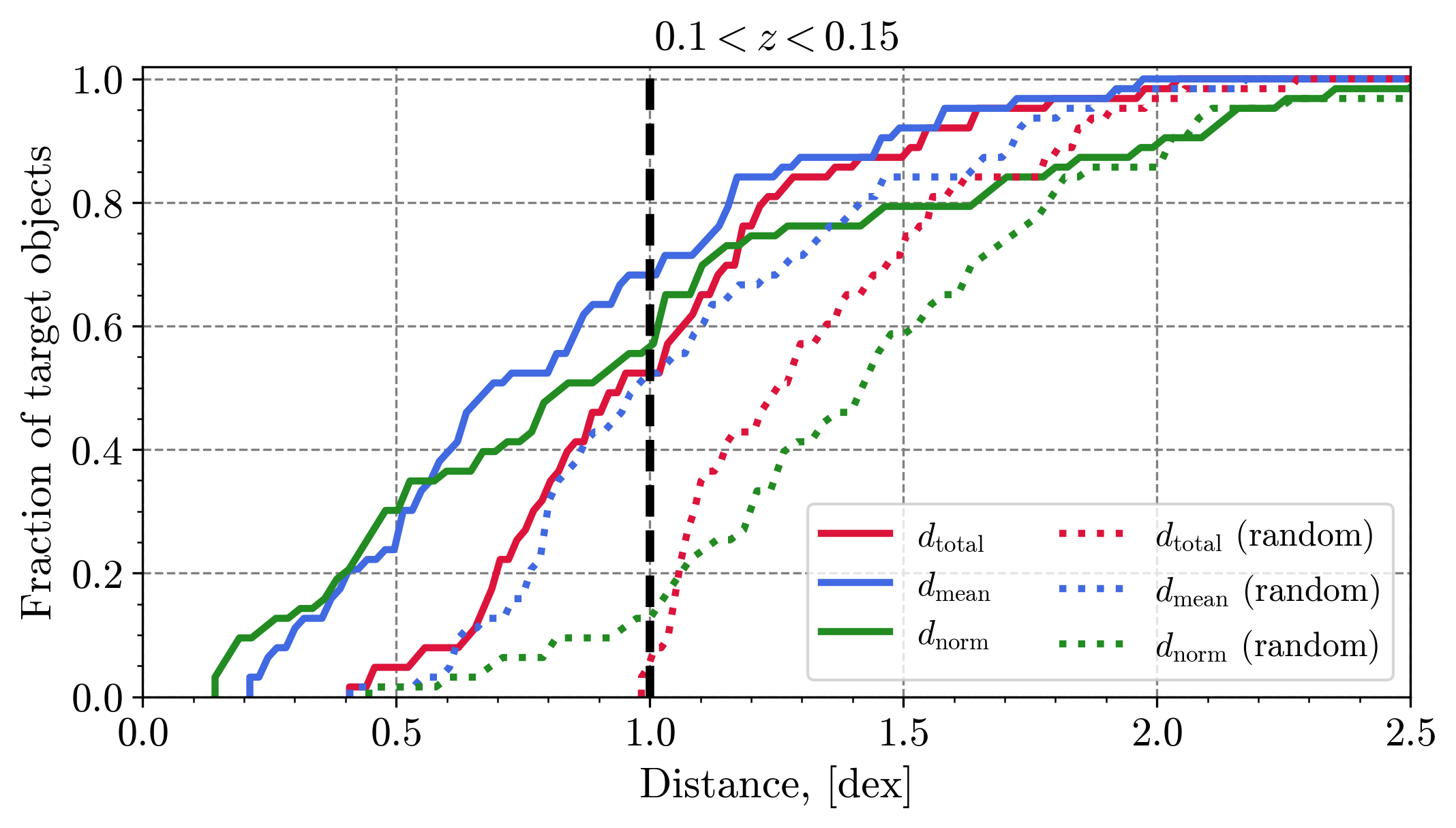}
    \includegraphics[width=.44\textwidth]{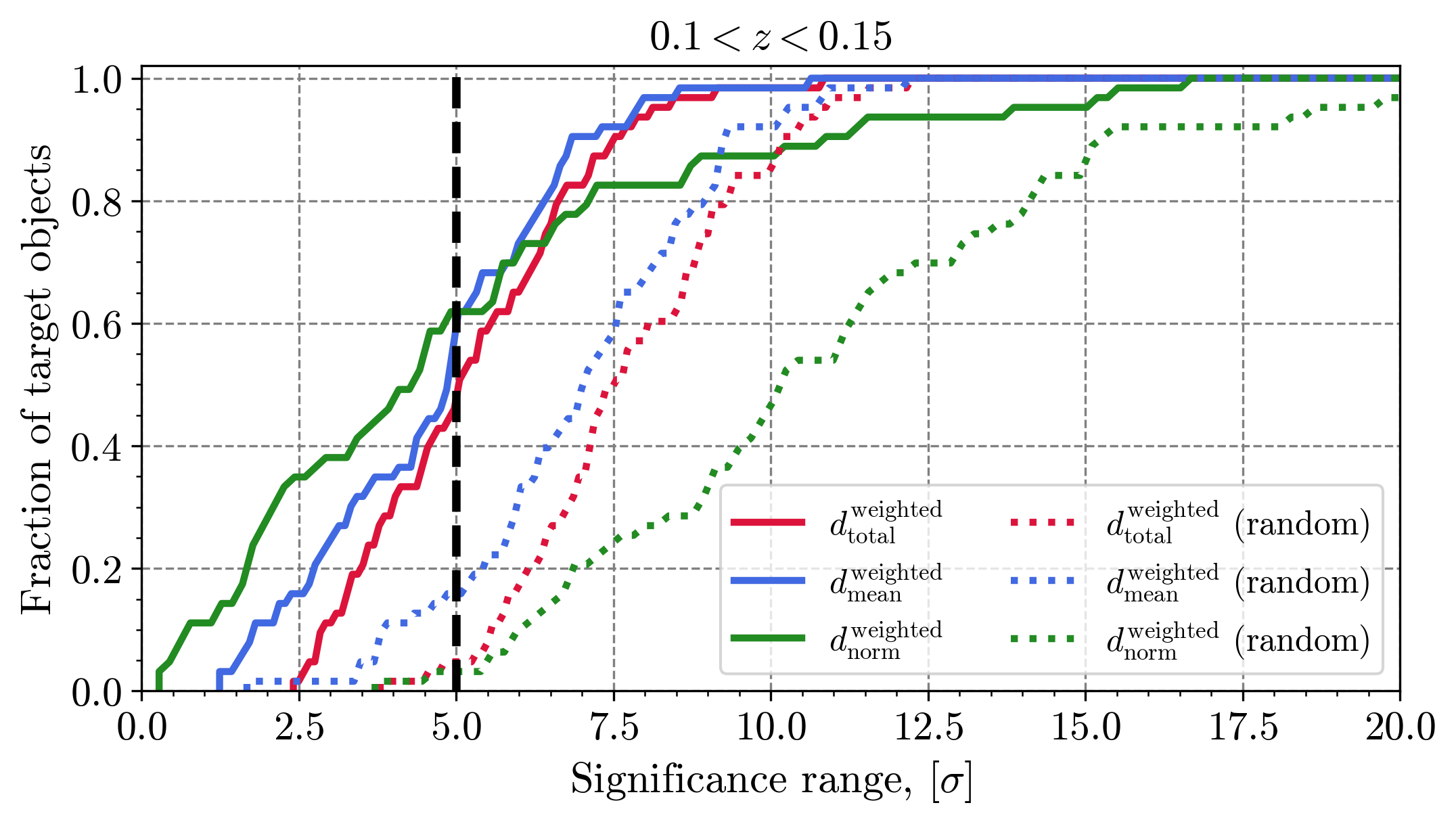}\\

    \caption{The same as in Fig.\,\ref{fig:hist-dist0} for target objects within three redshift ranges.}
    \label{fig:hist-dist3}
\end{figure*}

\begin{figure*}
    \centering
    \includegraphics[width=.45\textwidth]{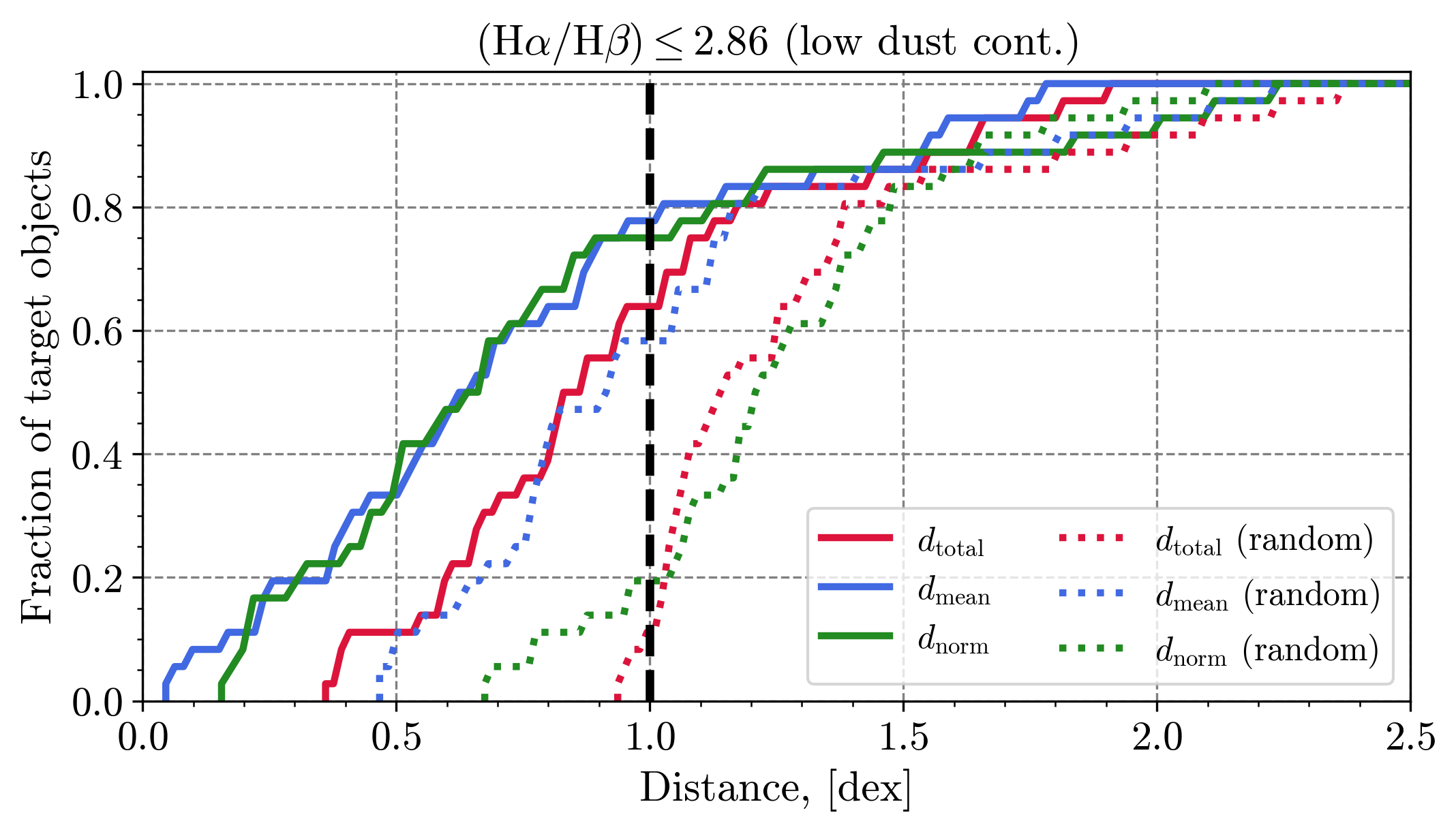}
    \includegraphics[width=.45\textwidth]{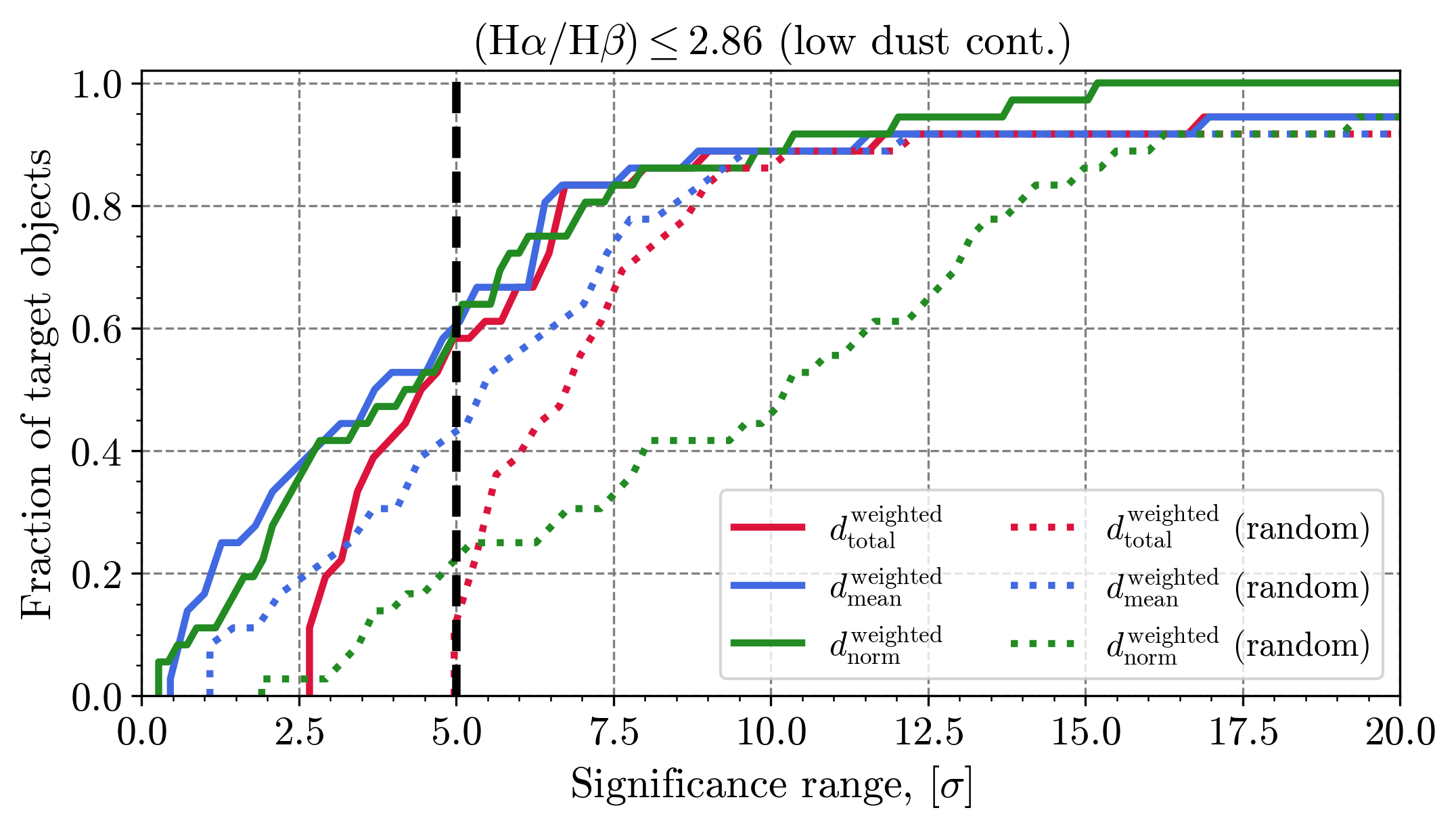}\\

    \includegraphics[width=.45\textwidth]{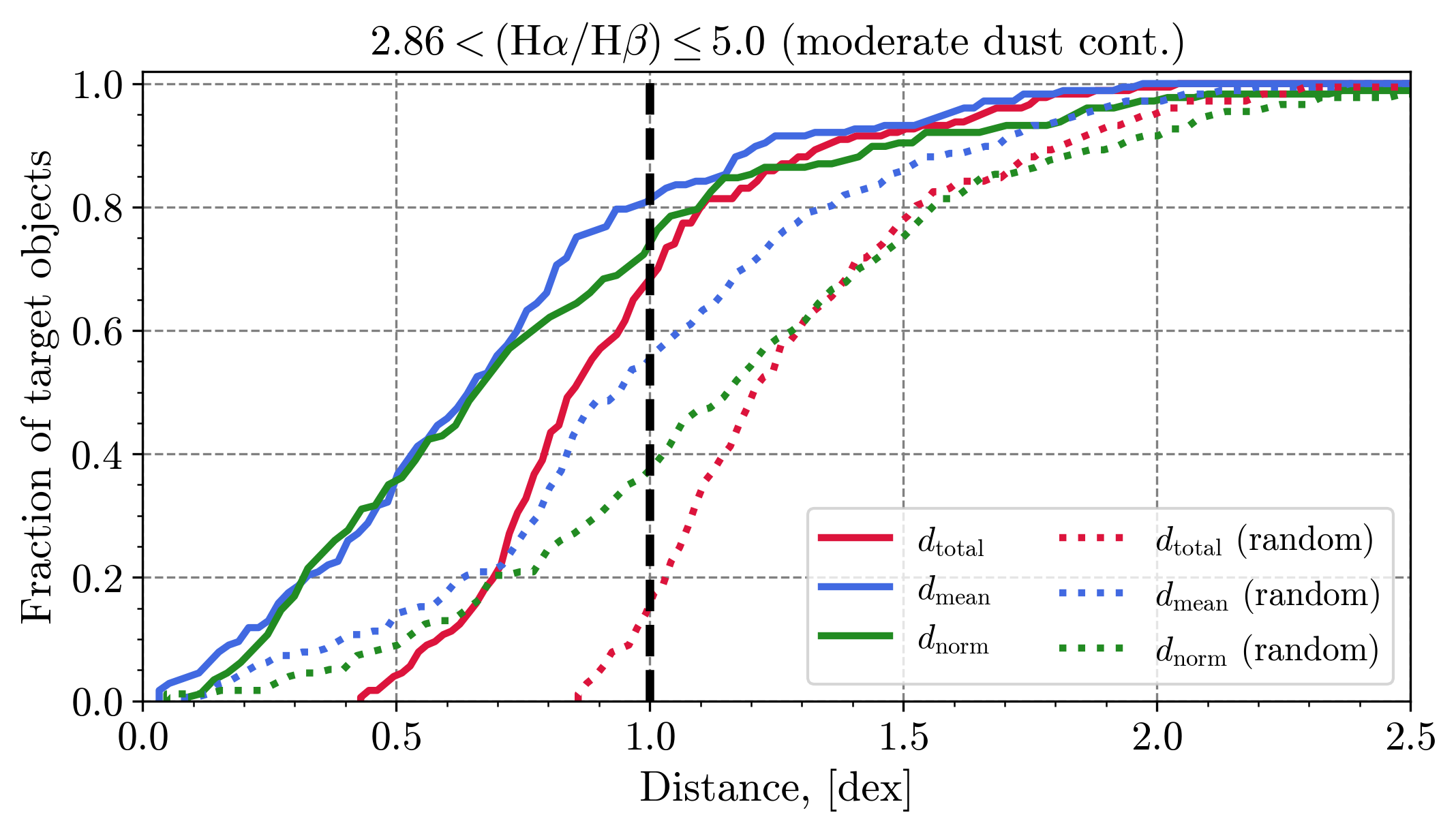}
    \includegraphics[width=.45\textwidth]{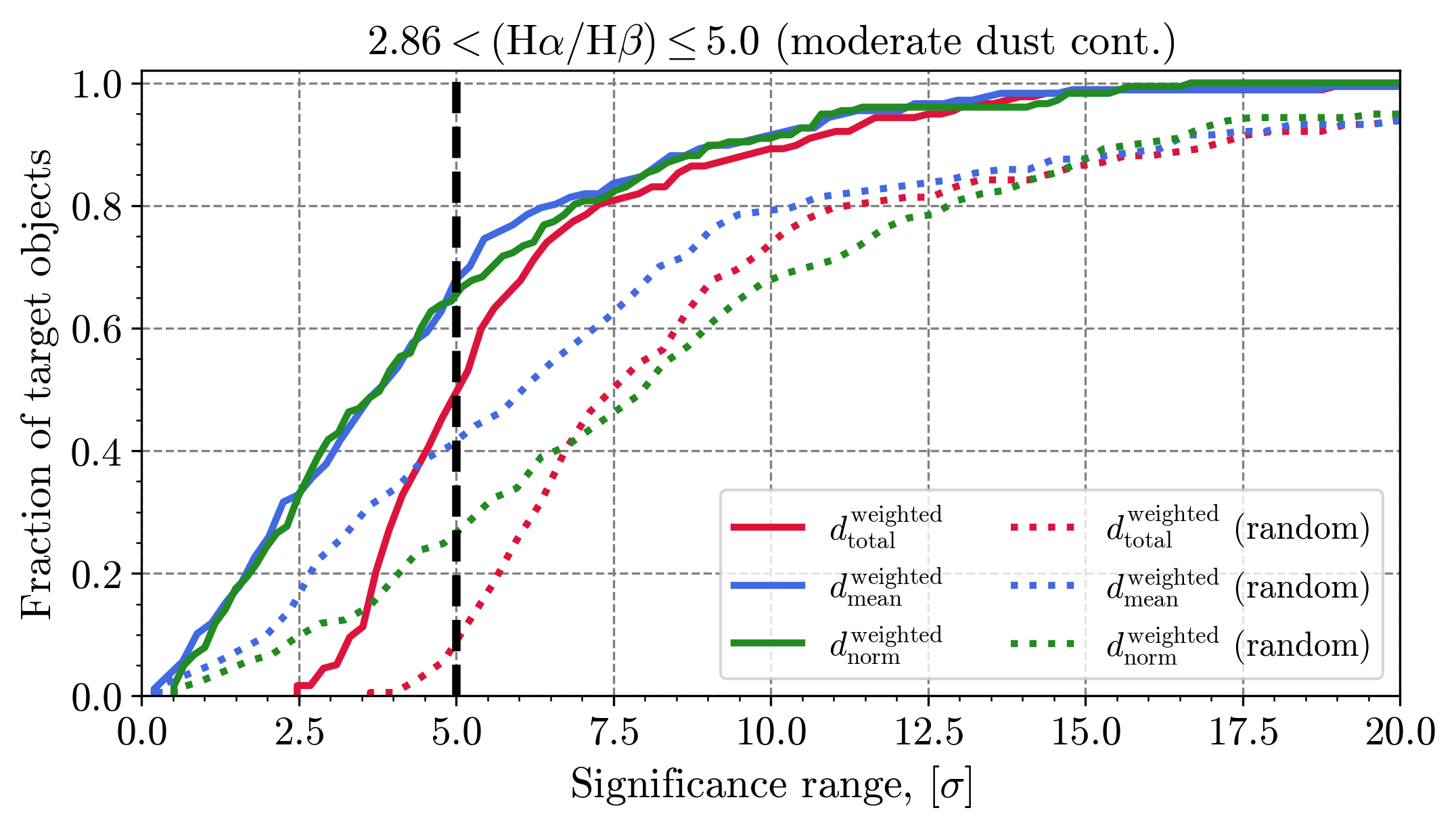}\\

    \includegraphics[width=.45\textwidth]{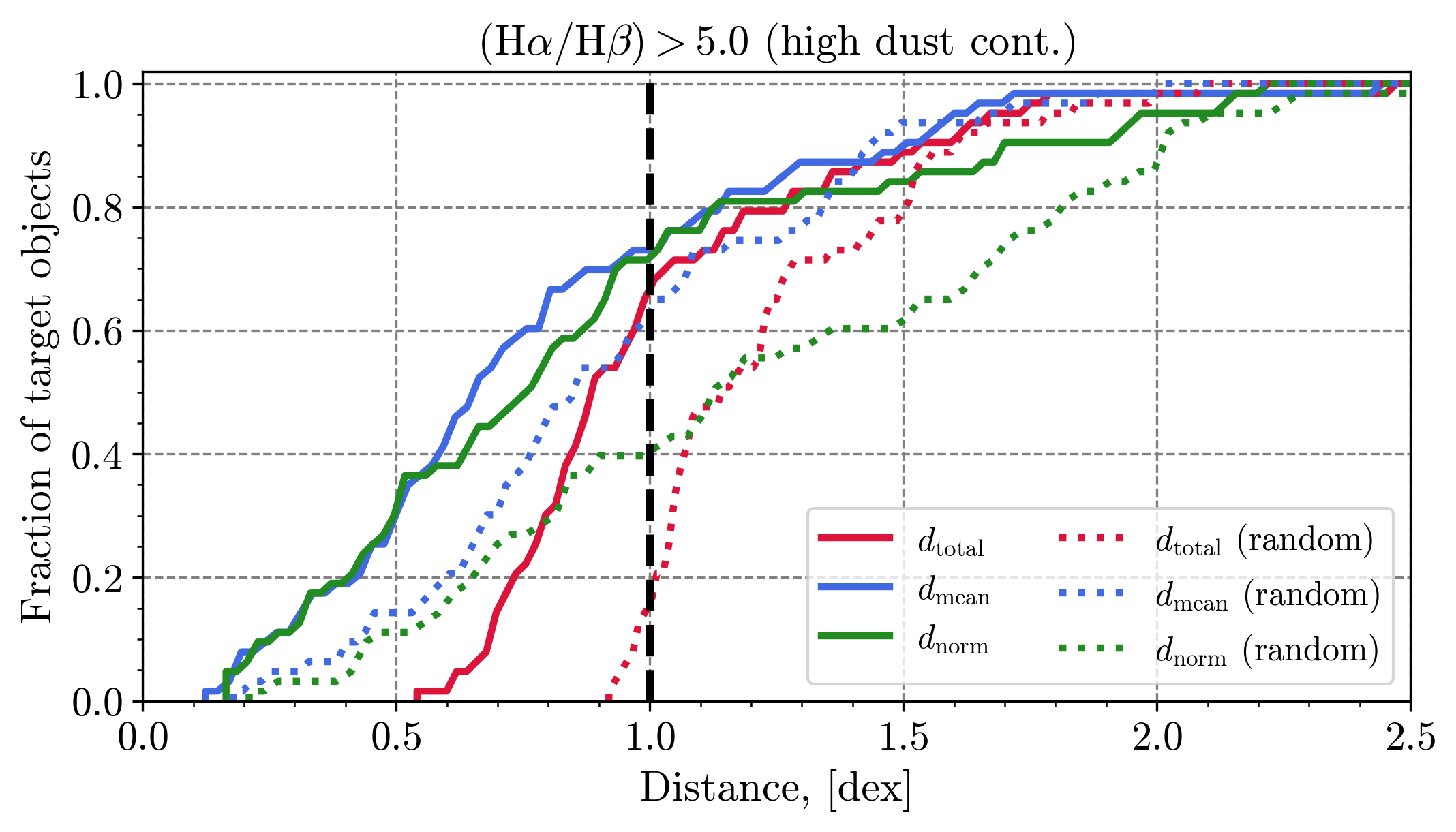}
    \includegraphics[width=.45\textwidth]{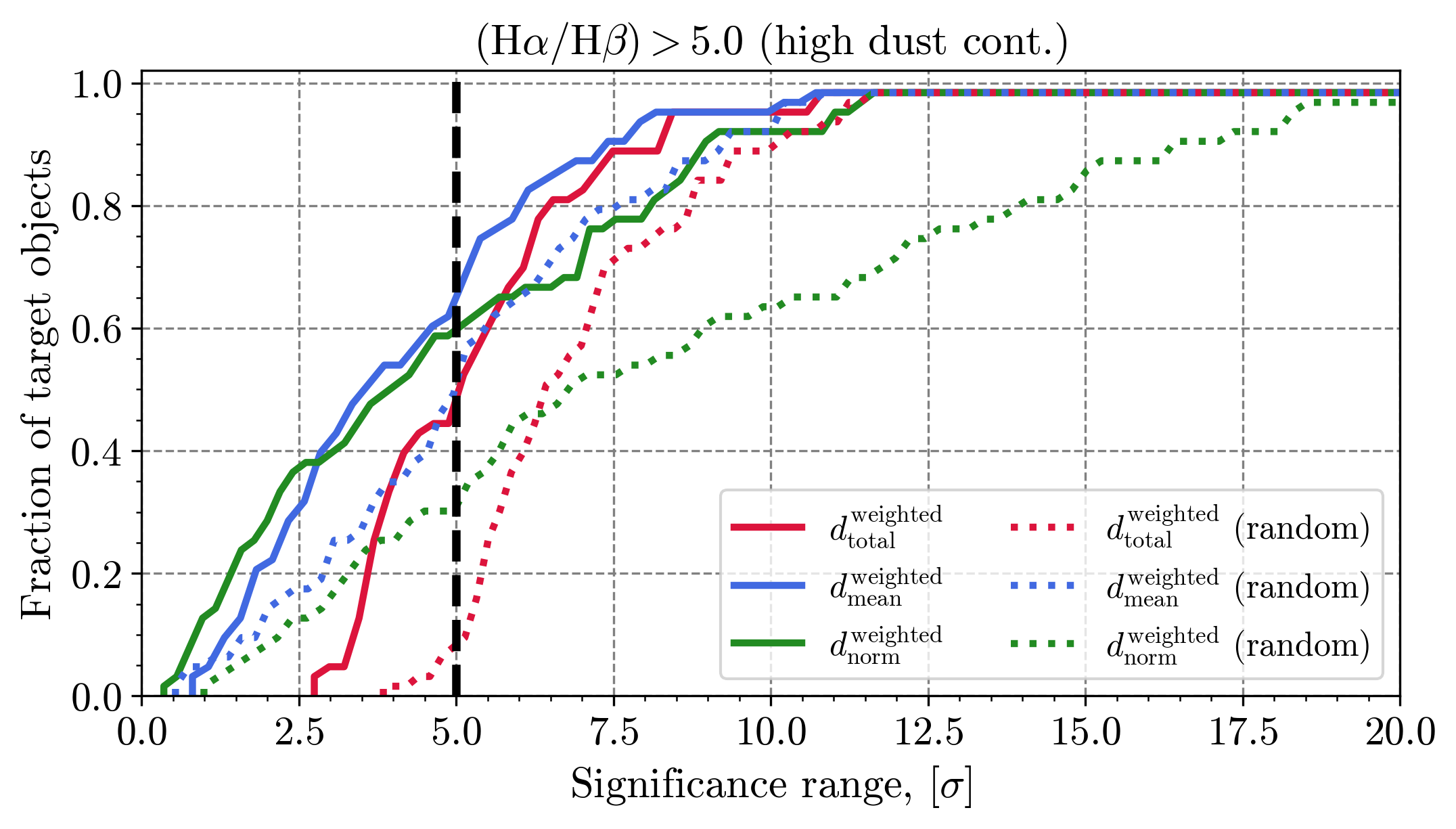}\\

    \includegraphics[width=.45\textwidth]{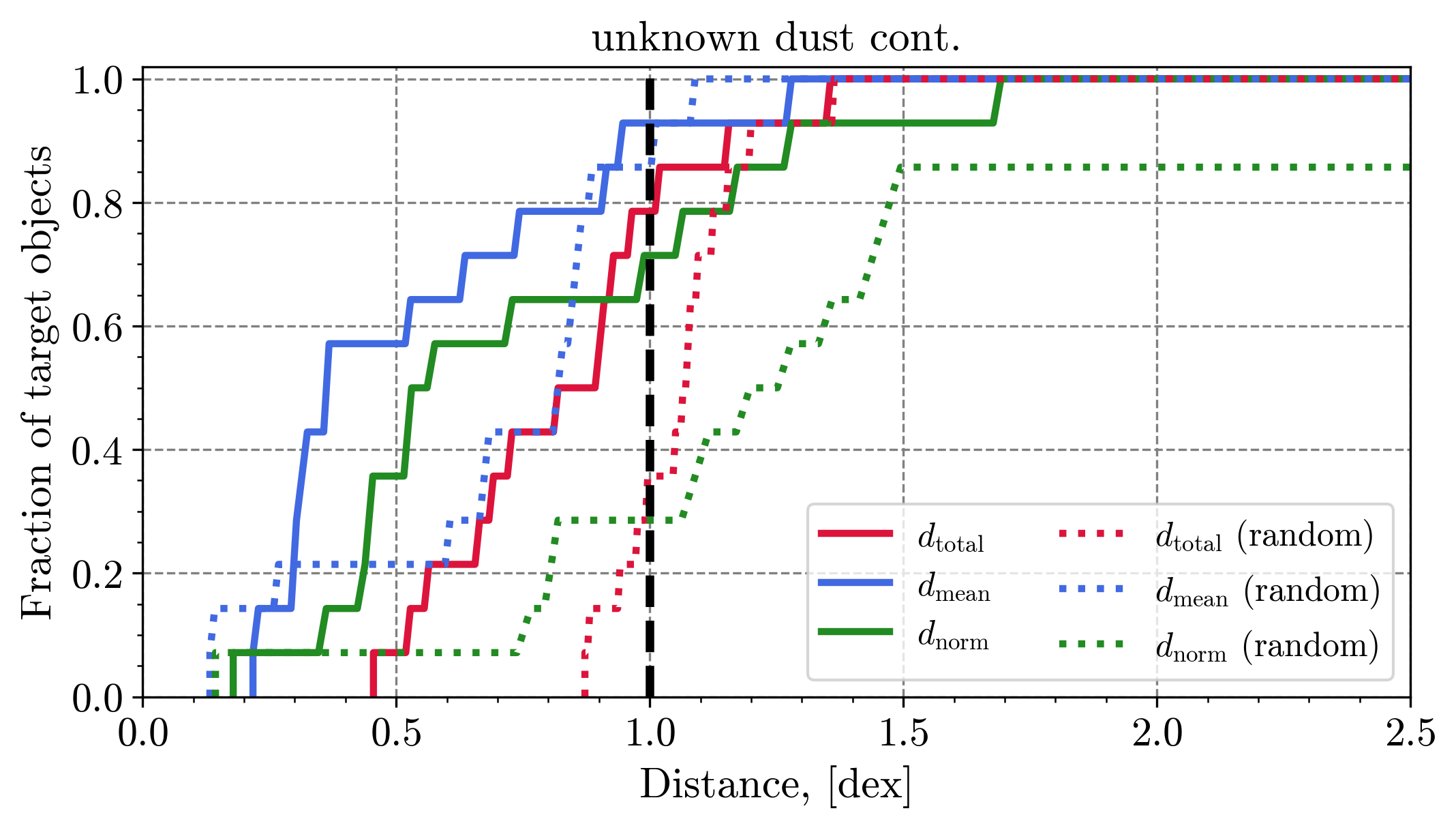}
    \includegraphics[width=.45\textwidth]{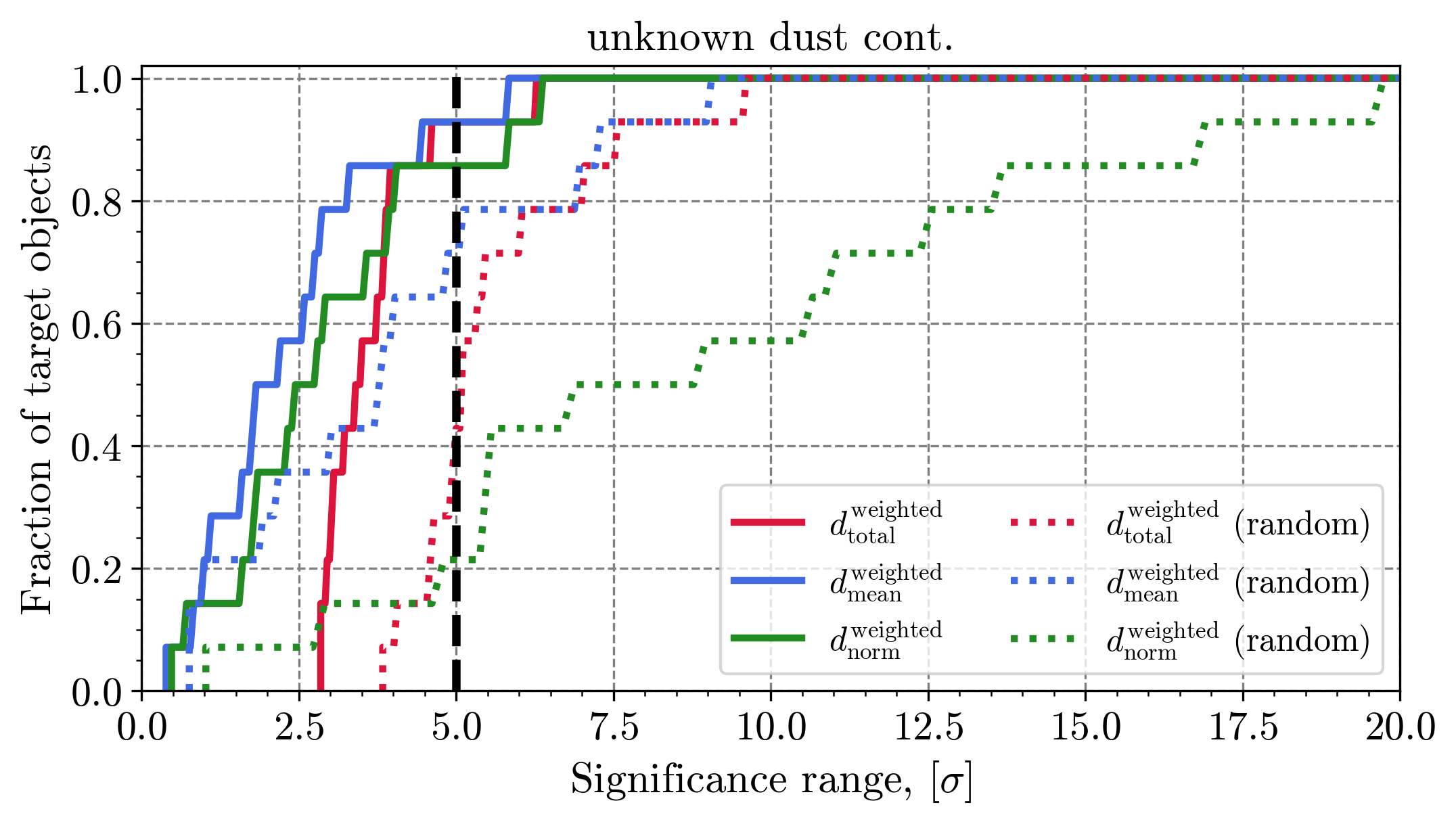}\\

    \caption{ The same as in Fig.\,\ref{fig:hist-dist0} for target objects with low (${\rm H}\alpha/{\rm H}\beta \leq 2.86$), moderate ($2.86 < {\rm H}\alpha/{\rm H}\beta \leq 5.0$), high (${\rm H}\alpha/{\rm H}\beta > 5.0$) and unknown dust content. }
    \label{fig:hist-dist4}
\end{figure*}

\clearpage
\section{The distribution of the retrieved neighbours in the SFR (or $\mathcal{M}_{\ast}$) space with respect to the target objects.}\label{sec:appendix2}
\begin{figure*}[bh!]
    \centering
    \includegraphics[width=.47\textwidth]{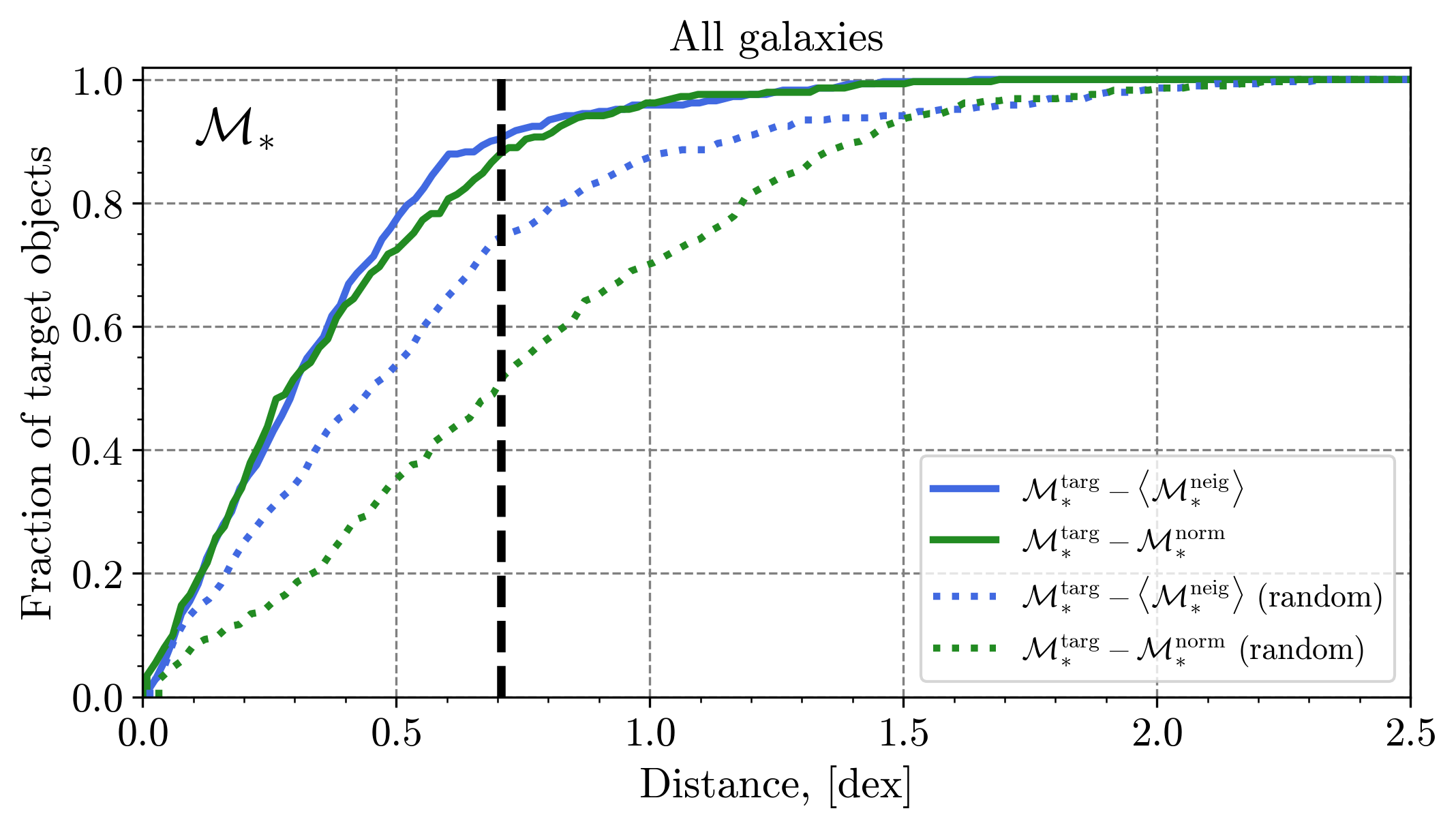}
    \includegraphics[width=.47\textwidth]{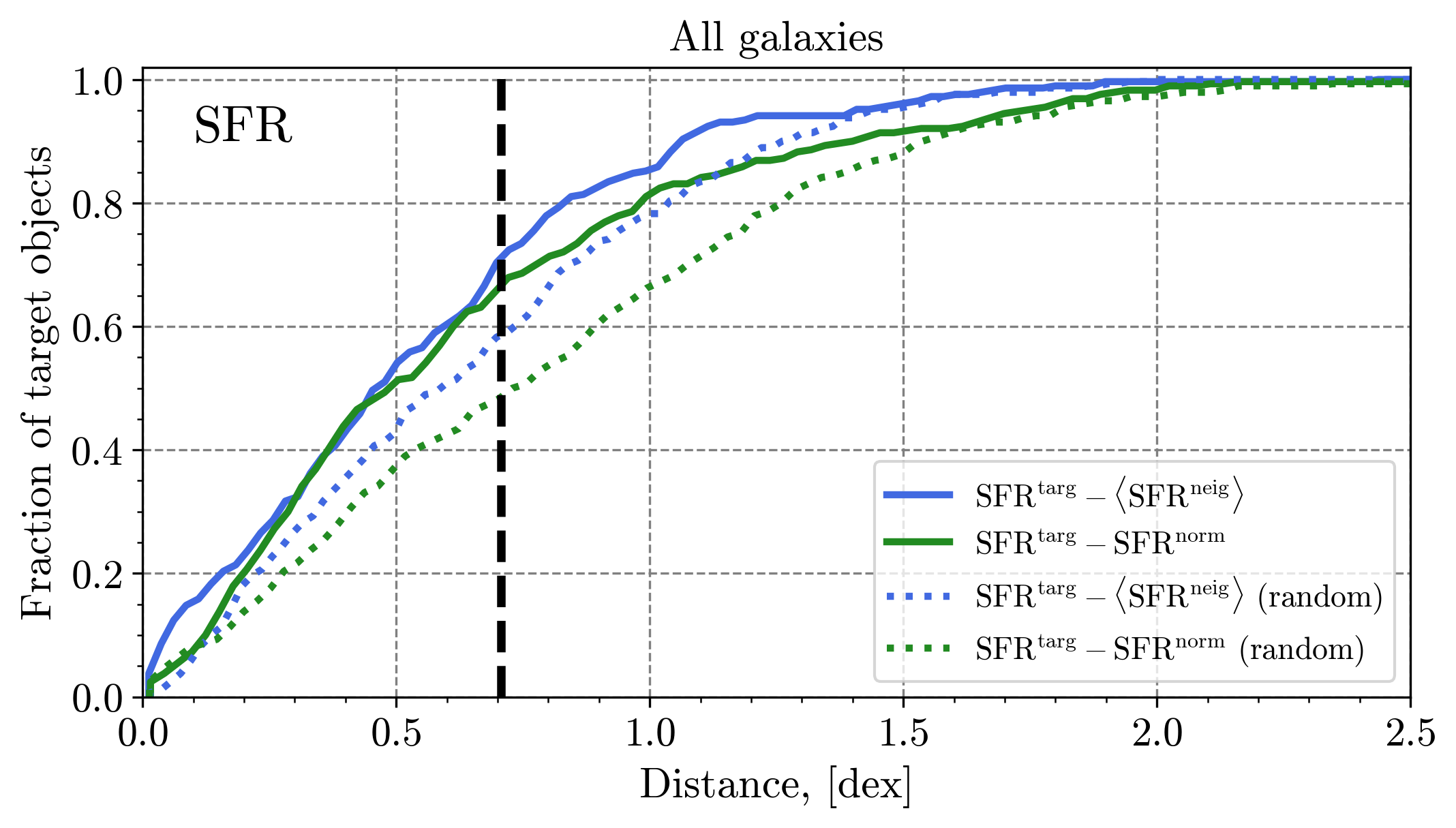}\\
    
    \includegraphics[width=.47\textwidth]{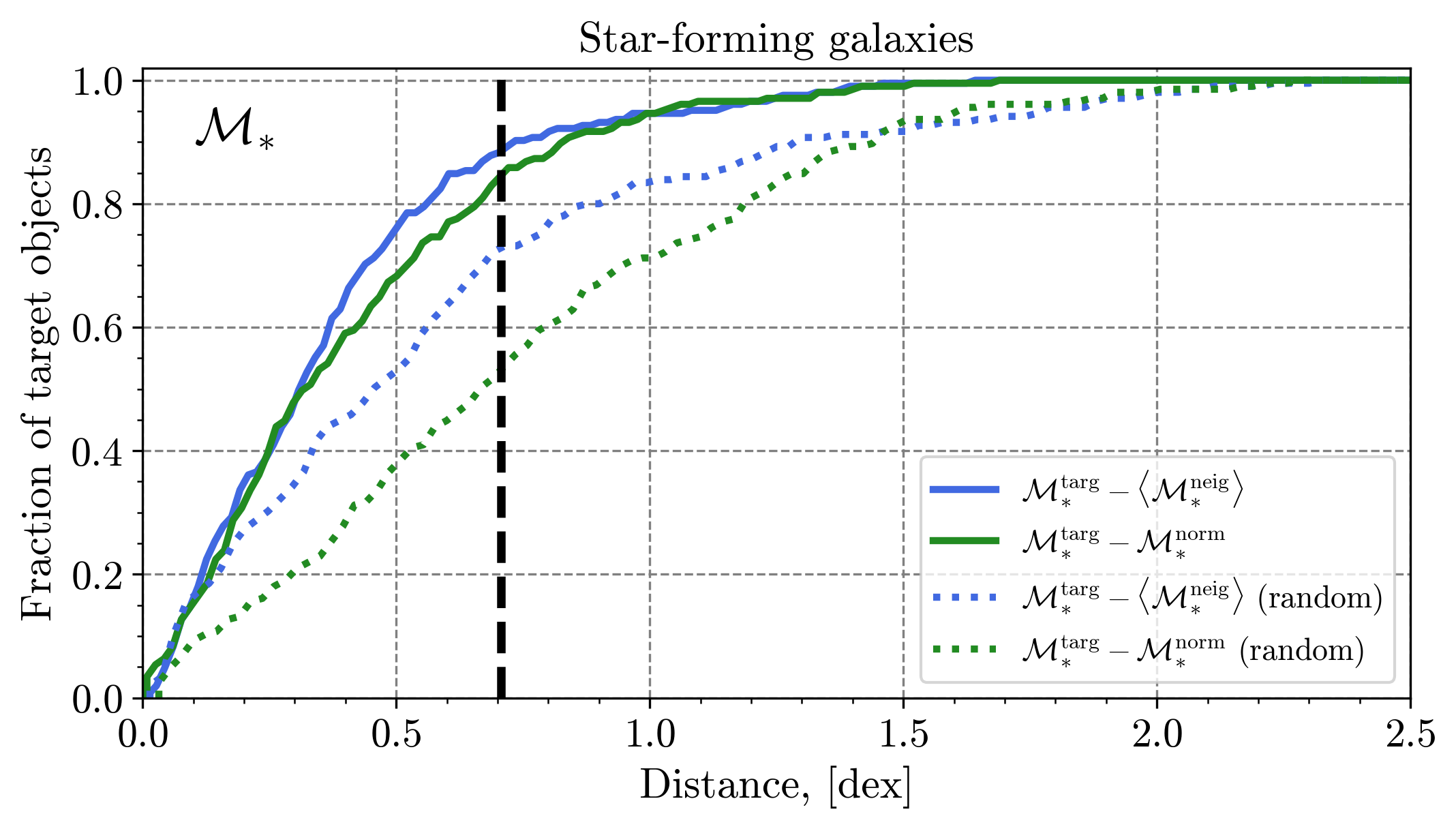}
    \includegraphics[width=.47\textwidth]{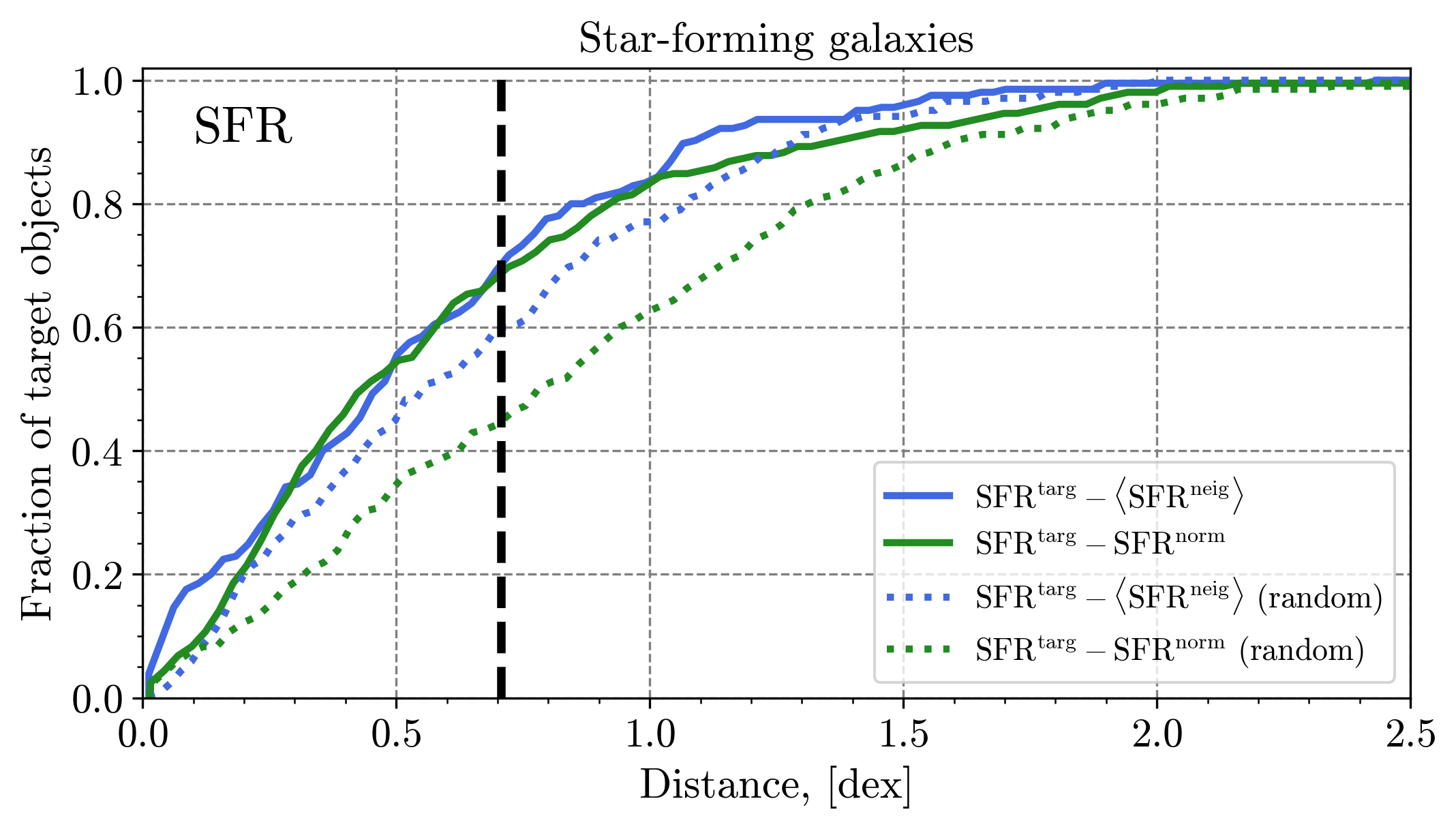}\\

    \includegraphics[width=.47\textwidth]{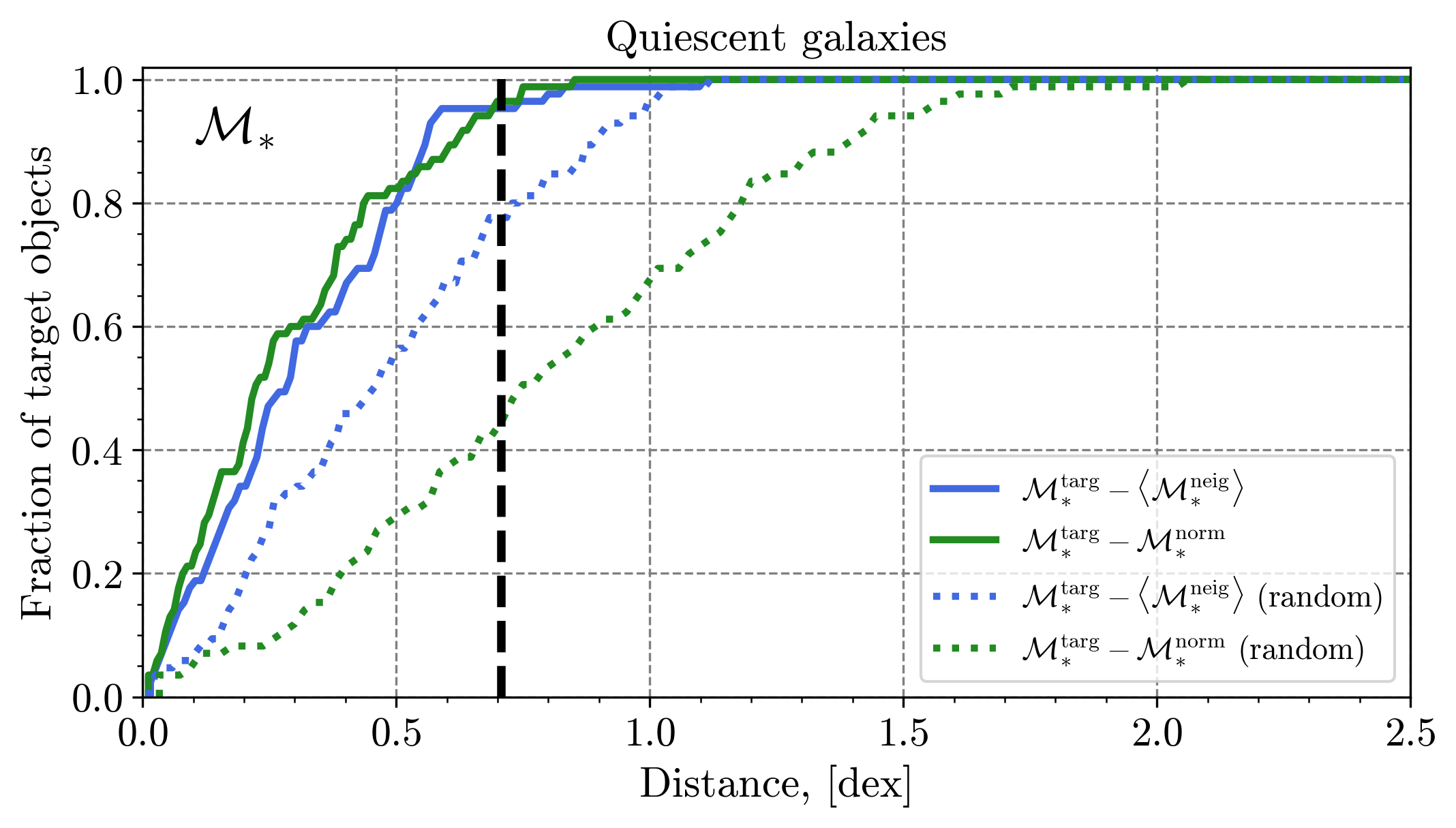}
    \includegraphics[width=.47\textwidth]{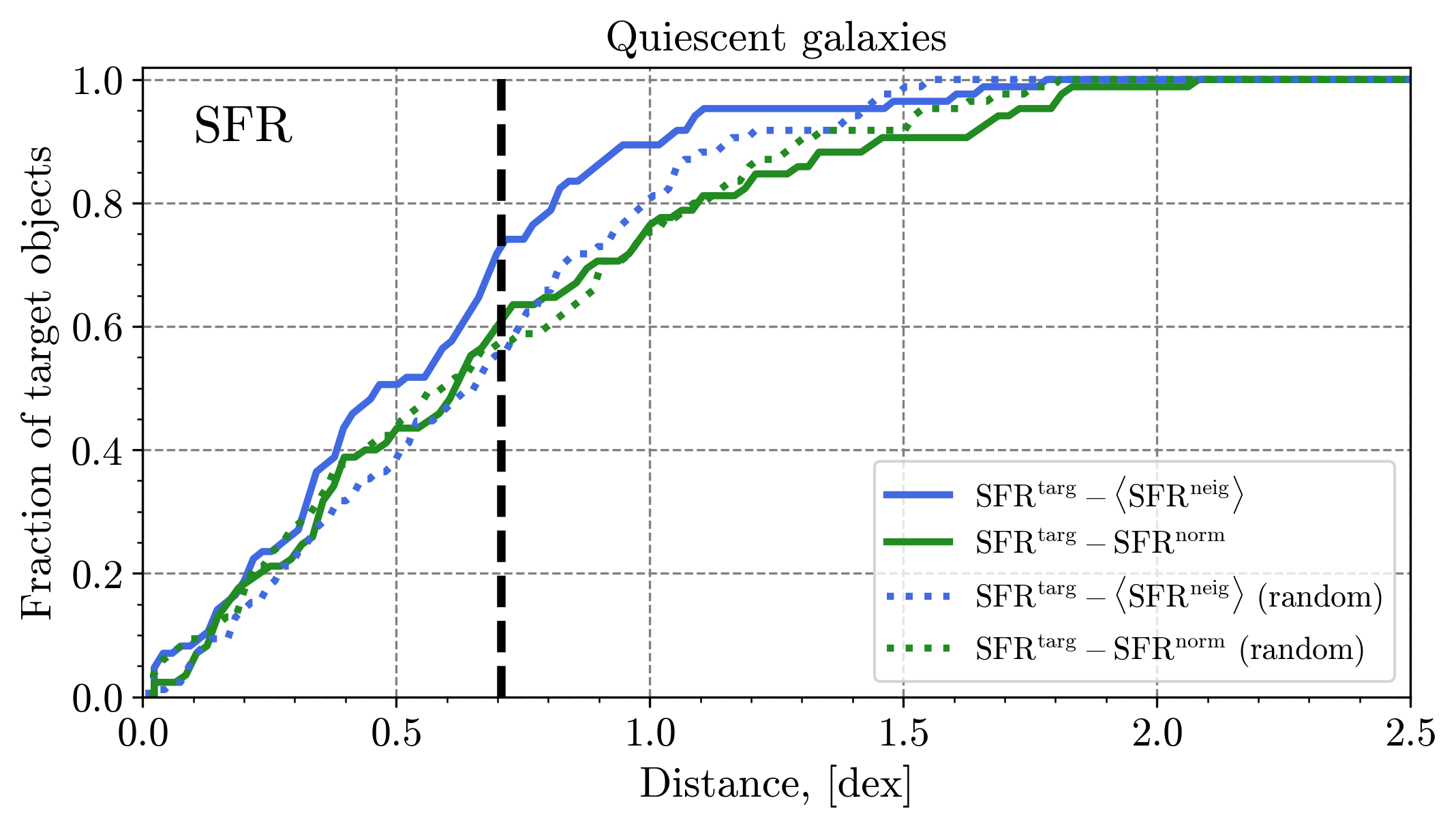}\\

    \caption{{\it Left panel:} The distributions of distances between the mean and the normalised mean $\mathcal{M}_{\ast}$ for the retrieved neighbours and the target object, i.e. $\mathcal{M}_{\ast}^{\rm targ} - \langle\mathcal{M}_{\ast}^{\rm neig}\rangle$ (by blue colour) and $\mathcal{M}_{\ast}^{\rm targ} - \mathcal{M}_{\ast}^{\rm norm}$ (by green colour), for {\it all, star-forming} and {\it quiescent} subsamples of target objects. The black dashed line represents the cut at 1\,dex\;$/\sqrt{2}$ used to define \ULISSE efficiency (see description in Section\,\ref{sec:results-histo}). {\it Right panel:} The same as on the left panel, but for ${\rm SFR}^{\rm targ} - \langle{\rm SFR}^{\rm neig}\rangle$ and ${\rm SFR}^{\rm targ} - {\rm SFR}^{\rm norm}$.} \label{fig:hist-dist-M-SFR}
\end{figure*}

\clearpage
\section{Statistical properties of neighbour distributions retrieved by \ULISSE}\label{sec:appendix3}
\begin{figure*}[bh!]
    \centering
    \includegraphics[width=.8\textwidth]{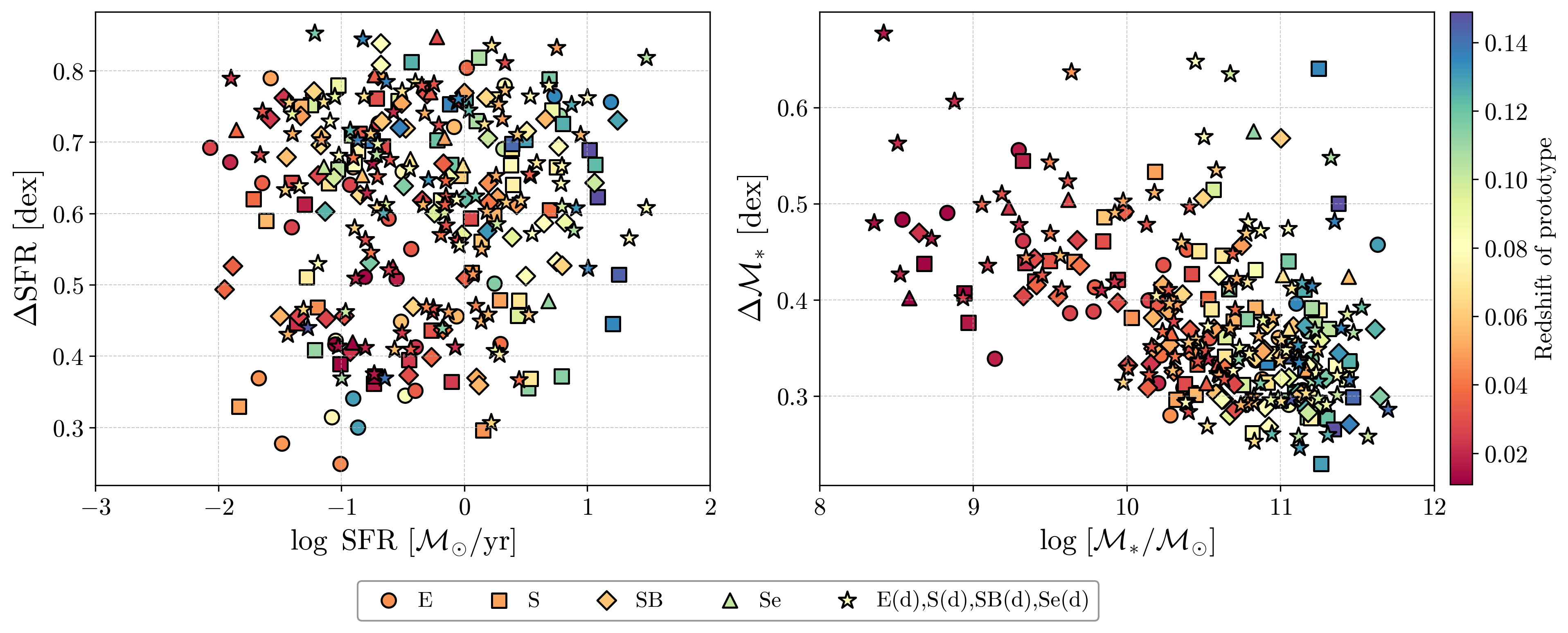}
    \caption{ { Scatter of retrieved neighbours in SFR ($\Delta\mathrm{SFR}$, \textit{left panel}) and $\mathcal{M}_{\ast}$ ($\Delta\mathcal{M}_\ast$, \textit{right panel}) as a function of the target properties for all 290 target objects analysed in this work. Marker shapes indicate different morphological classes based on the GZ2 catalogue, and the colour scale reflects the redshift of the target object. }}
    \label{fig:dev_Mass_SFR}
\end{figure*} 

\vspace*{-0.3cm}
\begin{figure*}[h!]
    \centering
    \includegraphics[width=.4\textwidth]{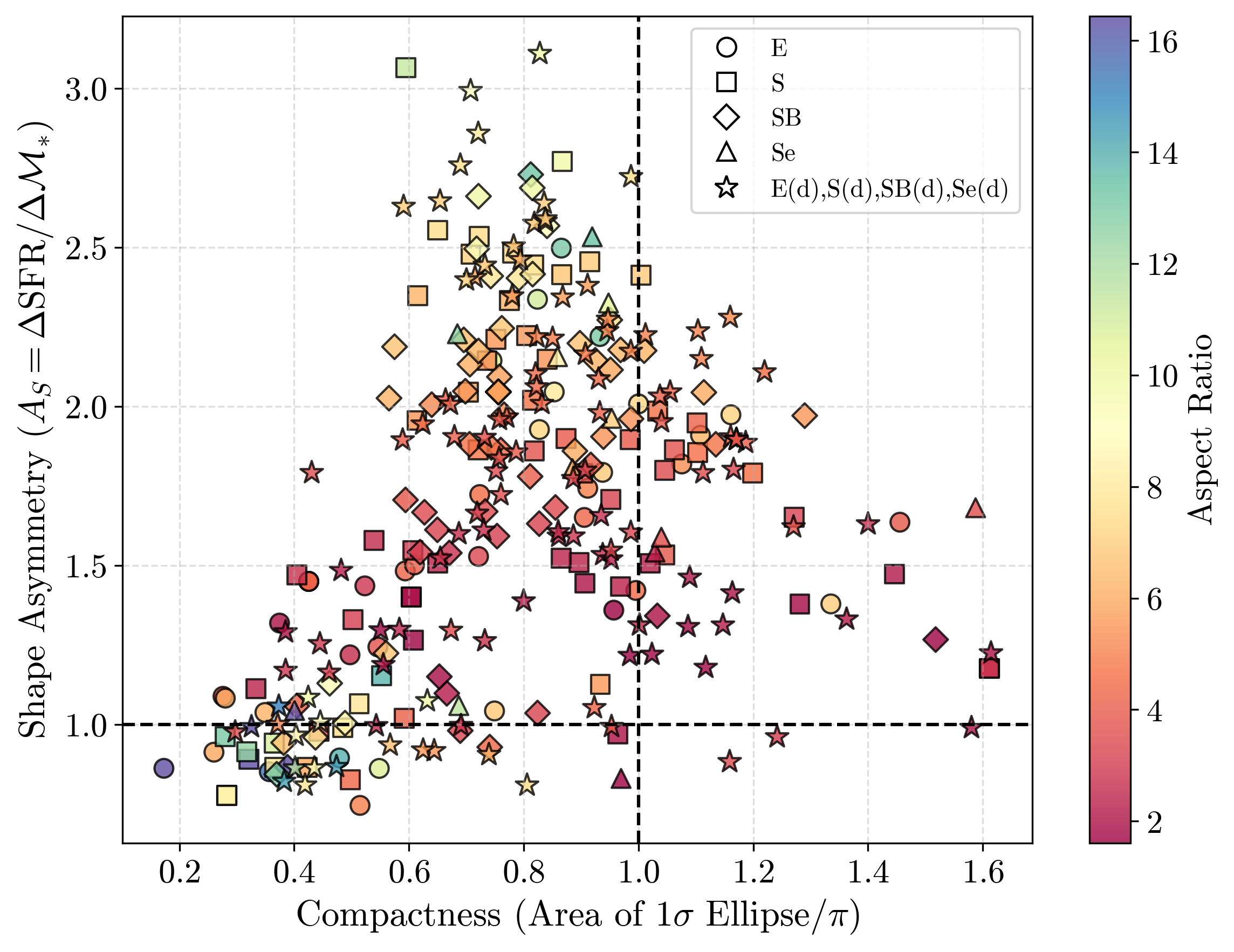}
    \includegraphics[width=.51\textwidth]{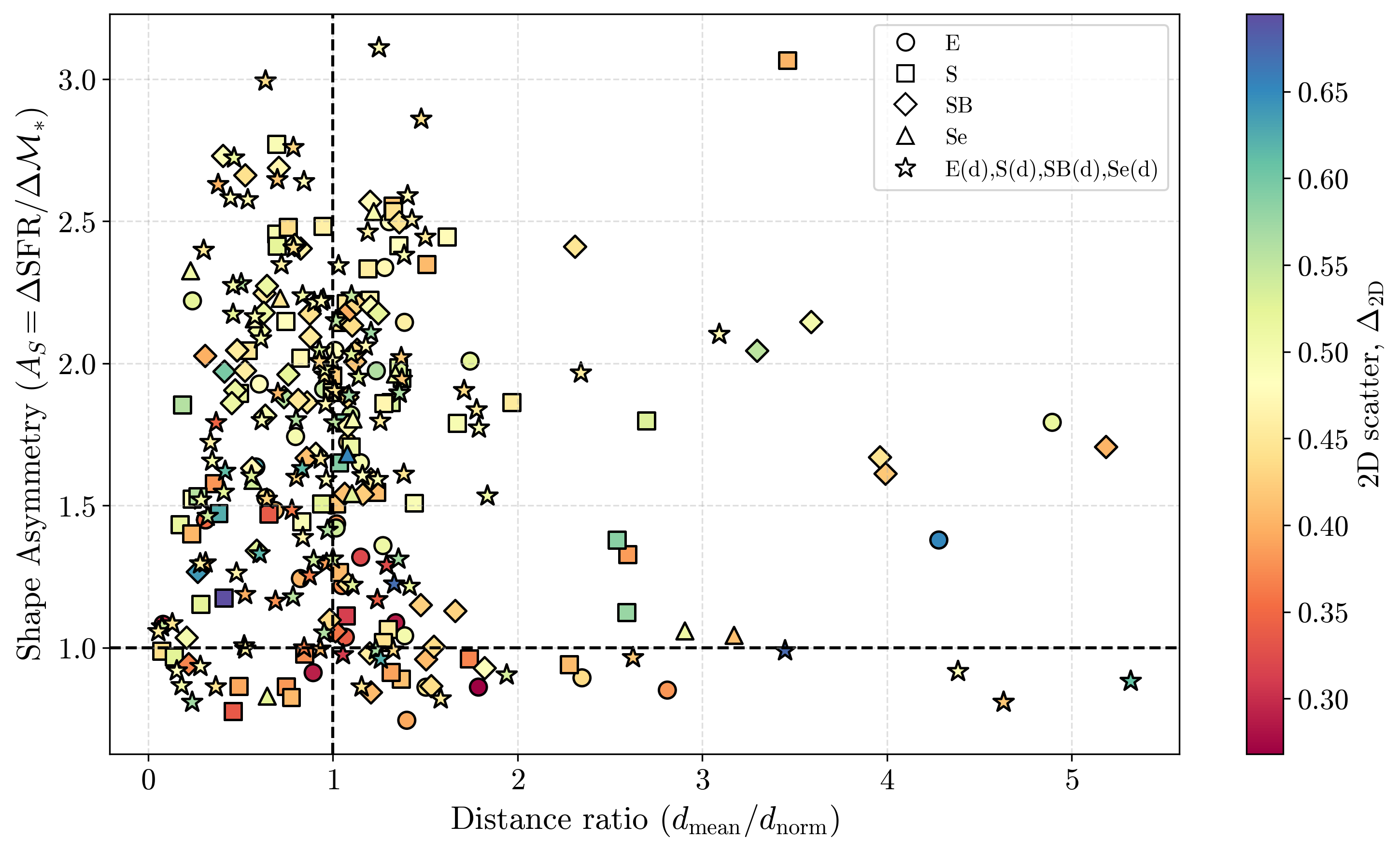}
    \caption{ { \textit{Left panel:} Shape asymmetry $A_{S}$ versus compactness $\mathcal{C}$ of the neighbour distributions retrieved by \ULISSE for all 290 target objects. The colour scale represents the aspect ratio $\mathcal{R}$ of each corresponding neighbour distribution. The black dashed lines corresponds to key values: compactness $\mathcal{C} = 1$ separates compact ($\mathcal{C} < 1$) from dispersed ($\mathcal{C} > 1$) distributions, while shape asymmetry $A_{S} = 1$ corresponds to equal scatter in $\mathcal{M}_{\ast}$ and SFR. Deviations from unity indicate dominant spread along either $\mathcal{M}_{\ast}$ ($A_{S} < 1$) or SFR ($A_{S} > 1$).
    \textit{Right panel:} Shape asymmetry $A_{S}$ versus the ratio between mean and normalised mean distances of the neighbour distributions. Colour scale shows the two-dimensional scatter $\Delta_{2D}$ of the neighbours. The black dashed line indicates $d_{\rm mean}/d_{\rm norm} = 1$, i.e., cases where the normalisation to the ‘rarity’ of the target object in the sample has no effect on the mean position of the retrieved neighbours. Values significantly above 1 reflect targets located in sparsely populated regions of parameter space, where $d_{\rm mean}$ is biased due to the limited availability of similar objects. Marker symbols in both panels indicate the GZ2 morphological class of each target galaxy.} }
    \label{fig:comp-As-R-morph}
\end{figure*} 

\vspace*{-0.3cm}
\begin{figure*}[h!]
    \centering
    \includegraphics[width=.8\textwidth]{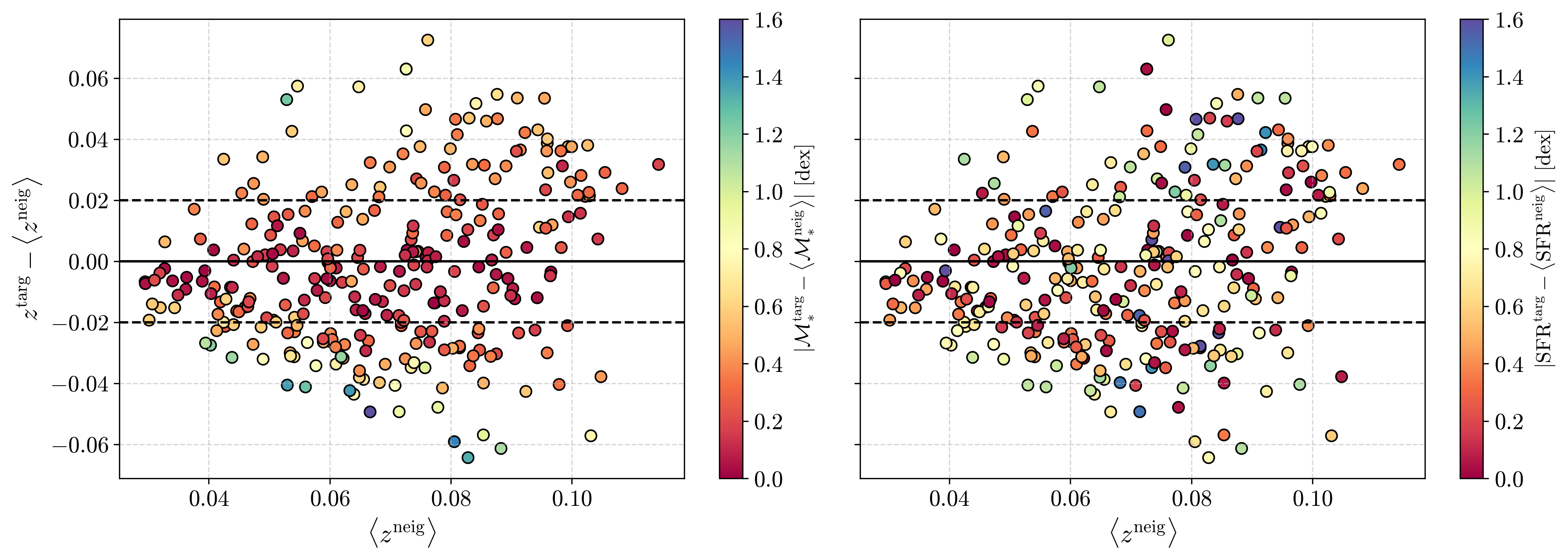}
    \caption{ { Relative redshift offset $z^{\rm targ} - \langle z^{\rm neig} \rangle$ as a function of the average redshift of retrieved neighbours. The colour shows the difference between the target and averaged values for neighbours in $\mathcal{M}_{\ast}$ (\textit{left panel}) and SFR (\textit{right panel}), respectively.}}
    \label{fig:mean-Z}
\end{figure*}

\end{appendix}


\end{document}